\documentclass[conference,11pt]{IEEEtran}

\usepackage[english]{babel}
\usepackage[utf8x]{inputenc}

\usepackage{graphicx}
\graphicspath{{./graphics/}}

\usepackage{listings}

\usepackage{hyperref}

\usepackage[cmex10]{amsmath}
\usepackage{booktabs}
\usepackage{lscape} 
\usepackage{longtable}
\usepackage{multirow}
\usepackage{rotating}
\usepackage[table]{xcolor}

\usepackage{cite}
\usepackage{balance}

\usepackage{amssymb}

\title{A Survey on Honeypot Software and Data Analysis}
\author{
\IEEEauthorblockN{Marcin Nawrocki\IEEEauthorrefmark{1}, Matthias W\"ahlisch\IEEEauthorrefmark{1}, Thomas C. Schmidt\IEEEauthorrefmark{2}, Christian Keil\IEEEauthorrefmark{3}, Jochen Sch\"onfelder\IEEEauthorrefmark{3}}
\IEEEauthorblockA{\IEEEauthorrefmark{1} Freie Universit\"at Berlin, Berlin,
Germany, \\ Email: \{marcin.nawrocki, m.waehlisch\}@fu-berlin.de}
\IEEEauthorblockA{\IEEEauthorrefmark{2} HAW Hamburg, Hamburg, Germany,
\\ Email: t.schmidt@haw-hamburg.de}
\IEEEauthorblockA{\IEEEauthorrefmark{3} DFN-CERT Services GmbH, Hamburg, Germany,
\\ Email: \{keil, schoenfelder\}@dfn-cert.de}
}

\usepackage{pifont}
\newcommand{\cmark}{\ding{51}}%
\newcommand{\xmark}{\ding{56}}%

\usepackage{xspace}
\newcommand{\eg}{\textit{e.g.}~}

\newcommand{\etc}{\textit{etc.}~}
\newcommand{\one}{({\em i})\xspace}
\newcommand{\two}{({\em ii})\xspace}
\newcommand{\three}{({\em iii})\xspace}

\newcommand{\colLow}{\cellcolor[HTML]{F5F5F5}}
\newcommand{\colHigh}{\cellcolor[HTML]{FAFAFA}}


\begin{document}

\maketitle

\begin{abstract}
In this survey, we give an extensive overview on honeypots.
This includes not only honeypot software but also methodologies to analyse honeypot data.
\end{abstract}

\section{Introduction}

Effective network security administration depends to a great extent on the
understanding of existing and emerging threats on the Internet. In order to
protect information systems and its users it is of crucial importance to collect
accurate, concise, high-quality information about malicious activities
\cite{yegneswaran2005using, leurre2008long}. The fact that cyber attacks are a
present threat is confirmed by recent statistics such as the Symantec Internet
Security Threat Report \cite{symantecReport} or the attacks report by 
ATLAS \cite{AtlasAttacks}. The discovery of vulnerabilities
such as Heartbleed, ShellShock, and Poodle, and 
their wide-spread prevalence across a number of operating systems draw the public
attention to system security. As observed with Heartbleed, attackers were much
faster in exploiting the vulnerabilities than vendors could create and roll out
patches. Relying only upon tradition lines of defence such as Intrusion
Detection Systems (IDSs) and dynamic firewalls alone does not provide a holistic
coverage on detecting novel and emerging patterns of attacks
\cite{vasilomanolakis2015taxonomy}. Honeypots are decoy computer resources whose
value lies in being probed, attacked or compromised
\cite{Spitzner:TrappingHackers:DefHoneypot}. They complement the traditional
detection mechanisms \cite{vasilomanolakis2015honeypot}. Honeypots are able to
spot zero-day attacks and give insights on attackers actions and motivation.

The field of honeypot research consists of two main pillars: a) the development
of honeypot software and its effective deployment b) the analysis of the 
acquired log data in a structured manner.

Honeypot surveys and software comparisons have been presented before, however an
up-to-date comparison and classification of honeypot software 
does not exist according to our current knowledge. Moreover, none of the 
surveys focus on creating an overview of related data analysis techniques.
Based on questions imposed by the data analysis, a practice-oriented review
of these techniques will be presented. Eventually, the question  will be 
elaborated which honeypot should be deployed and how the log records
should be analysed.

First surveys in the field of honeypot research presented in 2003 include only a
small subset of meanwhile available software and are by this time outdated
\cite{pouget2003white}. Unfortunately, papers tend to only discuss a small
subset of honeypot software or mention examples which denies a holistic view
\cite{ikinci2008monkey, AccPointHP2009Survey, almutairi2012survey,
Kambow2014:IndianSurvey3}. Mairh et al. \cite{mairh2011honeypot} presented in
2011 a honeypot survey, which illustrates the different types of honeypots and
suggests to use honeypots in educational environments. Deployment of honeypots 
is only done by the example of HoneyD. In 2012, Bringer~\cite{bringer2012survey} published the by far most exhaustive survey on recent
advances in the field of honeypots by ordering a large amount of honeypot
related papers by contentual category. However, it rather only considers new
types of honeypot software and misses the historical development of honeypot software.
Also the software release date and maintenance time spans remain unanswered, which are
an indicator for the current deployability.

\begin{figure}
  \includegraphics[width=1.0\columnwidth]{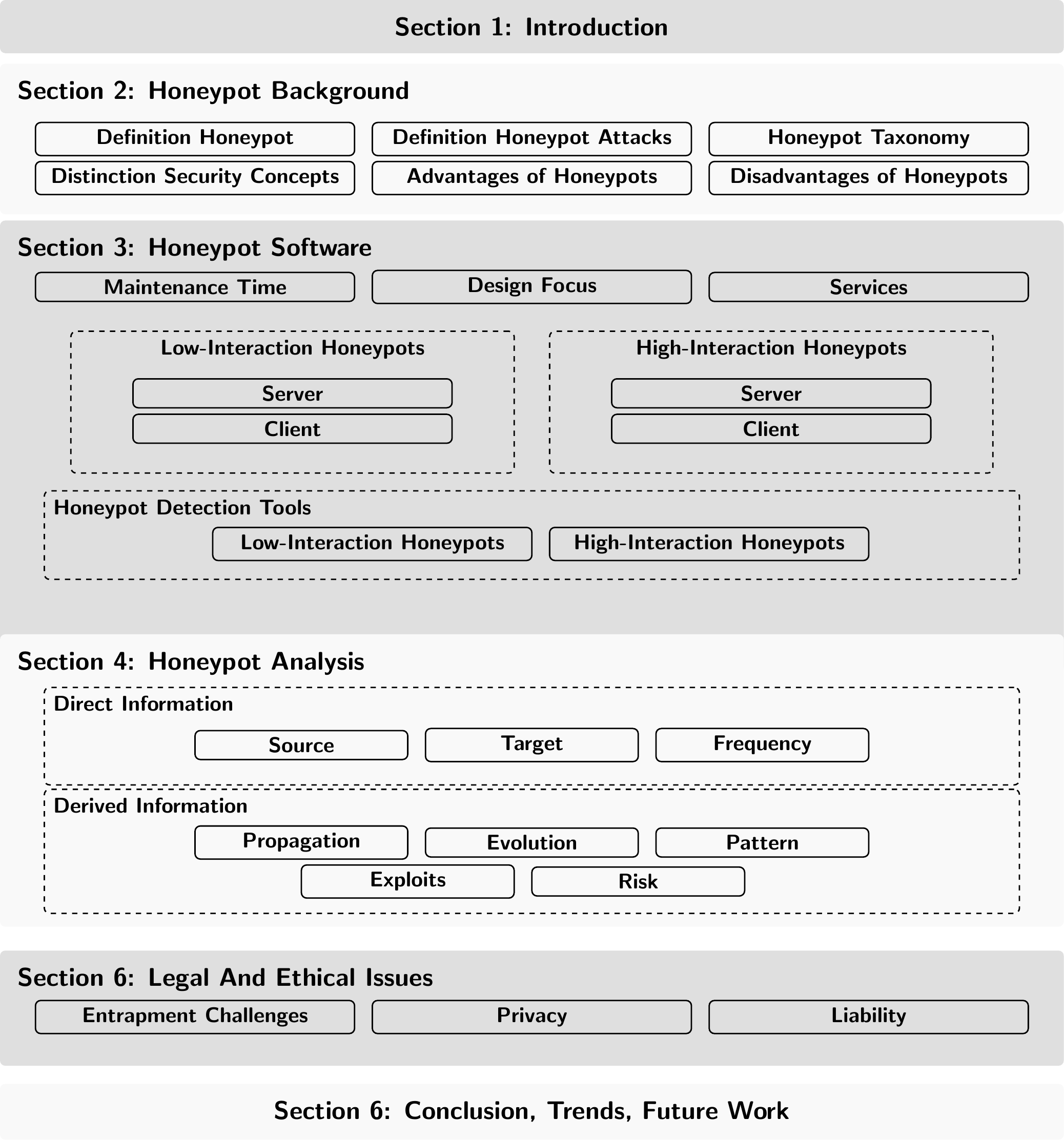}
  \caption{Categorisation of honeypot literature.}
  \label{fig:paperoutline}
\end{figure}

In this survey, we give an extensive overview about honeypot software as well as methodologies to analyze honeypot data.
We briefly discuss ethical and legal concerns.
Figure~\ref{fig:paperoutline} shows an overview about the specific topics of this paper, including guidance of the paper structure.

\section{Basic Background About Honeypots}
\label{sec:honeypotDefinition}


The first academic publication including explicitly the keyword honeypot dates back to the beginning of 2000~\cite{firstHPsansDef, SpitznerHoneypotPartOne}. 
One of the first broadly accepted formal definitions was introduced by Spitzner \cite{Spitzner:TrappingHackers:DefHoneypot}: \textit{A honeypot is decoy computer resource whose value lies in being probed,
attacked or compromised.}
However, the concept of honeypots is not new and was already used implicitly in the 1990s in the field of information protection and network defences \cite{Stoll:CuckoosEgg, Cohen:FirstHoneypot}, but was named differently, for example a (chroot) jail \cite{Cheswick:Berferd}.
Later, multiple honeypots have been interconnected, called honeynet.

At the beginning of the year 2000, highly effective Unix- and Windows-based worms spread exponentially \cite{pathan2014:wormHistory}.
Despite the extensive distribution, capturing and analysing this incident was fairly difficult as the worms only resided in the systems memory or masqueraded themselves.
One of the first honeypots dedicated to a known threat was used to analyse the sub7 malware.
The honeypot emulated a Windows system infected by the sub7 trojan by responding on port 27374, which many worms used for subsequent attacks.
The SANS Institute \cite{Homepage:SANS} captured the W32/Leaves worm within minutes.
Hereafter, in the year 2002, a honeypot captured the first unknown threat, the CDE Subprocess Control Service buffer overflow.
This vulnerability allowed to gain access to any Unix-based system \cite{Cert:FirstUnknownMW}.
A possible defence was introduced quickly.
Since then honeypots have been established as an effective concept against vulnerabilities.

The general objective of a honeypot is either to distract attackers from their actual target, or to gather information about the attackers and attack patterns, such as the set of popular target hosts and the frequency of requests-responses.
However, it is worth noting that honeypots should not be considered as an implementation to solve a certain problem but rather as a generic concept.
Regardless of where and how honeypot are deployed (\eg a router, scripts emulating specific services, virtual machines, or a standard physical system), honeypots contribute by being exposed to attacks.

\subsection{Distinction to Other Security Concepts}
\label{sec:DistinctionHoneypots}

Security concepts can be classified by the areas of operation which are 
prevention, detection, and reaction \cite{Schneier:SecurityDefinition}.
\textbf{Prevention} can be defined as any undertaking which \one discourages 
intruders and \two makes breaches in the system infeasible.
\textbf{Detection} is the process of identifying the presence of actions 
that harm the systems \one confidentiality,  \two integrity and \three  availability.
\textbf{Reaction} describes the execution of reactive measures after detecting
harmful actions. Ideally, reaction reinforces prevention and improves 
future detections. Security concepts are designed to accomplish best
results in their area of operation, which leads to inferior results in
the remaining areas.

Besides honeypots, a combination of firewalls, intrusion detection systems (IDS), 
anti-virus (AV), intrusion prevention systems (IPS), and log-monitoring can be used 
to improve network security. It is important to differentiate these concepts 
in order to use them to their full potential \cite{mairh2011honeypot}. 
Traditionally, \textbf{firewalls} monitor and control network traffic based 
on predetermined, static security rules. They act as a barrier between secure
and insecure networks. Advanced firewalls filter not only by network attributes,
but also protocol states and application layer rules.
\textbf{IDS} analyze whole packets, both header and payload, looking for malicious 
signatures. When a known event is detected an alert is generated. \textbf{IPSs} 
can be understood as an  extension of IPS which not only issue alerts, but 
actively reject packets and dynamically add new rules. \textbf{AV} software is 
also signature based, but operates on a file-level and tries to capture 
infections by a trojan or virus. Complete \textbf{log-monitoring} of application 
logs, service logs, system logs \etc increases the awareness of network
infrastructure or availability problems. Problems are detected by evaluating
the log files statistically for known text patterns that indicate important events.
\textbf{Cyber security standards} prevent attacks by defining security guidelines
for the software and infrastructure, such as a clean configuration, hardened
kernel and recent updates.
Nowadays a clear distinction between these concepts became difficult, as security
solutions are hybrids that combine several concepts in order to minimize 
the individual trade-off.

Honeypots add little value directly to prevention as they do not protect against security breaches.
At most, they inhibit attacks because attackers are concerned
about wasting time and resources attacking honeypots instead of the actual
target: production systems. However, they fail against automated toolkits and
worms, which form one of the most common attack types and which scan, probe and
try to exploit every network participant they can find. This also holds for
intrusion detection systems, they might deter, however they will not actively
prevent automated attacks. Firewalls are the best concept for prevention, as
they hide services and block communication on unused ports or from suspicious
IP-address-ranges, so that intruders are kept out or do not find any services to
communicate with. Best prevention is achieved by good cyber security standards 
and practices, such as disabling unnecessary or insecure services, 
frequent update and patch policies, strong authentication mechanisms and so on.

If detection is concerned, honeypots add extensive value. It is very difficult to
detect attacks on production systems because the attacks simply submerge in the
vast amount of production activity. Honeypots can simplify the detection
process. Since honeypots have no production activity, all connections to the
honeypot are suspect by nature and therefore detect an unauthorized probe, scan,
or attack with almost no false positives and negatives. As the name suggests,
intrusion detection systems are designed to detect attacks. However, their
effectiveness is highly dependant on the signature quality. Administrators can
be overwhelmed with false positives, become numb and start to ignore those
messages. On the other hand, false negatives are also possible if no appropriate
signature exists as new exploits are used. Snort developed an effective signature-based recognition after years of
community contribution.
If we consider a traditional firewall concept without the abilities such as flow statistics (e.g., packet count), a firewall is significantly limited in the detection of attacks.

The reaction to attacks can be accelerated with the help of honeypots. The
analysis of the attack is substantially easier, as attack data is not mingled
with production activity data. Furthermore, in contrast to production systems,
honeypots can be taken completely off-line for the analysis, which enables a
proper and full forensic analysis. The insights can then be used to clean the
production systems and understand the exploit, which is the first step to patch
the corresponding vulnerabilities.
Intrusion detection systems are primarily
passive, that means that they analyse packets and create alerts.
Traditionally, they do not alter the network packet flow.
They only inform administrators about malicious activities.
The reaction to new attacks might also include the creation of new signatures.
As firewalls do not detect attacks, they cannot react dynamically to attacks. 
However, administrators can define new firewall rules manually, which are usually simpler then the rules from an
intrusion detection system and are based rather on packet meta information than
payload and packet flow \cite{SpitznerHoneypotPartOne}.

\begin{table}[t!]
\centering
	\caption[Distinction between security concepts.]{Distinction between
		security concepts \\based on areas of operation.}
	\label{tab:distinction_hp_table}
	\begin{tabular}{llll} 
		\toprule
		Objective & Prevention & Detection & Reaction \\ 
		\midrule
		Honeypot & + & ++ & +++ \\
		Firewall & +++ & ++ & + \\
		Intrusion Detection Sys. & + & +++ & + \\
		Intrusion Prevention Sys. & ++ & +++ & ++ \\
		Anti-Virus & ++ & ++ & ++ \\
		Log-Monitoring & + & ++ & + \\
		Cyber Security Standard & +++ & + & + \\
		\bottomrule
	\end{tabular}
\end{table}

Our discussion on the different security concepts shows that all concepts tend to have a specific objective. Primarily, honeypots implement reaction, IDSs implement detection, and firewalls implement prevention concepts. These concepts do not exclude but may complement each other. To achieve all objectives (\emph{prevention, detection, reaction}), hybrid solutions  such as IPSs have been developed.  We summarize our findings in \autoref{tab:distinction_hp_table}, the number of $(+)$ denotes the contribution level to the various objectives. However, this rating should be understood as a general trend of the overall capabilities, since individual deployment might shift the focus towards another objective.

\subsection{Taxonomy of Honeypots}
\label{sec:honeypotTaxonomy}

Marty Roesch, the developer of Snort, established one of the first
classifications of honeypots \cite{SpitznerHoneypotPartOne},
separating honeypots primarily based on their field of operation:

\begin{itemize}
	\item production honeypots
	\item research honeypots
\end{itemize}

Production honeypots are characterized by ease of deployment and utilization and
are meant to be used in production environments of companies. Their intended
purpose is to achieve a higher security level within the network of one specific
company by deflecting attacks. However, the production honeypots come with a
trade-off between ease of operation and the quantity of collected information.
By contrast, research honeypots provide comprehensive information about attacks,
but are more difficult to deploy. They are usually used by research
organizations and network forensics scientists in order to scrutinize attacks
and to develop general counter-measures against threats. Research honeypots help
to understand the motives, behaviour, tools, and organization of the black-hat
community.

To set an example: A company redirects incoming traffic to unused IP-addresses 
to a SSH-honeypot. The honeypot identifies the attackers by IP-addresses which
can be used in filter rules and block the access to the production systems. 
The functionality of the  honeypot can be limited, as it only has to recognize the traffic and its source.
A research organization however might be interested in studying the
details of the attack, that is which dictionaries were used to guess the passwords
in order to create a recommendation on password strength. Consequently,
they have to use a research honeypot, which is more sophisticated and 
provides extensive log records with SSH credentials.

Another widely used classification of honeypots is based upon the
characteristics of interaction, on the one hand considering the level of
interaction \cite{Mokube:Challenges:HighLowDef}:

\begin{itemize}
	\item low-interaction honeypots (LIHP)
	\item medium-interaction honeypots (MIHP)
	\item high-interaction honeypots (HIHP)
\end{itemize}

On the other hand, considering the direction of the interaction
\cite{HoneyC:HPClient}:

\begin{itemize}
	\item server honeypots
	\item client honeypots
\end{itemize}

Low-interaction honeypots simulate only a small set of services like SSH or FTP,
they do not provide any access to the operating system to the attacker.
LIHP produce minimal responses in order to allow protocol handshakes. As the
collection of information is limited, LIHP are used mainly for statistic
evaluation. However, they suffice to recognize peaks in the number of requests,
for example created by autonomous worms. LIHP tend to be production honeypots.

Medium-interaction honeypots are slightly more sophisticated than LIHP, however
they still do not provide any operating system functionality. The simulated
services are more elaborated and provide a higher level of interaction for the
attacker, as a matter of fact MIHP produce reasonable replies in hope of
triggering follow-up attacks. To be an attractive target, MIHP also 
emulate more services than LIHP. Because of the reduced interaction capabilities
LIHP and MIHP both have low chances of being compromised.

High-interaction honeypots are the most sophisticated honeypots. They are the
most complex to implement, deploy and maintain, as they provide to the attacker
a real operating system environment, which is not restricted. Furthermore, a
huge set of services is installed. HIHP collect the largest possible amount of
information, including complete attack logs, data access, traversing of file
trees, executed byte codes etc. Because of the high complexity, HIHP log
analysis is usually done by networks forensics scientists and less frequently
automatically. HIHP tend to be research honeypots.

An overview of the properties with respect to the level of interaction is
presented by \autoref{tab:honeypotTaxonomy}. However, it is important to point
out that this classification is quite academic and impractical. Over time many
different flavours of honeypots were created which are very difficult to divide
into specific categories. That is why it became common to only differentiate
between low and high honeypots, as recommended by Lance Spitzner
\cite{SpitznerLowHighOnly}. Eventually, all honeypots that are mere port
listeners or emulate services became low interaction honeypots (previously
presented as low and medium) and anything that provides real services and
aspects of an operating system a high-interaction honeypot
\cite{HoneyDBackground}.

Server Honeypots wait until the attackers initiate the communication, whereas
client honeypots actively search for potential malicious entities and request an
interaction. The most common field of application are web browsers: client
honeypots request a homepage and check for unusual activities. Client-LIHP
emulate components while client-HIHP use real web browsers. Traditional
honeypots are server based.

The last classification is based on the physicality of the honeypot
\cite{Provos2004HoneyD}:

\begin{itemize}
	\item physical honeypots
	\item virtual honeypots
\end{itemize}

Evidently, a physical honeypot is a real machine on the network. A virtual
honeypot is simulated (virtualized) by a host machine that forwards the network
traffic to the virtual honeypot. Multiple virtual honeypots can be simulated on
a single host. Virtual honeypots are usually high-interaction honeypots.

\begin{table}[t!]
	\centering
	\caption[Properties of honeypot interaction levels]{Properties of honeypot
		interaction levels \cite{Kambow2014:IndianSurvey3}.}
	\label{tab:honeypotTaxonomy}
	\begin{tabular}{l|ll|l} 
		\toprule
		& LIHP & MIHP & HIHP \\ 
		\midrule
		real operating system & \xmark & \xmark & \cmark \\
		risk of compromise & low & mid & high \\
		wish of compromise & \xmark & \xmark & \cmark \\
		information gathering & low & mid & high \\
		knowledge to deploy & low & low & high \\
		knowledge to develop & low & high & high \\
		maintenance time & low & low & very high \\
		\bottomrule
	\end{tabular}
\end{table}

\subsection{Specification of Honeypot Attacks}
\label{sec:honeypotAttacks}

A technological attack is defined as an attempt to destroy, expose, alter,
disable, steal or gain unauthorized access to or make unauthorized use of an
asset; where asset is defined as anything that has value to an organization
\cite{defAttackComputer}. However, honeypots are not directly of value, their
value lies in being attacked. In order to define what in particular a honeypot
attack is, firstly one has to accept the fact that honeypots are attacked
inadvertently, they are exposed to attacks which are originally directed at
production systems. That means that attacks on honeypots are no different to
other systems as long as the attacker does not recognize the presence of the
honeypot. Second, one has to differentiate between \textit{anomalous} and
\textit{normal} behaviour. As honeypots are controlled environments, one can tag
any anomalous behaviour as an attack \cite{mairh2011honeypot}. However, this
specification requires the differentiation between server and client honeypots.
Server honeypots are completely passive, therefore all incoming requests form an
anomaly and are by definition an attack. Client honeypots actively search and
contact communication partners, therefore client honeypots must discern which
communications comprise an anomaly. Heuristics usually verify this by looking
after uncommon modifications.

\subsection{Advantages and Disadvantages of Honeypots}
\label{sec:advDisadvHoneypots}

Short and comprehensive lists presenting the advantages and disadvantages of
honeypots have been assessed by several researchers \cite{SpitznerHoneypotPartOne,
Mokube:Challenges:HighLowDef, kaurcomparison}. We summarize the different
aspects as follows.

\smallskip
\noindent The advantages of honeypots include:

\begin{LaTeXdescription}

\item[Valuable Data Collection] Honeypots collect data which is not polluted
with noise from production activities and which is usually of high value. This
makes data sets smaller and data analysis less complex.

\item[Independent from Workload] Honeypots do only need to process traffic which
is directed at them or originates from them. This means, that they are
independent from the workload which the production systems experience. 

\item[Zero-Day-Exploit Detection] Honeypot capture everything that is used
against them, this means that also unknown strategies and zero-day-exploits will
be identified.

\item[Reduced False Positives and Negatives] Any activity with server-honeypots
is an anomaly, which is by definition an attack. Client-honeypots verify attacks
by detecting system state changes. These procedures result in reduced false
positives and false negatives.

\item[Flexibility] Honeypots are a very flexible concept as can be seen by
vast amount of different honeypot software. This means, that 
well-adjusted honeypot tools can be used for specific tasks, which furthermore
might reduce redundant load.

\end{LaTeXdescription}

\smallskip
\noindent The disadvantages of honeypots include:

\begin{LaTeXdescription}

\item[Limited Field of View] Server-honeypots have one common problem: they are
worthless if no one attacks them. As long as attackers do not send any packets
to the honeypot, it will be unaware of any unauthorized activity on production
systems.

\item[Being Fingerprinted] Low-interaction honeypots emulate services, that
means that their services might behave different than the real services, which
can be used for fingerprinting honeypots and consequently detect them.

\item[Risk to the Environment] If honeypots get exploited, they can introduce a
risk into the user's environment. As discussed, the higher the interaction
level, the higher the possible misuse.

\end{LaTeXdescription}




\section{Honeypot Software}
\label{sec:honeypotSoftware}

This chapter introduces honeypot software, classified into the interaction level
and their type of communication architecture. This means that honeypots will be
introduced as either client or server honeypots and assigned either as low or high
with respect to their interaction level. Hereby, this overview focuses on 
available and deployable software as we hope that this will be help the reader to choose the
proper honeypot for his needs.
However, we also include outdated honeypot implementations that had a significant impact on the research and were seminal during the release time.
To clarify the state of maintenance we explicitly note release dates.
This overview will not only illustrate how versatile honeypots are but also in which different application scenarios it can be deployed.
Finally, we present honeypot related tools as well as mechanisms that detect the presence of honeypots.

\subsection{Low Interaction Server Honeypots}
\label{sec:low-server-hp}

\subsubsection{Common Internet Services}

The first publicly available honeypot software was released in 1997, the \textbf{Deception Toolkit(DTK)}~\cite{FirstHoneypot:DTK}. It is written in C and Perl, and emulates vulnerabilities of well-known services of Unix
systems. The developers explain: \textit{``We use deception to counter attacks
[...] the deception is intended to make it appear to attackers as if the system
running DTK has a large number of widely known vulnerabilities.''}~\cite{FirstHoneypot:DTK}.

In 1998, \textbf{Back Officer Friendly (BOF)}~\cite{Zhang2003HP:bof} was
developed, originally to notice attacks by the remote administration tool Back
Orifice. BOF has an outstanding ease of use considering the early stages of this
research area, is runnable on Unix- and Windows based systems, and provides an
emulation for several well-known services including Telnet, FTP, SMTP, POP3, and
IMAP2. The main objective of BOF is to waste the time of intruders.
Consequently, connection attempts are merely logged and a plain response is created.

One year later the first commercial honeypot was released: \textbf{CyberCop Sting}
\cite{honeypotSoftware:Cybercop}. CyberCop Sting creates a virtual network on a
single host and simulates different types of network devices, including Windows
NT servers, Unix servers, and routers. Each virtual network device is connected to the Internet via a public IP
address. It thus can receive as well as send genuine-looking packets. The simulated
systems include decoys for Telnet- and Nmap-based fingerprinting.

\textbf{HoneyBOT}~\cite{HoneyBotPage} is a low-interaction
honeypot for Windows, which opens a range of roughly 1,000 listening sockets
mimicking vulnerable services. The log tool is quite sophisticated, as raw
packet level data, keystrokes, and malware (trojans, rootkits) are saved for
analysis. Furthermore, a report to a centralized collection point via the syslog
utility is optional. It is closely tied with a GUI, which offers a
classification of log events based on port number or attacker IP address.

\textbf{HoneyD}~\cite{Provos2004HoneyD} is one of the best-known and most seminal
honeypot implementation, in particular concerning the
virtualization of hosts. HoneyD is a small daemon that simulates thousands of
virtual hosts at the same time, class B sized networks has been tested
successfully. The hosts can be configured to run arbitrary services, and their
fingerprint can be adapted so that they appear to be running certain operating
systems. Different operating systems are simulated on the TCP/IP stack level by
learning TCP fingerprints from reading nmap/ xprobe fingerprint files, which
results in an effective deception of these common tools. Not only hosts are
simulated, but also any arbitrary routing topology including dedicated routes
and routers. Moreover, the routes can be annotated with latency and packet loss
to improve realistic characteristics of the topology. It is possible to ping or to
traceroute all virtual machines.

\textbf{HOACD}~\cite{HOASAnnouncement} bundles HoneyD, OpenBSD, and Arpd on a bootable CD.
The objective of HOACD is to provide a ready to use honeypot system.
It stores logs and configuration files on a local hard disk.

\textbf{Honeyperl}~\cite{HoneyPerl} is a honeypot written completely in Perl. Its
prominent feature is the ability of being extendable by Perl modules. Many
plugins exist for simulation of Telnet-sessions, SMTP etc.

\textbf{Impost}~\cite{ImpostHP} is a network security auditing tool designed to analyse the forensics behind compromised and/or vulnerable daemons.
Impost supports two modes:
It can either act as a low-interaction server honeypot, which allows dynamic communication patterns by Perl scripts; or it can operate as a packet sniffer and monitor incoming data to pre-defined destination port.
It is an early honeypot implementation, which combines honeypots with full packet sniffing functionality.

\textbf{KFSensor}~\cite{KFSensorHP} is a Windows based commercial honeypot.
It is designed for deployment in a Windows based corporate environment in order to improve the network security of an organization.
It includes IDS functionality by (i) providing a Snort compatible signature engine, (ii) emulating Windows networking protocols (e.g., NetBios) and vulnerable system services or trojans.
KFSensor resists denial of service and buffer overflow attacks.

\textbf{LaBrea}~\cite{LaBreaHP} is the first honeypot incorporating the tarpit
techniques, sometimes described as sticky honeypots.
LaBrea implement two functionalities.
(i) It takes over unused IP
addresses by replying to unanswered ARP requests in a network, and (ii) emulates hosts which answer to SYN packets with a SYN/ACK.
However, LaBrea is not intended to support full TCP-connections.
When the attacker sends a TCP ACK and data packets, LaBrea will not reply.
Following TCP retransmission, the attacker will be delayed as he waits for the TCP timeouts and resends the data packets.
LaBrea is a cross-platform tool.

\textbf{NetBait}~\cite{chun2003netbait, NetBaitService} is a commercial solution that
implements honeypot farms. NetBait is available as a stand-alone product or as
a service. A redirector is deployed, which forwards
attacks on unused IP address ranges either to predefined honeypots within the
cooperate network or to a honeypot farm which is maintained by NetBait Inc.
outside the network. Users of NetBait only have to deploy a redirector in
their networks, no skilled network administrators and security specialists are
required, which saves costs. Moreover, NetBait is described as a distributed
query processing system on honeypot data, which has been collected by a set of cooperating
machines. NetBait supports up to 15,000 hosts per network. The company was offering their
their service from 2002 up to 2007.

\textbf{NetFacade}~\cite{PolymorphicWorms:HoneypotOverviewChapter} is one of the very
first commercial low-interaction server honeypots, which simulates an entire
class~C network with up to seven different operating systems. Its development
started in 1998 and was released in 2000. However, it has seen little
public exposure. Its developer, Marty Roesch,  gained valuable knowledge
and developed debug tools which led to the development of Snort.

A simplistic honeypot server written in Perl is
\textbf{single-honeypot}~\cite{SingleHoneypotHP}, which was designed using
the KISS-principle. It logs access on all ports without emulating services.

\textbf{Smoke Detector}~\cite{pouget2003white} by Palisade Systems Inc. is another early
commercial product. It distinguishes itself by not only providing a software
but also offering a hardware unit, which has Smoke Detector installed and
preconfigured. Smoke Detector mimics up to 22~services and emulates 19~distinct
hosts in a network. Access attempts are reported.
Complementary tools enable the analysis of logs.

\textbf{Specter}~\cite{SpecterHP} is one of the longest available commercial products,
being purchasable for 16 years, however no new major release has been seen since
2005. It is advertised as a smart honeypot-based intrusion detection system and
offers common Internet services such as SMTP, FTP, POP3, HTTP, and TELNET.
Furthermore, Specter supports the analysis of ICMP, TCP, and
UDP packets on all ports. It logs malicious activities and informs the
administrator automatically. Specter provides massive amounts of decoy content
including images, MP3 files, email messages, password files, documents, and various
types of software.

Also Symantec provided an extension to his anti-malware products by selling the
\textbf{Symantec Decoy Server}~\cite{SymantecHP}.
This honeypot acts as a fully functioning server, and can simulate email traffic between users in the organization to
mirror the appearance of a live mail server. It records every action
for further analysis.

\textbf{Tiny Honeypot (thp)}~\cite{TinyHoneypotGH} is a simple server honeypot
which listens on every TCP port not currently in legitimate use, logging all
activities, and providing enough replies and interaction to fool most automated
attack tools by a short fingerprint. 
The goal is to distract the attackers from real services by offering a large amount of open ports with fake services.
Netfilter / iptables rules are used to redirect any incoming connection requests to the thp listener. By
default a login banner and root shell including some simple emulation of tools
such as \texttt{wget} are provided. Thp can reside on production hosts with negligible
impact on performance.

\textbf{Nepenthes}~\cite{NepenthesHP, baecher2006nepenthes} is a low interaction
honeypot, which is designed to emulate vulnerabilities that worms misuse to spread, and
to capture these worms. Nepenthes provides a modular architecture which can be
easily extended to add new vulnerabilities. Other modules include functions such as
the download of files, submitting of the downloaded files, and a shellcode
handler. Albeit Nepenthes needs expert knowledge to emulate new vulnerabilities and to
conduct a successful conversation with malware, it captures new exploits for old
vulnerabilities. Unknown exploits are highlighted in the log files, an
information which can be used to build new modules or better dialogues to trigger more downloads.

\textbf{Dionaea}~\cite{DionaeaHP} is meant to be a Nepenthes successor overcoming some of
its shortcomings such as missing IPv6 support, multi-threading, TLS encryption for some
protocols, and switching from C++ to Python as the module scripting language in
hope of more contributors. The shellcode detection was extended by developing
\textbf{libemu}~\cite{libemuHP}, which detects shellcode not only by simple pattern
matching but by emulation. The libemu library receives a buffer and detects even
unknown shellcode fully automatically. Moreover, Nepenthes never supported
port 445 (SMB) because of too many different exploitable vulnerabilities,
Dionaea addresses this problem by emulating valid SMB sessions. Despite all the
improvements, it is worth noting that Dionaea is not an extension of
the Nepenthes code base, but rather the same developers based their implementation on the
the Nepenthes architecture.

\textbf{Honeytrap}~\cite{HoneytrapGit} is a low-interaction honeypot daemon which
distinguishes itself from other honeypots by implementing a \emph{dynamic} server
concept. It uses stream monitors to check the network stream for incoming
packets and starts appropriate listeners on demand. Each listener can
handle multiple connections and terminates after some idle time. This
concept targets completely unknown attacks which might occur on random ports and
unknown protocols---no predefined emulation is required. Service emulation is
not the main focus, however some basic emulation is provided. If no emulation
for a protocol exists, the default response is a single newline character,
which, according to the developer, is a simple but surprisingly successful
approach. Honeytrap distinguishes strictly between data capture, which is implemented
by the core system, and attack analysis, which is implemented by plugins. Honeytrap
supports several modes, including a mirror mode, which replies with exactly the
same packets as received, hence emulating services implicitly. If the connection
fails, default replies are used. Furthermore, Honeytrap can be used in proxy
mode (also called meta-honeypot), which relays incoming connections to other
hosts or services while still logging the communication. If ports are configured
in the ignore mode, honeytrap simply does not handle those ports.

\textbf{Mwcollect}~\cite{bacher2005know}, later known as
\textbf{Mwcollectd}~\cite{wicherski2010placing}, is
designed as a versatile malware collection daemon, attempting to combine the best
features of Nepenthes and Honeytrap. Mwcollect was actively developed during
2005-2006 and received a major update in 2009 after a long time of non-maintenance. Its
entire functionality is based on modules, which include \texttt{libcurl}
for HTTP and FTP downloads, TFTP, \texttt{dynserv-nfqueue} module, which implements the
Honeytrap principle of package mirroring and dummy newline replies, as well as
some Dionaea bindings, HTTP/SMB emulation, shellcode emulation via Libemu and
mwserv, which is the malware aggregation/submission service used by the
mwcollect Alliance.

\subsubsection{Dedicated SMTP Honeypots}

The Simple Mail Transfer Protocol (SMTP)~\cite{RFC-5321} defines an Internet protocol for \emph{sending} emails.
Mail Exchangers between email sender and receiver are usually determined via the DNS MX record.
SMTP honeypots may thus operate either independently of a DNS domain name, or linked to a specific domain name using this record.

\textbf{Jackpot}~\cite{grimes2005honeypots} is an SMTP relay honeypot to combat email spam. It is written in Java and comes with several configuration options to make trapping and tracking of spam as efficient as possible.
Spam is automatically classified by (i) using various antispam databases, (ii) automatic distinction between regular spam or relay test messages, and (iii) simulation of a very slow server by delaying replies.
Jackpot assumes advanced attackers that try to detect honeypots by sending test messages to their own email inbox.
To tackle honeypot detection, Jackpot delivers relay test messages to the inbox of the attacker but silently drops messages to inboxes of potential victims.
All spam messages and client IP addresses are logged.
Jackpot provides a HTML GUI for easy usage.

\textbf{SpamD}~\cite{SpamDBSD} is an email honeypot developed by the OpenBSD community.
It handles three types of hosts: blacklisted hosts, white-listed hosts, and grey-listed hosts.
\emph{Blacklisted hosts} are forwarded to SpamD and tar-pitted, i.e., the communication is manipulated with delays of 1~character per second during the complete dialogue, to waste time of the attackers.
Eventually emails are rejected with an error message.
A blacklisted host will never be able to talk to a real mail server.
In contrast to this, connection attempts of \emph{white-listed hosts} are sent to a real mail server.
New hosts are \emph{grey-listed} by default and forwarded to SpamD, which shows a temporary failure message when they try to deliver mail.
Additionally, grey-listed hosts will experience stutter during the first seconds of the dialogue.
The objective behind this artificial delay is to disturb spammers that are paid for emails per minute.
Those host will lose interest and thus may be identified as attacker.
It is worth noting that a real mail, which prioritize quality of service, will try to retransmit for a period of time.
SpamD is a lot more efficient compared to simple DNS lookups or spam blacklist checks.

Another historic SMTP honeypot, which does not have an official homepage any
more and was not maintained for many years, is \textbf{ProxyPot}~\cite{HoneypotNETOverview}.
It imitates an open proxy mail relay and is designed to solely log spam and record the sender identity.

Another simple SMTP honeypot is \textbf{SMTPot}~\cite{SMTPotMirror}, which is written in
Python in less than 300~lines of code. Being a simple program it pretends to be
an open mail relay and collects emails in mailbox files.

\textbf{Spamhole}~\cite{SpamHole} is another fake open mail relay.
It accepts any email messages that the client sends to it, however, rather than actually delivering the messages, it will silently drop them.

\textbf{Spampot}~\cite{Spampot} is a Jackpot clone.
In contrast to Jackpot, Spambot is written in Python.
Its author aimed at a higher support of different platforms. 
Spampot does not use any heuristics to rate mails, it simply stores 5\% of the incoming spam.

\subsubsection{Dedicated SSH Honeypots}

Originally developed in 1995, the Secure Shell (SSH) specifies a protocol for secure remote login~\cite{y-ssh-95}.
Today, SSH also supports tunneling and forwarding~\cite{RFC-4251}.
As such it is used in various application scenarios, e.g., secure copy.

\textbf{Kojoney}~\cite{KojoneySourceforge} is one of the very first dedicated low-interaction SSH honeypots.
\textbf{Kojoney2}~\cite{Kojoney2Github} is a major extension of the Kojoney code base, which was refined, expanded, and adjusted based on long-term deployment experiences of Kojoney. 
Due to the popularity of another SSH honeypot, Kippo, Kojoney2 incorporates many its most attractive features while still retaining the Kojoney core.

\textbf{Kippo}~\cite{KippoGithub} is designed to log
brute force attacks and the entire shell interaction performed by the attacker.
It provides a fake file system resembling Debian Linux with the ability to add
and remove files. Some basic tools such as \texttt{cat} are integrated, however, they are
configured to delude the attacker. Moreover, session logs are stored in an UML
compatible format, so that they can be replayed with the original timings of the
prompt. Kippo development has been continued by \textbf{Cowrie}~\cite{CowrieGithub},
which has already extended the software by SFTP/SCP support and additional
commands for example. Originally, Kippo was inspired by Kojoney.
All three SSH honeypots are written in Python and utilize bash scripts.

\subsubsection{Special HTTP and Web Honeypots}

\textbf{Glastopf}~\cite{GlastopfHomepage} is a modern, easy to deploy 
low-interaction web server. This honeypot tool collects information about web
application-based attacks such as local and remote file inclusion or SQL
injections. Furthermore, Glastopf downloads files from links that are included in incoming requests
and tries to respond such that the expectations of the attacker is satisfied and
subsequent attacks are initiated. Glastopf promotes itself by implementing a
vulnerability \emph{type} emulation instead of specific vulnerability emulation. Once a
vulnerability type is emulated, unknown attacks of the same type can be handled.
While the implementation may be tedious, on the long-term operators benefit from this abstraction.

\textbf{HoneyWeb} by Kevin Timm~\cite{pouget2003white, HoneyWebHTTP} is a HTTP-only low-interaction honeypot compatible with
HoneyD. HoneyWeb emulates various web server platforms, such as Apache,
Netscape, and Microsoft IIS. HoneyWeb looks at incoming URL requests, identifies
which platform they suit, and finally deceives intruders by emulating specific
HTTP headers.
To complicate the detection of the honeypot, HoneyWeb supports the dynamic assignment of a web server
platform and an attacker for a certain time frame.
If the same attacker makes a UNIX-style request, which is then followed by a Windows-style request, HoneyWeb will deliver an error page.
HoneyWeb is extensible by SSL.

Elasticsearch is a search server based on \textit{Lucene} and provides a
distributed full-text search engine with an HTTP web interface and
schema-free JSON documents. It became the most popular enterprise search
engine which led to new attacks for that service. That is why
\textbf{elastichoney}~\cite{elastichoneyHP} has been developed. It is a
simple Elasticsearch honeypot designed to catch attackers exploiting remote
code execution vulnerabilities in the Elasticsearch service. It is
distributed as a ready-to-deploy Docker file.

\textbf{Google Hack Honeypot (GHH)}~\cite{GoogleHackHoneypot} was introduced in 2005, in order to combat search engine hackers.
Those attackers carefully craft search terms (e.g., title, body text, filetype) to explore vulnerable websites, using search engines.
For example, \texttt{"\# -FrontPage-" inurl:service.pwd} will search for web pages that have been created with a misconfigured FrontPage HTML editor.
The search term will reveal user names and passwords, which will give login credentials to the corresponding web servers.
GHH was used to learn more about this threat by emulating a vulnerable web application referenced by a transparent hyperlink, which is hidden from common page viewers but is found using a crawler.

\textbf{Honey Accounts} \cite{oms-whapu-16} is the first honeypot framework to monitor the activity of compromised webmail accounts.
The honeypot consists of Gmail accounts in combination with Google Apps Scripts to perform time-triggered and event-triggered tasks.
The scripts inform the honeypot operator when an email was read, sent, or starred, by sending status messages to a separate email address under the control of the operator.
To make cybercriminals aware of the honey accounts, the operator needs to leak account credentials (e.g., on paste sites).

\subsubsection{Telephone Honeypots}

\textbf{Sandtrap}~\cite{HoneypotNETOverview} is a historic honeypot, which addresses the problem of \emph{war dialing}, a technique of using a modem to automatically scan a list of telephone numbers in search for computer systems.
Sandtrap can log incoming calls of up to 16~lines and emulate reachable modems by answering with login prompts. 
The caller~ID (phone number) and any login attempts are logged, moreover an alert system warns administrators of suspicious activity in real time.

As the telephone-infrastructure is shifted towards the voice over IP (VoIP), this service is
increasingly under attacks, which is why recent honeypot analysis concentrate
around VOIP~\cite{gruber2015global}. \textbf{Artemisa}~\cite{carmo2011artemisa} is a pure
VoIP SIP-specific honeypot implementation. It registers multiple SIP accounts,
which do not represent real human subscribers, at one or more VoIP service
providers, and waits for incoming attacks. It includes conversation recording,
a protection against message flooding, correlation rules to infer sequential and 
stateful attacks as well as rule-based fingerprinting of known SIP attack tools.

\subsubsection{Wireless, Mobile, and Bluetooth Honeypots}

Physical and virtual honeypots have been studied in detail~\cite{ph-vhfbt-08}, however, there is only little work in the field of mobile or wireless related honeypots.
Mobile honeypots have to be distinguished from wireless honeypots, which focus on the attacks on the wireless technology.
The term \emph{mobile honeypot} usually refers to honeypots that focus on attacks on mobile devices.\footnote{Note that the term ``mobile honeypot'' is also used to describe other scenarios. Balachander Krishnamurthy \cite{k-mmhtu-04} uses it to describe prefixes of darknet address space that (1) are advertised to upstream ASes, making the information mobile, and (2) change aperiodically, moving the darknet in the address space.}
They can either be mobile themselves in running on the mobile device, or they run on common stationary hardware (e.g., desktops) that is connected to a network of a mobile operator \cite{wtkss-fimh-12,wvkss-diomh-13}.
The first class of approaches may complicate the measurement across different types of systems. 
In addition, they are only required if the hardware characteristics are relevant for the study.

Mobile honeypots in the sense of honeypots focussing on mobile devices are for example developed by the Chinese Chapter of the Honeynet Project \cite{honeynetproject}.
They are using prototype deployments of honeypots for Bluetooth, WiFi, and MMS.

\textbf{FakeAP}~\cite{Github:FakeAP} was developed as a proof of concept at the Def~Con~X~2002 hacking conference and introduced wireless honeypots. 
The key task of FakeAPs is to deceive attackers, not to log their actions,
as it rapidly generates 802.11b beacon frames with random ESSID, BSSID (MAC), and
channel assignments. This approach hides ones own access point from plain sight and
confuses wardriving tools such as Kismet and NetStumbler.

TJ OConnor and Ben Sangster built \textbf{honeyM}~\cite{cs-hfivh-10}, a framework for virtualized mobile device client honeypots, which emulates in particular wireless technologies. 

The bluetooth honeypot \textbf{bluepot}~\cite{bluepotGithub} was developed to capture
attacks on bluetooth devices. It is designed designed to accept and store any
malware sent to it and to interact with common bluetooth attacks such as
\textit{Blue Bugging} and \textit{Blue Snarfing}. This honeypot does not require
any specific device, it is written in Java and runs on any Linux machine, but it
does obviously demand from the user to possess at least 1 active bluetooth
interface. This honeypot has also some graphical dashboard which allows
monitoring of attacks and presents some simple graphs and lists.


\textbf{HoneySpot}~\cite{s-hwhma-07} summarizes the concepts of a \emph{Honey}pot and wireless hot\emph{spot}.
It offers 802.11 wifi access with the objective of being proved, attacked, or compromised.
It focuses on pure layer~2 wireless attacks that exploit weaknesses in the wireless technology, in particular subverting deployed security mechanisms.

In contrast to HoneySpot, \cite{wtkss-fimh-12} was designed to explore IP-level attacks on mobiles.
The authors deploy common honeypot tools (Kippo, Glastopf, and Dianaea) on a standard Linux PC that is connected to an UMTS network.

\textbf{HoneyDroid}~\cite{mulliner2011poster} is a prototype of a
honeypot especially designed to run on mobile Android devices. For this purpose the
Android smartphone has to be rooted and extended by Galoula, which makes common
UNIX (BusyBox) services and file systems available on Android. This step makes
the installation of Kippo and Honeytrap possible. Kippo has been adjusted to
resemble the behaviour of the Android OS. Furthermore, an Android app is
included, which reads and visualizes the different log file results and
transmits them eventually to a centralized data collection point.

\subsubsection{IoT Honeypots}

The Internet of Things (IoT) is the network of physical devices.
In terms of hardware and software, the IoT is not well-defined.
Devices usually range from smartphones down to very constrained embedded hardware with network connectivity \cite{RFC-7228}.
As those devices can be queried and controlled remotely, attacks on those devices
emerged. \textbf{IoTPOT}~\cite{IoTPOTPaper} was used by a research team to analyze the
increasing threats. It analyzes Telnet-based attacks against various IoT devices
running on different CPU architectures such as ARM, MIPS, and PPC. Another
proof of concept in this field is \textbf{honeypot-camera}~\cite{cameraHP}, which
emulates an openly accessible webcam, including some fake images and device
specific deceptions such as watermarks and daylight intensity.

The honeypot \textbf{Shockpot}~\cite{ShockpotGithub} is designed around a single,
critical vulnerability called Shellshock/ CVE-2014-6271 \cite{ShellshockCVE} for
the bash shell. As the vulnerability was very far-reaching, the study of its
exploitation became interesting. The pure python implementation emulates a
Apache-server, that appears to processes trailing strings after function
definitions in the values of environment variables. This allowed remote
attackers to execute arbitrary code via a crafted environment in the original
vulnerability.

\subsubsection{ICS/SCADA Honeypots}

Another possible deployment place for honeypots are SCADA or ICS systems, which
are systems that monitor and control industrial processes in the physical world,
often such systems are part of critical infrastructures and therefore
particularly endangered. \textbf{Conpots}~\cite{conpotHoneypot} goal is to collect
intelligence about the motives and methods of adversaries targeting industrial
control systems. It is written in Python and speaks several common Internet
protocols such as HTTP, but also some ICS-specific protocols such as kamstrup,
BACnet, and mosbus.

\textbf{HoneyPoint}~\cite{HoneyPointServer} is a honeypot which comes in several license versions
and provides several sub-tools (HPPE, HPSS, HPSC, HPNTA) because of commercial
reasons. Its target platform is Windows. HoneyPoint offers fake network services
and web applications, including manipulated documents which allow for tracking down an
attacker every time the document is opened after the download. Furthermore,
emulated devices in Industrial Control Systems (ICS) such as SCADA are supported.

\subsubsection{Further Special Purpose Honeypots}

\textbf{HoneySink}~\cite{HoneySinkHP} is a honeypot which
specializes on sinkholing. This allows security researchers to monitor the
communication within a botnet and to prevent interaction between bots and their
command and control servers. HoneySink allows its user to sinkhole any number of
domains to it and configure the emulation of DNS, HTTP, FTP, IRC on a basis of
protocol-domain combinations. The authors envision two main use cases: First,
deployment in internal networks where self-maintained DNS servers
redirect traffic to known blacklisted URLs to HoneySink. This helps to detect
infected machines within your own networks. 
Second, HoneySink can be configured globally to respond by its own DNS functionality to requests for
domains which have been taken over by law enforcement. This prevents criminals
to maintain control over their bots. HoneySink is the first freely distributed
network sinkhole software, which aims for being a generic framework for analysing various
botnets.

\subsection{High Interaction Server Honeypots}
\label{sec:high-server-hp}

\textbf{Argos}~\cite{Argos:2006Qemu} was released in 2006 and as it is based on Qemu, it
focuses on efficient x86 emulation. A high-interaction honeypot environment is
provided, which aims to automatically identify and produce remedies for zero-day
exploits. Upon attack detection an intelligent process- or kernel-aware logging
is executed, furthermore own forensics shellcodes are injected allowing in-depth
analysis. Argos distinguishes itself by applying memory tainting and creating
signatures for intrusion detection systems with very few false-positives. There
is also a support for running as a client honeypot (using the same
infrastructure as the Shelia client honeypot), however this mode became 
neglected.

\textbf{Honeywall}~\cite{HoneywallCD} aims at making honeypot deployments simple and
effective. This CentOS-based live-CD utilizes Sebek as the dedicated honeypot
software and offers a GUI for system configuration, administration, and data
analysis. It features an architecture that allows you to deploy both 
low-interaction and high-interaction honeypots, but is designed primarily for 
high-interaction.

\textbf{High Interaction Honeypot Analysis Tool (HIHAT)}~\cite{genericPHPhoneypot,
homepagePHPhoneypot} transforms arbitrary PHP applications into web-based high
interaction honeypots in an automated fashion. HIHAT has been compatible with 4
major PHP frameworks during its active development time. Furthermore HIHAT
offers a graphical user interface which enables honeypot monitoring and
analysing the acquired data. Extensive statistics are generated and IP addresses
are mapped to a geographic locations. Automatic malware is attracted by
insertion of transparent links. Attack types, which has been spotted with HIHAT
include command injection, file inclusion and bot self-propagation and are
characterized by the HTTP GET request, the four different arrays provided by
PHP, and the data transferred. The tool automatically filters for attack
patterns via regular expression which were derived from an analysis of known
attacks against web applications.

\textbf{HoneyBow Sensor}~\cite{honeybowSensor} is 
based on VMWare images and consists of the following 3 components: The MwWatcher
malware collection tool, which monitors file system changes in real time and
catches potential malware on a Win32 guest system. MwFetcher malware collection
tool, which extracts potential malware from a VMware virtual disk by comparing
the infected file list with the clean file list. Finally, the MwSubmitter
malware submit tool, which submits potential malware samples collected by
MwWatcher and MwFetcher to the mwcollect Alliance. The host tools are Linux
based. Infected guests can be replaced with clean guests automatically after
malware extraction.

\textbf{Sebek}~\cite{SebekHP, SebekArticle} is a kernel module installed on high
interaction honeypots, usually virtual guest machines, for the purpose of
capturing attacker's activities such as as keystrokes or file uploads on Win32
and Linux systems. It works by monitoring system call activity and recording
data of interest and is based on two components: The client, which runs on the
honeypots, captures activities and sends the data to the second component, the
server, as stealthy as possible. The client masquerades its existence by using
early root-kit techniques. The server is the centralized data collection point
for all honeypots within the network.

Implementation such as \textbf{Canarytokens} \cite{CanarytokensGithub} 
show that honeypots are not necessarily computers, they can also be 
computer resources, which are called honeytokens \cite{HoneytokensSpitzner}. 
Honeytokens are an URL, domain, word or 
PDF document, Bitcoin wallet \etc an adversary might access. An access
usually indicates a system breach and hence will trigger a message to 
the admin. Canarytokens is also available as a pre-hosted service
\cite{CanarytokensService}, which minimises the deployment effort.

\subsection{Low Interaction Client Honeypots}
\label{sec:low-client-hp}

\textbf{PhoneyC}~\cite{phoneycPaper, phoneycHomepage} is a 
honeypot which mimics legitimate web browsers and enables the study of malicious
HTTP pages. It features the interpretation of remote links (hrefs, iframes \etc)
and scripting languages via spidermonkey. Moreover, specific ActiveX and PDF
vulnerabilities are emulated and heap spray and shellcode detectors are
included. By using dynamic program analysis, PhoneyC removes obfuscation from
many malicious pages. Different browser identities are supported. The data flow
of PhoneyC is as followed: One or more URLs are passed to the client, which
retrieves the content. The content is scanned by an anti-virus program if it is
a file. Valid HTML files are broken up by script code languages and interpreted
by the specific engines. Alerts are risen on suspicious behaviour.

\textbf{HoneyC}~\cite{HoneyCHome, HoneyC:HPClient} is a honeypot
which identifies malicious servers on the web by using different visitor
clients, search schemes, and analysis algorithms. It consists of three
components: the visitor, the queuer, and the analysis engine. The visitor
interacts with the potentially malicious servers, makes the requests and
processes the response. The queuer creates a queue of servers to visit. 
Self-evidently, the analysis engine is the component responsible to evaluate
whether security policies have been violated by deploying snort rule matching.
All modules use the pipe to communicate with each other.

\textbf{Thug}~\cite{ThugGithub} is a pure Python honeypot and
build upon the experiences which were gathered by the development of PhoneyC. It
aims at mimicking the behaviour of a web browser on a certain operating system
in order to detect malicious contents. Thug uses the Google V8 Javascript engine
and an own DOM tree implementation. This framework performs static syntax tree
and dynamic code analysis. Vulnerable modules are emulated like ActiveX, Flash,
Adobe Reader. Different personalities exist, which include various Windows
versions, Linux, iOS, MacOS and Android but also multiple browsers like Safari,
Chrome or Internet Explorer. Proxies are supported, which can be used to
anonymize the access to a malicious page by for example TOR.

\textbf{YALIH}~\cite{Paper:YALIH, YalihGithub} (Yet Another Low Interaction Honeyclient)
is designed to detect malicious websites by integrating a combination of
multiple antivirus engines and pattern matching using string or regular
expressions for detection. It is capable of extracting embedded JavaScript files
and performs de-obfuscation and de-minification of scripts. IYalih has an IMAP
plugin in, which scans an email inbox for spam, extracts the URLs and visit
these sites. Its emulated browser handles cookies and session, redirection,
referrers and different browser personalties. The developers compared its
effectiveness using a malware URL database and determined 15\% less false
negatives than Thug, 80\% than HoneyC and 35\% than Monkey-Spider while still
requiring only a moderate scanning time.

\textbf{SpyBye}~\cite{SpyByeHP} is developed by the same author as HoneyD, however it is
a low interaction client honeypot. Originally, one had to enter an URL into a
form and wait for the analysis to complete, similar to other honeypot clients.
However, SpyBye matured into an HTTP proxy, which intercepts all browser
requests while the user just simply visits a homepage. SpyBye intends to
determine if embedded links on your web page are harmless, unknown or even
dangerous and scan the content against the ClamAV engine. It is a reliable
indication for a compromise if administrators scan their homepages and detect
foreign URLs which were not set by them.

\textbf{Monkey-Spider}~\cite{ikinci2008monkey, MonkeySpiderHP} is a crawler based 
honeypot utilizing anti-virus solutions to detect malware. It
is claimed to be fast and expandable with other detection mechanisms. It uses
the well-known scalable Heritrix crawler to create arc files and then pass them
along to anti-virus scanners such as ClamAV and Avast. Monkey-Spider detection
is done solely by external signatures. URLs, binaries and malicious JavaScript
are extracted and saved in a threats database.

\textbf{ADSandbox}~\cite{dewald2010adsandbox} is a honeypot which
utilizes at its core the Mozilla JavaScript engine SpiderMonkey within a sandbox
and logs every action during the execution. After that a heuristic assesses
malicious behaviour. The heuristic performs static and dynamic analysis. The
user interacts with ADSandbox in two different methods: First, by means of the
browser helper objects, which hands over the URLs visited within the browser to
the analysis engine and displays custom error pages if necessary or in case of
no suspicion lets the user visit the page transparently. Second, the user can
manually supply an URL and some additional parameters using the shell. ADSandbox
is meant primarily to provide real-time protection for browser users, hoever it
never left the prototype status.

\subsection{High Interaction Client Honeypots}
\label{sec:high-client-hp}

\textbf{Capture Bat}~\cite{CaptureBat}  monitors the state of the Win32 operating systems
during the execution of applications and processing of documents. The client
connects to a central capture server that requests the client to visit an URL
with a specific browser. The client is run inside of a virtual machine so that
infections can be reverted by resetting the virtual machine.  Capture Bat
provides insights on how the software operates even if no source code is
available by observing state changes on a low kernel level (by the help of a
file system, registry and process monitor). Event noise that naturally occurs in
an operating system environment while idling or on standard execution of
applications is filtered, allowing analysts to easier spot and understand the
behaviour of for example malicious Microsoft Word documents.

\textbf{HoneyClient}~\cite{HoneyClientGit, wang2005Presi} is the first open source client
honeypot. It is state-based and detects attacks on VMWare Windows clients
running on Linux hosts by monitoring files, process events, and registry
entries. Its architecture is threefold: An Agent, Manager and Util Module exist.
The agent component is a SOAP server, running as a daemon within a cygwin
environment on the guest. It receives messages which trigger actions like
visiting a homepage with a specific web browser (Internet explorer, firefox are
supported) or performing an integrity check of the Windows files. The manager
module handles the guests on the host system by communicating with the agents.
The Util package provides the SOAP/http protocol integration and a configuration
possibility.

\textbf{Capture-HPC}~\cite{CaptureHPCHP} is a honeypot framework designed with efficiency
and scalability in mind. Capture-HPC consists of a Capture Server and Capture
Client. The Capture server is a simple server that manages various capture
clients and the VMware servers, which host the guest OS that run the Capture
clients. It distributes each URL it receives to the active clients. The Capture
client consists of 2 modules: a kernel driver which uses event-based detection
mechanisms for monitoring the system's state changes (file system, registry, and
processes that are running) and an user space process, which accepts action
requests from the Capture server, and communicates potential state changes back
to the server. An exclusion list is used to filter default events.	

\textbf{Strider HoneyMonkey}~\cite{HoneyMonkeyHP, wang2006automated} was developed by the
Microsoft Research Team and distinguishes 
itself by a narrow coupling with the Windows OS and by creating
heterogeneous virtual hosts, which differ by their patch level. The technology
was never provided publicly but was used only for internal purposes. HoneyMonkey
utilizes the Internet Explorer as the Browser, however does not allow any 
pop-ups or installation of plugins or software. Any read or write which happens
outside of Internet Explores file directories are considered malicious and
highlighted for manual analysis. Furthermore, new spawned child processes of the
Internet Explorer are observed. Upon detecting an exploit, the monkey notifies
the Monkey Controller on the host machine to respawn a clean HoneyMonkey virtual
machine, which then continues to visit the remaining URLs.

\textbf{Trigona}~\cite{TrigonaHP} tries to improve the efficiency of high interaction
honeypot clients by combining the advantage of high interaction where no
emulation is used and low interaction, where a high throughput and a small
resource fingerprint is possible. As opposed to other frameworks Trigona does
not load several VMs or resets one VM repeatedly for each URL, but initiates
one VM, which access multiple, for example 200, URLs at once. The network 
traffic is packet captured for analysis at a later stage. The VM is reverted
after a group of URLs. Exploit Kits and Malware Binaries can then be extracted
from the packet captures files, the state information of the operating system
are not analysed.

\textbf{HoneySpider}~\cite{HoneyspiderNetwork} is hybrid honeypot framework, which is
based on fairly up-to-date high and low interaction honeypot tools and
integrates a crawler application specially designed for the bulk processing of
URLs. The framework focuses primarily on attacks involving the use of web
browsers and their plug-ins by detecting drive-by downloads and malicious
binaries including 0-day exploits. HoneySpider automatically obtains and
analyses the malware. The first version included a simple web-client, JavaScript
analysis and Capture-HPC. It evolved and possesses now a shellcode and 
anti-virus scanner; PDF, SWF and office documents analyser, moreover some 
bindings for Thug, Cuckoo and couchDB.

\textbf{SHELIA}~\cite{ShielaHP} focuses on
the analysis of URLs and attachments received in emails. It supports IMAP/POP,
so it can read messages from an email in-box, however direct input of links and
files is also possible. SHELIAs design philosophy is that false positives are
much more important than false negatives because they are possibly more harmful
in cases of signature creation, therefore they are avoided. This is done by
detecting intrusions not by verifying modifications to the file system or
registry after visiting a website, but by tracking the caller of sensitive
operations. More precisely, whenever a call modifies the registry, the file
system, or network activity, Shelia tracks whether the call is coming from an
unauthorized area which is not supposed to contain code. Shelia runs in a
virtual machine which is reset every n checks to prevent infections to survive
if they are not detected by Shelia. Upon exploit detection extensive analysis
and logging to a database takes place.

\textbf{UW-Spycrawler}~\cite{UW2006spycrawler} was
used for research projects, however never made publicly available as the authors
focussed on Internet research rather than developing a general user-friendly
framework. It focusses on the detection of malware using one of the two attack
methods: Piggy-backed malware code, which comes with legitimately looking
executables or drive-by download attacks, which exploit a vulnerability in the
browser to install malware. UW-Spycrawler supports automatic software
installation by automatically accepting the EULA and other installation steps.
Piggy-backed spyware is then recognized by using a signature-based anti-spyware
program (AdAware). Drive-by attacks are assumed to escape the browser-sandbox
and modify the system, so URLs are marked as malicious if one of the event
triggers is detected, like file and registry writes, process creation, browser
crashes.

\textbf{Web Exploit Finder (WEF)}~\cite{WEFhomepage} 
is designed to detect drive-by-download attacks and consists of a
VMware virtualization layer, specially crafted Windows guests and an user
dashboard which communicates with the guests controls the action of the web
browsers. Attacks are detected by checking for modifications to the operating
system by evaluating the relevant system calls. As such changes to the operating
system are not designated, a deep integration and interaction with the kernel
was  deployed by the so-called rootkit module.

\textbf{HoneyIM}~\cite{xie2007honeyim} utilises Capture-HPC in a new context. 
The open-source instant messaging (IM)
client Pidgin is used to extend the basic functionalities and to create decoy IM
users. A legitimate IM user has to add the decoy IM user to its contact lists,
if some IM malware compromises and tries to propagate, it will contact all users
in the contact list, therefore the decoy will also receive malicious URLs etc.
and thus will be registered by the HoneyIM system. Furthermore, the malware can
then be analysed in the Capture-HPC environment. This helps to recognize
infections in IM networks early, as HoneyIM delivers attack information to
network administrators in real-time so that system quarantine and recovery can
be quickly performed.

\textbf{PwnyPot}~\cite{pwnypotGit} is a honeypot, which does not
detect malicious servers based on system changes but tries to identify the
malware already during the exploitation stage, that means before any changes to
the system or infections have occurred. This approach makes recognition of 
zero-day exploits easier as no signatures are required. Supported programs 
include among others Internet Explorer, Firefox, Office Products, Adobe Acrobat
Reader and Flash. The main features of PwnyPot are general protections like heap
spray mitigation and null page allocation prevention, return oriented
programming (ROP) detection and ROP gadget dumps, moreover detection of possible
DEP and ASLR bypasses. Shellcode is recognized and analysed dynamically. PwnyPot
offers bindings for Cuckoo, one of the leading open source automated malware
analysis systems, which can perform an automatic analysis.

\subsection{Honeypot Related Tools}
\label{sec:tools-hp}

The following tools extend the functionality of honeypots or are meant to be 
used simultaneously with honeypots, for example by making managing tasks easier
or detection executables automatically.

\textbf{Bait-n-Switch}~\cite{baitNswitch2003} was not a honeypot technology as such,
however it was one of the first attempts to multiplex hostile and regular
traffic between production systems and honeypots. It is a system which reacts to
malicious intrusion attempts by redirecting all hostile traffic to a honeypot.
Bait-n-Switch is realized by as a Snort extension, based on linux' iproute2 and
netfilter. The honeypot software can be chosen arbitrarily. The same developers
also  developed a low-interaction honeypot called \textbf{BigEye}~\cite{pouget2003white},
which only emulated FTP and HTTP services.

\textbf{Honeynet Security Console (HSC)}~\cite{HoneynetSecurityConsole} is an analysis
tool to view events on your personal honeynet. It focuses on visualization and
grouping of of events from Snort, TCPDump, Firewall, Syslog and Sebek log files.
Moreover, it correlates information between those different log file types, so
that analysis can be done with a more holistic approach.

\textbf{GSOC-Honeyweb}~\cite{Honeyweb2009} manages client honeypots via a user-friendly
web interface. This application is threefold: The front-end, providing a
standardized interface for various client honeypots; a business layer,
communicating with a Java wrapper and a back-end, providing the data persistence
to collect, store and aggregate client honeypot results. GSOC-Honeyweb should
not be confused with HoneyWeb by Kevin Timm \cite{pouget2003white,
HoneyWebHTTP}.

\textbf{Honeysnap}~\cite{honeysnapHP} is a diagnostic tool which can be used to perform a
number of diagnostics on data which was collected by a server honeypot. The
primary intention is to provide an analysis on a directory full of pcap data. It
decodes and analyses a variety of protocols supporting: outgoing packet counts
and binary extraction for telnet, ssh, http, https, ftp, smtp, and IRC; incoming
and outgoing connection summaries; word based inspection of IRC traffic for
basic keyword profiling. In addition, it focuses on honeypot specific data sets
such as Sebek keystroke data.

\textbf{PE Hunter}~\cite{PEHunterHP} is a plugin for snort for extracting Windows
portable executables from the network stream and is meant to be used in front of
honeypots, which trigger the transfer of the executables. It works by spotting a
PE header, using a simple heuristik to calculate the file length and finally
dumping the corresponding bytes to a file.

\textbf{HoneyMole}~\cite{HoneyMoleHP} incorporates Capture-HPC and supports
administrators to deploy and distribute sensors worldwide which tunnel traffic
in a transparent way to a centralized farm of honeypots. Sensors can be
understood as simple, encrypted ethernet bridges over TCP/IP. The idea here is
that sensors require minimal maintenance efforts, which saves time for
administrators. Moreover, data about attacks is collected in one point, which
saves time for analysts.

\textbf{TraCINg}~\cite{dionaeaDetection} can be described as a cyber incident monitor,
which can receive data from arbitrary honeypots as long as it is well-structured
with JSON. Currently, only a Dionaea-plugin exists. TraCINg collects data from
several honeypot sensors and tries to correlate attacks in order to find
emerging worm outbreaks. It considers mutual attack sources as well as timing 
properties in its analysis.

\subsection{Honeypot Detection Tools}
\label{sec:detection-hp}

Not only do tools exist which extend the functionality but rather are
adversaries of honeypots: honeypot detection tools. This class of tools is able
to detect low- as well as high-interaction honeypots.

Low-interaction honeypots are detectable because of the service emulation, which
will never be able to behave like the real service because of the nature of
emulation and security concerns. This means that specific actions trigger
different responses, as has been shown for example for Kippo and OpenSSH
\cite{kippoDetection}. Specially crafted messages trigger a characteristic
response, which often contains a specific string or number, often called
\textit{magic numbers}. This magic number identifies Kippo. Nmap also detects
some of the Dionaea services as being part of the honeypot and string 
obfuscation is necessary to overcome the signature-based detection
\cite{dionaeaDetection}.

High-interaction honeypots use real services in
constraint environments, therefore fingerprinting them is based on detection of
unusual additional libraries or debuggers and characteristics of virtualization
software \cite{fu2006recognizing}. Holz \cite{holz2005detecting} presented
several techniques, for example VMware uses only a specific range of MAC-
addresses for its virtual network interfaces, chroot and jail environments are
fingerprintable by special \emph{ls} calls and even the presence of debuggers such as
\emph{ptrace} can be detected by simple system calls. It has been
shown with the help of a timing analysis, that honeypots, especially those
running in virtualized environments, respond slower than real services
\cite{mukkamala2007detection}. This fact could also be used for detection,
however has to be used with caution as the response time is also highly
dependant on other factors like network load, routing \etc. Zou
\cite{zou2006honeypot} and Wang \cite{wang2010honeypot} demonstrated that
botnets can be designed to be aware of honeypots. Their work is based on the
following assumption: Honeypots must not participate in real (or too many real)
attacks because of legal constraints. Attackers can detect honeypots in their
botnet by verifying whether the compromised machines can successfully send out
unmodified malicious traffic to attackers' sensors. Sebek was detectable by the
relative address space positions of the \emph{write()} and \emph{read()} calls, as the
integration of Sebek positions these two farther away from each other
\cite{phrackHoneyDetection}. This approach of verifying address space positions
is used even today but for different elements, for example the interrupt
descriptor table register (IDTR) \cite{redPillDetection}.




\subsection{Summary of Honeypot Software}

An overview of available honeypot software, its classification, and
publication details is shown by \autoref{tab:honeypotSoftwareServer} and
\autoref{tab:honeypotSoftwareClient}. One of the first findings is that
different honeypots exist which are applied to different protocols and network
types. This higlights the universality of the concept of  honeypots. Another
finding is that certain honeypot software overlap in their field of operation.
In this cases, the quality and maintenance life time of the honeypot influence
the success.

Despite many different honeypots and related tools, the general trend is
clear: Simple proof of concepts developed into complex honeypot tools which
are designed to be deployed for a long time. As the intended analysis is
getting more complex, management (e.g. GSOC-Honeyweb, Vagrant) and analysis
tools (TraCINg) emerge and even combinations with intrusion and malware
detection tools are considered  (see \autoref{sec:DistinctionHoneypots},
\autoref{sec:tools-hp}).

\autoref{tab:honeypotSoftwareServer} clarifies that the  first honeypot and also
the majority of available honeypot software are low-interaction server
honeypots. The reason for this observations might be the fact, that on the one
hand server honeypots require less implementation effort than client honeypots
as they do not have to have a sophisticated crawler engine and on the other hand
the emulation of services requires less maintenance effort than providing real
services on  high-interaction honeypots. This circumstance led to the situation,
that many developer initially released small proof of concept honeypots, which
should just introduce the concept. Hence, such honeypots (e.g. BOF, 
single-honeypot) had a rather short maintenance life time. After the approval of the
honeypots effectiveness, reliable deployment was necessary, which led to more
complex honeypots which were longer maintained (e.g. HoneyD, SpamD). The
nowadays broadly applied state-of-the-art honeypots (Kippo, Nepenthes-Dionaea,
Honeytrap) distinguish themselves with long maintenance life times. Such life
times  decrease deployment efforts for administrators and foster the development
of  community based plugins. However, it is difficult to determine if the the
long development life time is responsible for success of those tools or vice
versa.  Another general trend is also the focus on the service, low-interaction
server honeypot tend to either specialize in the emulation of on one or a few
well-known services, or  simply perform default answers on all ports. It is also
noticeable, that high-interaction server  honeypots were developed later and
only few exist - Sebek and Argos has been here predominant. Although being a
mere data capture  tool designed to capture attacker's activities, Sebek has to
be highlighted for its pioneering influence. However, only Argos has received
recent updates for newer version of Windows. One of the reasons of Argos
success might be its modern memory taint analysis, which is is very good in the
detection of recent attack types (compare \autoref{sec:ExploitDetection}).

As shown by \autoref{tab:honeypotSoftwareClient}, client honeypots appeared
almost 5 years later than server honeypots. Client honeypots mimic the behaviour
of users in order to rate the risk imposed by the Internet, therefore the list
of emulated and examined software consists of common www-technologies such as
browsers and their plugins, Flash or PDF viewers (Browsers are the primary user
interfaces to the World Wide Web and because of that arguably the most
frequently used program by the common user). As this complex browser
environments are more difficult to emulate, we see the ratio of high and low
honeypots shift towards high-interaction honeypots. Client-honeypot rather
implement less services than server-honeypots, however this is completely
legitimate, as client honeypots actively influence which application 
are  required. As client honeypots do not improve the security of production
networks directly, low-interaction client honeypots usually are published in the
context of Internet-wide research of several years (HoneyC, Monkey-Spider,
PhoneyC etc).  This is why this type of software has relatively long maintenance
time compared to server honeypots. Besides the complete emulation of services
the current  trend in low-interaction client honeypots is the performance, so
that a larger and faster view of the Internet is possible (Thug, YALIH). Another
difference to server honeypots can be observed: For client honeypots, low as
well as high interaction honeypots appeared at the same time. Furthermore, 
high-interaction client and server honeypots appeared almost at the same time.
HoneyClient is the first high-interaction client honeypot and stands out with
his long maintenance. However, it should not be deployed any more as the only
currently maintained honeypot (even with an highly scalable associated project
network) is HoneySpider.

This overview is limited to the classification, maintenance time and the focus
on services, software architecture and its application area. Although this
properties are already enough to assess the deployability to a specific scope,
future work might consider more quality measures such as robustness, quality of
collected data and its ease of analysis, actual containment and  detection
precision. However, an elaborated long-term deployment test of each honeypot
would be necessary, hence it is recommended to narrow down the choice by the
presented overview.

\begin{table*}
	\caption{Chronological overview and classification of server honeypot software by their interaction level type.
		 (+) indicates some	additional services, (++) indicates many additional 
		 services, (*) marks vague timestamps.}
	\label{tab:honeypotSoftwareServer}

\begin{tabular}{clllcp{4.5cm}p{5.000cm}} 
	\toprule

	\multirow{3}{*}{\rotatebox{90}{Type}} & 
	\multirow{2}{*}{Software} &
	\multicolumn{2}{l}{Maintenance} &
	\multirow{2}{*}{Free} &
	\multicolumn{2}{l}{Focus} \\

	\cmidrule(lr){3-4} \cmidrule(lr){6-7}
	& & First & Last & & Services / Applications & Design / Details \\ 
	\midrule

	\colLow & DTK \cite{FirstHoneypot:DTK} & 1997 & 1999 & \cmark &
		SMB, SSH, DNS, FTP, Netstat(++) & 
		implement many known vulnerabilities \\

	\colLow & BOF \cite{Zhang2003HP:bof} & 1998 & 1999 & \cmark &
		Back Orifice, Telnet, SMTP(+) & 
		waste intruders time, easy deployment \\

	\colLow & NetFacade 
		\cite{PolymorphicWorms:HoneypotOverviewChapter} & 1998 & 2002* & 
		\xmark & \textit{not specified} & class C network emulation \\

	\colLow & CyberCop String \cite{honeypotSoftware:Cybercop} & 
		1999 & 1999 & \xmark & 
		Telnet, FTP, SendMail, SNMP &
		emulating different network devices \\

	\colLow & Specter \cite{SpecterHP} & 1999 & 2005 & \cmark &
		SMTP, FTP, HTTP and Telnet(+) &
		commercial deployment, decoy files \\

	\colLow & Sandtrap \cite{HoneypotNETOverview} & 2002* & 2002* & \xmark & 
		dialup modem & war dialing trapping \\

	\colLow & single-honeypot \cite{SingleHoneypotHP} & 2002 & 2002 & \cmark & 
		\textit{all ports, but no emulation} & 
		mere logging, KISS architecture \\

	\colLow & HoneyWeb \cite{HoneyWebHTTP} & 2002 & 2003 & \cmark &
		HTTP & 
		various web server header emulation \\

	\colLow & LaBrea \cite{LaBreaHP} & 2002 & 2003 & \cmark & 
		\textit{all ports, but no emulation} & 
		simple TCP tarpit by SYN/ACK \\

	\colLow & SMTPot \cite{SMTPotMirror} & 2002 & 2003 & \cmark &
		SMTP & 
		spam accumulation, KISS \\

	\colLow & THP \cite{TinyHoneypotGH} & 2002 & 2003 & \cmark &
		SSH (shell), HTTP, FTP  & 
		coexistence honeypot and real services \\

	\colLow & Jackpot \cite{grimes2005honeypots} & 2002 & 2004 & \cmark & 
		SMTP &
		delay spam, utilizing spam databases \\

	\colLow & FakeAP \cite{Github:FakeAP} & 2002 & 2005 & \cmark &
		802.11b AP beacons  &
		p.o.c wireless honeypots\\

	\colLow & HoneyBot \cite{HoneyBotPage} & 2002* & 2007* & \cmark & 
		SSH, SMTP, FTP, HTML(++) &
		windows vulnerabilities and GUI \\

	\colLow & BigEye \cite{pouget2003white} & 2003 & 2003 & \cmark & 
		HTTP, FTP & 
		emulation of different web servers \\

	\colLow & Spamhole \cite{SpamHole} & 2003 & 2003 & \cmark &
		SMTP &
		silent dropping of emails \\

	\colLow & Spampot \cite{Spampot} & 2003 & 2003 & \cmark & 
		SMTP &
		platform independence \\

	\colLow & HoneyPerl \cite{HoneyPerl} & 2003 & 2003 & \cmark & 
		HTTP, FTP, SMTP, Telnet(+) & 
		extensibility by modules \\

	\colLow & Decoy Server \cite{SymantecHP} & 2003* & 2003 & \xmark & 
		SMTP, POP3 & 
		fake email server traffic \\

	\colLow & Smoke Detector \cite{pouget2003white} & 2003* & 2004* & \xmark & 
		FTP, HTTP, IMAP, SSH, SMB(++) &
		honeypot as a hardware \\

	\colLow & NetBait \cite{NetBaitService} & 2003 & 2007* & \xmark & 
		\textit{not specified} & 
		honeypot as a service \\

	\colLow &  HoneyD \cite{Provos2004HoneyD} & 2003 & 2008 & \cmark & 
		HTTP, POP3, SMTP, FTP(+) & 
		emulating heterogeneous networks \\

	\multirow{-15}{*}{\colLow \begin{sideways}low\end{sideways}}
		& KFSensor \cite{KFSensorHP} & 2003 & 2015 & \xmark & 
		HTTP, SMTP, MSSQL, FTP(+)
		& commercial deployment of honeypots \\

	\colLow & SpamD \cite{SpamDBSD} & 2003 & 2015* & \cmark & 
		SMTP & 
		tarpit against spam \\

	\colLow & HOACD \cite{HOASAnnouncement} & 2004 & 2004 & \cmark & 
		\textit{compare HoneyD}&
		live bootable CD (HoneyD, Arpd) \\


	\colLow & ProxyPot \cite{HoneypotNETOverview} & 2004* & 2004* & \cmark & 
		SMTP &
		email spammer identification \\

	\colLow & Impost \cite{ImpostHP} & 2004 & 2004 & \cmark & 
		\textit{all ports, but no emulation} &
		full packet sniffing \\

	\colLow & Kojoney \cite{KojoneySourceforge} & 2005 & 2006 & \cmark & 
		SSH (shell activity) &
		first dedicated SSH honeypot \\

	\colLow & Mwcollect \cite{wicherski2010placing} & 2005 & 2009 & \cmark & 
		\textit{compare Nepenthes, Honeytrap} &
		merging Nepenthes and Honeytrap \\

	\colLow & Nepenthes \cite{NepenthesHP} & 2005 & 2009 & \cmark & 
		FTP, HTTP, TFTP, MSSQL(++)
		& capture worm payload \\

	\colLow & GHH \cite{GoogleHackHoneypot} & 2005 & 2013 & \cmark & 
		HHTP-Apache, PHP, MSSQL & 
		crawler and search engines \\

	\colLow & Honeytrap \cite{HoneytrapGit} & 2005 & 2015 & \cmark & 
		HTML, FTP(+), \textit{dyn. emulation} &
		attacks via unknown protocols \\

	\colLow & HoneyPoint \cite{HoneyPointServer} & 2006 & 2014 & \xmark & 
		\textit{not specified} &
		ICS/Scada, back tracking intruders\\

	\colLow & Dionaea \cite{DionaeaHP} & 2009 & 2013 & \cmark & 
		SMB, FTP, SIP, MYSQL(++) & 
		nepenthes successor, capture payload \\

	\colLow & Kippo \cite{KippoGithub} & 2009 & 2014 & \cmark & 
		SSH (shell activity) & 
		emulate entire shell interaction\\

	\colLow & Artemisa \cite{carmo2011artemisa} & 2010 & 2011 & \cmark & 
		VoIP, SIP & Bluetooth Malware \\

	\colLow & bluepot \cite{bluepotGithub} & 2010 & 2015 & \cmark & 
		Bluetooth & Bluetooth Malware \\

	\colLow & HoneySink \cite{HoneySinkHP} & 2011 & 2011 & \cmark &
		DNS, HTTP, FTP, IRC &
		bot sink holing \\

	\colLow & HoneyDroid \cite{mulliner2011poster} & 2011 & 2014* & \cmark & 
		\textit{compare Kippo, HoneyTrap} &
		p.o.c Android OS honeypot \\

	\colLow & Glastopf \cite{GlastopfHomepage} & 2011 & 2015 & \cmark & 
		HTML, PHP, SQL &
		web applications, vulnerability types \\

	\colLow & Kojoney2 \cite{Kojoney2Github} & 2012 & 2015 & 
		\cmark & SSH (shell activity) &
		applying Kojoneys lessons learned \\

	\colLow & Conpots \cite{conpotHoneypot} & 2013 & 2015 & \cmark & 
		kamstrup, BACnet, mosbus & ICS and SCADA architectures \\

	\colLow & IoTPOT \cite{IoTPOTPaper} & 2014* & 2015 & \cmark & 
		telnet & IoT (ARM, MIPS, and PPC) \\

	\colLow & honeypot-camera \cite{cameraHP} & 2014 & 2015 & \cmark & 
		HTTP & Tornado Web, Webcam Server \\

	\colLow & Shockpot \cite{ShockpotGithub} & 2014 & 2015 & \cmark & 
		Apache, Bash & Shellshock vulnerability \\

	\colLow  & Cowrie \cite{CowrieGithub} & 2014 & 2015 & \cmark & 
			SSH (shell activity) & Kippos successor \\
			
	\colLow & Canarytokens \cite{CanarytokensGithub} & 2015 & 2016 & 
		\cmark & URLs, bitcoin, PDF & honeypot tokens \\

	\colLow \multirow{-17}{*}{\begin{sideways}low\end{sideways}} & 
		elastichoney \cite{elastichoneyHP} & 2015 & 2015 & \cmark & 
		elasticsearch & elasticsearch RCEs \\

	\addlinespace

	\colHigh & Sebek \cite{SebekHP} & 2003 & 2011 & \cmark &
		Win32 and Linux systems & 
		attackers OS activities, state-based \\

	\colHigh & Honeywall \cite{HoneywallCD} & 2005 & 2009 & \cmark & 
		\textit{compare Sebek}, CentOS &
		live bootable CD \\

	\colHigh & HoneyBow \cite{honeybowSensor} & 2006 & 2007 & \cmark &
		Win32 Systems & 
		extraction of malware, state-based \\

	\colHigh & Argos \cite{Argos:2006Qemu} & 2006 & 2014 & \cmark & 
		Linux, Windows XP-7 &
		0-day exploits identification, tainting \\

	\multirow{-5}{*}{\colHigh \begin{sideways}high\end{sideways}} & 
		HIHAT \cite{genericPHPhoneypot} & 2007 & 2007 & \cmark &
		php-BB,-Nuke,-Shell,-Myadmin &
		PHP framework extension, state-based \\

	\bottomrule
\end{tabular}
\end{table*}


\begin{table*}
	\caption{Overview and classification of client honeypot software by their interaction level type. (+) indicates some additional services, (++) indicates many additional services, (*) marks vague timestamps.}
	\label{tab:honeypotSoftwareClient}
\begin{tabular}[th!]{clllcp{4.8cm}p{5.2cm}p{2cm}}

	\toprule

	\multirow{3}{*}{\rotatebox{90}{Type}} & 
	\multirow{2}{*}{Software} &
	\multicolumn{2}{l}{Maintenance} &
	\multirow{2}{*}{Free} &
	\multicolumn{2}{l}{Focus} \\

	\cmidrule(lr){3-4} \cmidrule(lr){6-7}
	& & First & Last & & Services / Applications & Design / Details \\ 
	\midrule

	\colLow & HoneyC \cite{HoneyCHome} & 2004 & 2007 & \cmark &
		HTTP &
		identify malicious servers with snort\\

	\colLow & SpyBye \cite{SpyByeHP} & 2007 & 2007 & \cmark & 
		HTTP &
		proxy, URL check by ClamAV \\

	\colLow & Monkey-Spider \cite{MonkeySpiderHP} & 2007 & 2009 & \cmark & 
		HTTP, JavaScript &
		threat database creation, several AV \\

	\colLow & PhoneyC \cite{phoneycHomepage} & 2007 & 2011 & \cmark & 
		HTML, JavaScript, PDF, ActiveX(+) &
		browser identities, dyn. analysis \\

	\colLow & ADSandbox \cite{dewald2010adsandbox} & 2010 & 2010* & \cmark & 
		HTML, JavaScript & 
		transparent protection, stat/dyn. analysis \\

	\colLow & Thug \cite{ThugGithub} & 2011 & 2015 & \cmark & 
		HTML, JavaScript, PDF, Flash(+) & 
		complete emulation, stat/dyn. analysis \\

	\colLow \multirow{-7}{*}{\begin{sideways}low\end{sideways}} &
		YALIH \cite{YalihGithub} & 2014 & 2015 & \cmark & 
		HTML, JavaScript, (IMAP) &
		precise by combining analysis methods \\

	\addlinespace

	\colHigh & HoneyClient \cite{HoneyClientGit} & 2004 & 2010 & \cmark & 
		Windows (Firefox, IE) &
		proof of concept, state-based \\ 

	\colHigh & Capture-HPC \cite{CaptureHPCHP} & 2004 & 2009 & \cmark & 
		Linux, Windows (Firefox, Office (+)) &
		efficiency, scalability, state-based \\ 

	\colHigh & UW-Spycrawler \cite{UW2006spycrawler} & 2005 & 2006* & \cmark &
		Windows (IE) &
		spyware detection, state-based \\ 

	\colHigh & HoneyMonkey \cite{HoneyMonkeyHP} & 2005 & 2007* & \xmark & 
		Windows (IE) & 
		IE vulnerabilities, state-based \\

	\colHigh & WEF \cite{WEFhomepage} & 2006 & 2007 & \cmark & 
		Windows (IE) &
		drive-by download attacks, state-based \\

	\colHigh & Capture Bat \cite{CaptureBat} & 2007 & 2007 & \cmark &
		Windows (Word, IE) & 
		state changes on a low kernel level \\

	\colHigh & HoneyIM \cite{xie2007honeyim} & 2007 & 2007* & \cmark & 
		\textit{compare Capture-HPC} & 
		instant messaging \\

	\colHigh & SHELIA \cite{ShielaHP} & 2008 & 2009 & \cmark &
		Windows (IMAP, POP) & 
		email malware, call-tracing \\

	\colHigh & Trigona \cite{TrigonaHP} & 2010 & 2010 & \cmark &
		Windows (Browsers) &
		high throughput, --- \\

	\colHigh & HoneySpider \cite{HoneyspiderNetwork} & 2011 & 2015 & \cmark & 
		\textit{Capture-HPC, THUG} &
		hybrid client honeypot framework \\

	\colHigh \multirow{-11}{*}{\begin{sideways}high\end{sideways}} & 
		PwnyPot \cite{pwnypotGit} & 2013 & 2013 & \cmark &
		Windows (Browsers+Plugins) &
		memory corruption, shellcode detection\\


	






	\bottomrule
\end{tabular}
\end{table*}

\section{Long-Term Honeypot Projects}
\label{sec:honeypotProjects}		

This section introduces long-term honeypot projects and 
meta-projects, which are responsible for the creation and publication of 
significant honeypot software, data and several research papers. Rather short 
projects and honeypot deployments are not enlisted here, but considered in the 
next section which presents metrics used in the field of honeypot data analysis.

The first honeypot project was the \textbf{Honeynet Project}, which started off
in 1999 and released a paper series which has been combined in the well-known book
\emph{Know Your Enemy} \cite{honeynet2004knowUrEnemy}. This project focuses on
investigating the latest attacks and is developing open source security tools to
improve internet security, recently also at the Google Summer of Code. Its
chapters are spread out around the world and use different tools to collect
informations, based on the needs client/server and low-/high-interaction
honeypots are used. The project is still alive and by far the largest honeypot
association, which comprises other smaller (regional) honeypot projects.

One of the very first large-scale worldwide honeypot projects was launched by
the Institut Eurocom in 2003 and is called the \textbf{Leurre.com} project. Its
last publication dates back to 2008 and describes the infrastructure and the
main insights. This project is based on a worldwide distributed system of 
low-interaction honeypots present in more than 30 countries. The main objective
is to get a more realistic picture of internet threats by collecting unbiased
quantitative data and create a long-term and location-independent perspective.
\cite{leurre2008long}.

The European Network of Affined Honeypots, 
also known as the  \textbf{NoAH-Project}, is a project
coordinated by the  Foundation for Research and Technology Hellas (FORTH) and 
several European academic and business partners. This three-year project 
gathered and analysed internet attacks by
deploying the high-interaction Honeypot Argos. The aim was to help NRENs
(National Research and Education Networks) and ISPs (Internet Service Providers)
to limit damage to their networks and to better assess threats. This project
started in 2005 \cite{NoahProject}.

Deriving from the mwcollect development, the \textbf{mwcollect Alliance} is a
non-profit community aiming primarily at malware collection. This closed
community deployed Nepenthes sensors on the internet. Analysis of the data was
performed on the Alliance's server in real-time. This project was mainly active
from 2006 to 2009. Members were chosen by a simple email verification, however
one of the requirements were frequent contributions of new samples, that means
active Nepenthes sensors. Mwcollect alliance consisted of not evenly distributed
sensors across European countries \cite{allianceMwcollectHP,
wicherski2006medium}.

\textbf{Telekom-Fr\"uhwarnsystem} was first presented in 2013, however early
cooperations with worldwide partners started in 2012. From the beginning, this
project was based on a multi-honeypot platform. Initially, partners had to set
up the honeypots by themselves, by this time T-Pot exists, which combines common
open-source low-interaction server honeypot tools (dionaea, glastopf, kippo,
honeytrap) and  reduces the maintenance for partners. Furthermore, the project
is now open for uncertified contributors, however this data and its analysis are
marked and separated as the community data set \cite{sicherheitstacho}.



An overview of the various honeypot projects is shown in
\autoref{tab:honeypotProjects}.

\begin{table*}
	\caption[Long-term Honeypot Projects]
		{Overview of long-term honeypot projects and alliances.}
	\label{tab:honeypotProjects}
	\centering
	\begin{tabular}{llcc} 
		\toprule
		Project & Topology & Begin & Duration \\ 
		\midrule
		Honeynet Project \cite{honeynet2004knowUrEnemy} & Multi-Honeypot 
			Platform & 1999 & ongoing \\
		Leurre.com \cite{pouget2005advantages:seq, leurre2008long} & HoneyD 
			Sensors & 2003 & 5 years \\
		NoAH-Project \cite{NoahProject} & Argos Sensors & 2005 & 3 years \\
		mwcollect Alliance \cite{allianceMwcollectHP} & Nepenthes Sensors &
			2006 & 3 years \\
		Telekom-Frühwarnsystem \cite{sicherheitstacho} & Multi-Honeypot
			Platform & 2012 & ongoing \\

		\bottomrule
	\end{tabular}
\end{table*}


\section{Honeypot Data Analysis}
\label{sec:honeypotMetrics}

In the following, we describe the process of analysing honeypot data in a step-by-step
manner by means of analysis questions, which gives an overview about different
methods and metrics in the field of honeypot data analysis.

The deployment of honeypots creates log records which describe the occurred
incidents. The possible incidents are dependant on the type of honeypot
deployed. Therefore it is important to define the problem statement that should
be investigated before the actual deployment. If honeypots should look actively
for communication partners, a client honeypot has to be used, a server honeypot
otherwise. If we rather investigate meta-information of protocols and
transactional data for specific services, low-interaction honeypots are the
right choice. High-interaction honeypots should be used, if content, shellcode
execution and the integrity of the operating system are of concern. A single
problem statement can be answered by various metrics, each with its own approach
and accuracy.

Hence, the overview in this section will follow the presented honeypot
classification and will be additionally divided into problem statements and the
associated metrics. However, this section depicts merely metrics used in data
analysis, performance metrics which are usually applied to benchmark a honeypot
are not described, as they are mostly self-explanatory and just measure CPU,
RAM, HDD load or scalability. 

In general, no research presents any in-depth honeypot data analysis before the
year 2004. Prior to this, analysis consisted of mining IDSs alarms. However,
IDSs can only be seen as complementary analysis tool. Honeypots can bring more
information than simply provided by IDSs, especially if static signature based
IDSs are used \cite{pouget2004honeypot}.

\subsection{Attack Profile}
\label{sec:attackProfile}

McGrew \cite{mcgrew2006experiences} suggested that an attack profile which
provides useful information about an honeypot attack should contain information 
about the following attributes:
\begin{LaTeXdescription}
	\item[Motivation] Motivation describes the reason of the attack. 
	\item [Breadth/Depth] The breadth of the attack is described by the number
			of machines affected and the depth is the degree to which a specific
			target was analysed or how large the impact of the attack was on the 
			system.
	\item [Sophistication] Sophistication characterises the level of expertise
			required to perform a specific attack. 
	\item [Concealment] Concealment measures the quality of hiding the 
			evidence of the attack.
	\item [Attacker Source / Root Cause] Attackers should be, as far as 
		possible, identified or the root of the attack like a specific worm 
		depicted. 
	\item [Vulnerability] Vulnerability is the flaw in the system/protocol that
		allowed the attack to take place.
	\item [Tools] Also the tools involved on a high interaction attack like 
		root-kits or back-doors should be documented.
\end{LaTeXdescription}

Discussions about attacks on honeypots should always have such an 
profile as their basis. Motivation often can only be guessed, however actions
on high-interaction honeypots might reveal some insights. The breadth and depth
can be derived from the attack frequency, the attack propagation and on 
high-interaction honeypots by the degree of the infection. Concealment on 
low-interaction honeypots is depicted by slow, not invasive but long attacks and
on high-interaction honeypots by the quality of installed back-doors and 
root-kits. The attack source can usually be determined by transactional meta
information, however the root cause of the found attack might be more difficult to identify,
as it tries to explain the actual observations. Vulnerabilities are usually 
identified by exploit detection techniques. These characteristics are now 
discussed in the following subsections.

\subsection{Attack Sources} 
\label{sec:attackSources}

If attacks occur on a honeypot, one has to specify where the attacks came from.
The identification of an attacker is independent from the architecture or
interaction type of the honeypot and can be done with different granularities. 
\begin{eqnarray} 
\text{IP-Address or IP-Prefix} \\ 
\text{Autonomous System Number} \\ 
\text{Domain Name / URL / URL-Type} \\ 
\text{Country} \\ 
\text{UserID / Email} \\
\text{User Agent} \\ 
\text{Operating System} 
\end{eqnarray} 
However, in the case of the server honeypots one has to consider that they might
have received a spoofed IP-address. This might be a valid IP-address with a
reachable or unreachable host, or it is a martial IP-address which should not
have ever left the local segments like the broadcast 
address 0.0.0.0 \cite{francois2008graphs}. Client
honeypots usually are seeded with URL-lists and crawl for new URLs. The
resources behind those URLs might be off-line. Moreover, one has to acknowledge
that this identifiers are very  changeable. IP-addresses can move from host to
host because internet network providers use IP-address-pools. That is the reason
why some analysis combine IP-addresses with a timestamp and define an attack source
as an IP-address that targets the honeypot environment for example within one
day \cite{pouget2004honeypot}. IP-Prefix announcements from autonomous systems
change over time or might be even hijacked. The Domain Name System inherently
allows the abstraction from  IP-address and/or hosts, which might lead to
misleading results. Client honeypot such as Monkey-Spider
\cite{ikinci2008monkey} also perform some type of URL classification based on
the URL and page-content, e.g \textit{adult content}, \textit{pirate},
\textit{typos} etc.. The country might be extracted from the AS-registration
information or some (commercial) third-party product has used to be used to
retrieve the data \cite{kyotoDataset, mcgrew2006experiences}. However, trends
show that usually the top 3 countries cause 60\% of the traffic, which countries
are observed is depended on the geographical location of the honeypot node.
Another way used to identify spammers in instant messaging networks is the user
name or the advertised URL \cite{trivedi2007analyzing}. Spammers tend to create
a lot of accounts, which distribute many different URLs, which however
lead/redirect to only a small subset of websites. A minor correlation between
spim and spam senders exist. Research based on honeypots supporting the SIP/VoIP
protocol also use the name of the user-agent for the fingerprinting of the
attack source \cite{valli2010voip}. This information can be used by any protocol
with such protocol-tag, however one has to keep in mind that such information
can be omitted and easily spoofed.  In order to infer the operating system from
which the attacks originate from, usually additional
passive OS fingerprinting tools like p0f are used, which recognize the attacking
OS by analysing the packet composition as each OS creates the packets slightly
differently. Almost all attacking machines are Windows-based
\cite{leurre2008long}.

\subsection{Attack Target}
\label{sec:attackTarget}

If it is specified who attacks the honeypot, the next step might be to
characterize the attack, more precisely one has to determine the target of the
attack. Server honeypots classify the target by a specific service, which is
usually bound to a dedicated port. Port numbers are administered by the Internet
Assigned Numbers Authority (IANA) and viewable in official lists. However,
services might be bound to another port. Therefore it is important to
differentiate between ports and services, as an intruder might to brute force a
SSH-service on another port, which can be recognized by incoming valid 
SSH-packets on non-default ports. Most of the time, services are bound to 
default ports to improve reachability, that is why many researches treat a port
as representative for a service \cite{pouget2004honeypot, mcgrew2006experiences}.
If a honeypot monitors a whole network each 
individual IP-address can be seen as a target identifier. Such networks can be
classified in detail into campus, enterprise, serive provider networks. 
Measurements indicate differences between these classes of networks 
\cite{yegneswaran2004design}.
\begin{eqnarray}
\text{IP} \\
\text{Port} \\
\text{Service}
\end{eqnarray}
Client honeypots use a software client, which accesses a potentially malicious
remote service. Therefore the target is usually the specific client software. It
might be an emulated web-browser for low-interaction honeypots or a real one
with plugins like Flash for high-interaction honeypots.
\begin{eqnarray}
\text{Software Client} \\
\text{Software Plugins}
\end{eqnarray}
Moreover, high-interaction honeypots (client as well as server) allow 
modifications to the operating system. Therefore OS-specific changes have to be
analysed, which might differ across systems. For Linux it usually means the
loading of some hidden kernel-modules and new cron jobs, for Windows Systems 
changes in the registry, system files and auto start entries. So an analysis
might examine which OS is preferably attacked.
\begin{eqnarray}
\text{OS, OS component} 
\end{eqnarray}

\subsection{Attack Frequency}
\label{sec:attackFrequency}

One of the fundamental questions which has to be answered while deploying
honeypots is if and how often honeypots are attacked? Interestingly, honeypots
are exposed to attacks minutes after they have been activated
\cite{mcgrew2006experiences}. However, this hold only true if the honeypot is
accessible from the internet, if a firewall blocks all incoming connections to
the firewall and only  internal communication is allowed attacks are observed
rarely as they would have to come from infected hosts from within this specific
network or because of a   local misconfiguration.
\begin{eqnarray}
\text{Time until First Attack}
\end{eqnarray}

Yegneswaran et al. \cite{yegneswaran2005using} defined three metrics to describe
the source arrivals in order to find differences for the events of
misconfiguration, bot-attacks and worms-attacks. These are \one the temporal source
counts, \two the arrival window and \three the interarrival distribution. The first is
analysed by the number of sources per time interval and shows distinctive
patterns. Worms show a logistic growth with a steep begin and end, as they
propagate very fast and autonomously and are shut down abruptly by a patch. Bots
show similar characteristics, however bots usually apply a poll and pull
communication pattern with their C\&C server using a wake up time every $x$ 
seconds, which results in less steeper curves. The arrival window checks how
many new sources have arrived in a specific time frame. 
Using a cumulative distribution function (CDF) plot no 
differences betweens these events could have been spotted. To evaluate the source
interarrival characteristics, the data is broken up in successive intervals,
each with an equal number of sources (e.g., 10 intervals each with 10\% of new
sources). Then the distribution of interarrival times is plotted. Bot-attacks
and worm-attacks show exponential interarrivals. Moreover, the source-net 
dispersion can be interesting. 
Worm outbreaks have a much higher dispersion than
bot nets and misconfigurations. A histogram can be computed on the count of
sources seen from each /8 address aggregate if IP-addresses are considered,
however other aggregates can be used.
\begin{eqnarray}
\text{Number of Sources per Time Unit} \\
\text{Number of new Sources per Time Unit (CDF)} \\
\text{Interarrival Time Distribution for }\nonumber \\
\text{equally-sized  Source Intervals} \\
\text{Number of Sources in specific IP-Aggregates}
\end{eqnarray}

Another metric combining the attack sources and frequency is relating the number
of IP-addresses as a function of the number of attacks for each address
\cite{kaaniche2007empirical}. This histogram follows the power-law distribution.
\begin{eqnarray}
\text{Number of Sources per Number of Attacks}
\end{eqnarray}

As we already have clarified the term attack is dependent on the honeypot type
used. Server honeypots do assess any communication as malicious, hereby 
low-interaction server honeypots describe the attack frequency necessarily based
on network properties like:
\begin{eqnarray}
\text{Received Packets per Time Unit} \\
\text{Received Data (kB) per Time Unit}
\end{eqnarray}
Measurements \cite{yegneswaran2004design} show, that if TCP is the dominant
procotol, packet sizes are relatively constant and hence the ration between
received packets and data per time unit is predictable.
Attack frequencies usually show specific peaks, instant massaging spam for 
example shows two daily peaks and one if observed on a weekly scale 
\cite{trivedi2007analyzing}. Moreover, peaks in attack frequencies can usually 
be linked to a single service, worm-activity etc. \cite{kyotoDataset}, which is
heavily exploited at this specific point of time. Instant messaging and email
honeypots can use the following metrics:
\begin{eqnarray}
\text{Messages/Emails Received per Time Unit} \\
\text{URLs / Attachments Received per Time Unit} \\
\text{Received Data (kB) per Message}
\end{eqnarray}

For high-interaction server honeypots the same metrics apply, however they can
be extended by a OS-specific metric:
\begin{eqnarray}
\text{Exploitations per Time Unit}
\end{eqnarray}
Client honeypots only count, independent from their interaction level, the 
exploitations per time unit and do not consider network features for the
attack frequency as they actively begin the communication. That means, that it 
is preconfigured at which rate a client honeypot makes communication requests.
So only exploitations are considered.

Another procedure used is the sessionization \cite{kaaniche2007empirical} of the
data. All the packets received from the same source within a time frame or
without triggering a timeout are supposed to belong to the same attack session.
24 hours frames or 30 minutes timeouts are common. Furthermore, the time between
the occurrence of an attack and the next attack can be examined. The probability
density function (PDF) for this metric follows a heavy-tailed power-law and can
be modelled by mixture of a Pareto and an exponential distribution. The lifetime
of a source can be described as the complete time we see a source active on a
honeypot \cite{yegneswaran2005using}, that means it is the time span from the
source's first occurrence up to is lasts activity and might include several
sessions. Botnets and misconfigurations cause short lifetimes, however worms
prove to be persistent as they often miss a mechanism to stop scanning.
If a specific source is observed regularly (that means it has frequent sessions
or one long ongoing session),  then it has a long source lifetime.
Additionally, Song et al. \cite{kyotoDataset} differentiate their sessions based
on a IDS  classification: All traffic data to the honeypot which triggers an IDS
alert is labelled as a known attack session, all traffic data which ends with
the transmission of shellcode but does not trigger an IDS alert is an unknown
attack session. The time between sessions is also worth considering, as it 
displays the pause between active sessions.
\begin{eqnarray}
\text{Number of Sessions per Time Unit} \\
\text{Session Duration} \\
\text{Time between Sessions} \\
\text{Source Lifetime} \\
\text{(Un-) Known Attack Sessions per Time Unit}
\end{eqnarray}

Song et al. \cite{kyotoDataset} demonstrated a plot, which visualizes the
frequency of the most targeted port per day. They plot the destination port as
a function of time using a log-scale for the ports. The log-scale is an 
advantage because most of the attacks happen on the smaller well-known ports. The
curve shows jumps across discrete levels representing ports of well-known ports
for SSH, SMB and so on.

Similar to the notation of sessions are the flows
\cite{almotairi2008principalComp}. The basic flows are based on the basic 
IP-flow and described by a 5-tuple consisting of source and destination 
IP-address, source and destination port, protocol type. The attack frequency can
also be described by the occurrence of basic flows: If a packet differs from 
another packet by any key-field or arrives after a time-out, it is considered to
belong to another flow. Therefore flows are a more strict requirement than 
sessions. Activity flows are an aggregation of basic flows based on the source
IP-address only with a timeout for the inter-arrival time between basics flows.
Hence, they resemble the definition of session.
\begin{eqnarray}
\text{Number of Basic Flows per Time Unit} \\
\text{Number of Activity Flows per Time Unit}
\end{eqnarray}

\subsection{Attack Evolution}
\label{sec:attackEvolution}

If we observe certain temporal patterns for a specific source, port, country
\etc, it may be important to detect unusual behaviour for it automatically,
because those anomalies might mark important events. That means, we want to
learn, what normal behaviour is and to spot if this normal behaviour changes.

One possible method is to calculate ratios for different time aggregates and
compare those values for different days or to the average ratio. This method is
useful to recognize temporal trends which are only visible on a specific time
scale \cite{thonnard2008framework}. The choice of a good time granularity
depends on the kind of attack phenomena which is investigated: For short 
high-intensity attacks, like botnet probes or flash worms, it may be more useful 
to apply smaller time units, while for worms with a stealthier propagation 
scheme a larger time unit should be used. 

%
%

Francois et al. \cite{francois2008graphs} demonstrated that graph intersections
can be used to analyse distributed honeypot platforms. This method allows to
highlight changes in the relationships between honeypots, for example if the 
percentage of mutual attacking IP-addresses changes for two nodes. Their 
research is based on two metrics, which create one value for the complete 
honeynet and not one value for each honeypot, which makes the analysis more 
simple. First, the maximal locality statistic, which is strongly related to the
centrality measurement of graphs.
$\psi_k(v)$ denots the number of arcs of the subgraph of neighbours of $v$
at a max. distance $k$.
\begin{eqnarray}
M_k = \max_{v \in \text{\textit{nodes}}} \psi_k(v)
\end{eqnarray}
Second, the standardized locality statistics at time $t$ of the distributed
system, which is calculated with respect to previous values of a sliding window
with a size of $\tau$.
\begin{eqnarray}
\tilde{\psi}_{k,t}(v) & = & \frac{\psi_{k,t}(v)-\hat{\mu}_{k,t,\tau}(v)}
	{\max(\hat{\sigma}_{k,t,\tau}(v), 1)}
\end{eqnarray}
with common average value
\begin{eqnarray}
\hat{\mu}_{k,t,\tau}(v) & = &
	\frac{1}{\tau}*\sum_{t'=t-\tau}^{t-1} \psi_{k,t'}(v)
\end{eqnarray}
and variance
\begin{eqnarray*}
\hat{\sigma}_{k,t,\tau}(v) & = & \frac{1}{\tau-1} 
	\sum_{t'=t-\tau}^{t-1} (\psi_{k,t'}(v)-\hat{\mu}_{k,t'}(v))^2  \\
\tilde{M}_{k,t} & = & \max_{v \in \text{\textit{nodes}}} \tilde{\psi}_{k,t(v)}
\end{eqnarray*}
The standardized locality statistic of a node is nearing zero if its number
of edges remains stable. High positive or negative values point out
significant changes in the graph structure. However, this is not a necessary
conditions. The standardized locality and especially the maximum
standardized locality may not differ after a change, if the number of
edges for individual nodes did not change.

The first analysis was done to identify honeypots which capture unique attacking
IP-addresses. Two nodes in the graph are linked only if the intersection between
the corresponding sets represents less than a threshold $\alpha$ of the union of
addresses. Therefore, central nodes (high locality value) capture unique 
IP-addresses. When the maximal standardized locality statistics is low, no
changes have occurred, however a high value indicates a major topology change: 
the relationships of attacking IP-address differ to the previous time instances.
A similar analysis can be done for ports, an edge connects to nodes only if the
set intersection of their attacked ports is lower then a threshold $\beta$.

Another way to distinguish between known and new patterns and highlight their
occurrence was presented by Yegneswaran et al. \cite{yegneswaran2005using}. They
deployed a combination of honeypots and the intrusion detection system Bro.
Therefore their events are based on Bro-profiles. However, their methodology
can also be applied to pure honeypot  data. They use a deviation value $\beta$
to detect large-scale and unusual  events. $\beta>10$ indicates botnet-waves and
fast-scanning worms, $\beta>3$ slow-scanning worms.
\begin{eqnarray}
\beta_{p_i} & = & m p_i / \sum^{i-1}_{j=0}{p_j}
\end{eqnarray}
where $p_i$ denotes the number of sources triggering profile $p$ in time interval $i$,
$m$ denotes the number of intervals prior to $i$ where $p$ was observed.

Bro was also used to create profiles for new events, however one of the findings
concludes that usually new events are just new minor variations of known
activity. 

Kaaniche et al. \cite{kaaniche2007empirical} presented a time evolution model
created by linear regression. They examined, if a model based on observations
restricted to attacks originating from a specific country can describe the
complete data set reliably. Surprisingly, they found a strong correlation
between the models for single countries and the overall data set, independently
from the countries proportion of the total number of attacks (some of the best
fit countries account only for 2\%-20\% of attacks). The linear regression model
is defined by:
\begin{eqnarray}
Y^*(i) & = & \sum{\alpha_j X_j(t)} + \beta,\quad j = 1, 2, ..k
\end{eqnarray}
where $X_j(t)$ denotes observed attacks from country $j$,
$\alpha_j$ and $\beta$ are best fit linear model parameters.

The correlation between models is measured by the correlation factor:
\begin{eqnarray}
R^2 = \sum{(Y^*(i)-Y_{av})}^2 / \sum{(Y(i)-Y_{av})}^2 \text{ where} \nonumber\\
Y(i) \,\text{observed attacks},\,
	Y_{av} \,\text{average number of attacks}
\end{eqnarray}
Such correlation factors can be used to rate if the observed events are expected
or vary substantially, that means have a high similarity distance to the model.

\subsection{Propagation of Attacks}
\label{sec:attackPropagation}

Besides analysing the attack activities in a isolated manner, one should also
try to identify the propagation of attacks across several honeypots if a large
honeynet is deployed. Propagation takes place, when one attacking IP-address is
observed on one platform, then subsequently on another
\cite{kaaniche2007empirical}. Because of the IP-address-pools, this check for
reoccurrence should happen within a specific time-frame for more precise
results. Already early distributed honeypot analysis show that it is beneficial
to deploy a large amount of honeypots from different IP-subnets and different
geographical positions \cite{chun2003netbait}, as it is more likely to spot an
attack, local events can be characterized as such and the propagation of an
attacker across targets can be depicted.

Propagation can be modelled by a propagation graph \cite{kaaniche2007empirical},
where nodes represent the individual honeypots and the edges $(i,j)$ describe
the probability to discover a seen IP-address at node $i$ also at node $j$.
However, nodes tend to show low propagation value, if not in the same subnet.
\begin{eqnarray}
\text{Propagation Graph}
\end{eqnarray}

Similarly, Vasilomanolakis et al. \cite{vasilomanolakis2015honeypot} describe
the propagation of attacks by \textit{single-dimen\-sional correlation} and
\textit {two-dimensional correlation}. 

Single-dimensional correlation aggregates attacks from the same origin, if the
attack origin has been observed on at least 2 sensors. This correlation
is analysed by two visualisations. First, a directed graph is created with nodes
for all honeypots and observed attackers. A directed edge represents an attack to a 
honeypot, that means that multiple edges to different sensors mark the attackers
presence on several honeypots. Second, the ratio of unique attackers that have
been observed on various honeypots is calculated. Vasilomanolakis findings suggest
that the presence on more than two sensors is very unlikely for attackers. 
Two-dimensional correlation includes
time as an additional dimension, which means that a mutual attack has to be
observed on at least two sensors within a specific time-frame. As already
discussed, this time frame should be below one day because of IP-address
pooling. Vasilomanolakis argues reasonably that internet-wide scanning
has been significantly improved in the last years: 
Publicly available tools like ZMap \cite{durumeric2013zmap} 
are capable of performing a complete scan of the IPv4
address space for one port with one probe per host in about 1 hour. Therefore,
in order to find strong relations one can set the time frame to 1 hour and even
lower. Lastly, they use a scatter plot, which uses time slots of one hour and
plots the number of unique attackers present on several sensors, in 
addition a colour signifies on how many sensors the attackers were present. 
Their observations suggest, that at least one unique attacker is targeting 
multiple sensor per time slot. 
\begin{eqnarray}
\text{Attack Graph} \\
\text{Ratio of Unique Attackers in } \nonumber \\
\text{Relation to Number of Targeted Sensors} \\
\text{Number of Unique Mutual Attackers in} \nonumber \\
\text{Relation to Targets per Time Slot}
\end{eqnarray}

A phase plot can also be used to visualize the successive targets, showing the
next target as a function of the last target for a specific amount of attack 
samples \cite{yegneswaran2005using}. Sequential IP-address scans will appear
as a straight diagonal line in this visualization. Phase plots can also 
visualize the coverage, which is the number of probed honeynet IP-addresses 
by a specific source. A full coverage can be recognized as a horizontal line.
\begin{eqnarray}
\text{Phase Plot: Successive Destination Targets} \\
\text{Phase Plot: Destination Targets for specific Sources}
\end{eqnarray}
%
%

Attacks on different targets can be visualized by a destination-net scan 
footprint \cite{yegneswaran2004design} which is a plot counting the number of 
attack sources over all targets they have attacked. Obviously, this
visualization works best if many honeypots are deployed or whole subnets are
redirected to a honeypot. Misconfigurations tend to show hot spots, worms and
bot cause an evenly distributed pattern. Furthermore, the first destination 
preference might be interesting to analyse, as this might reveal some ordering 
in the scanning of subnets by worms and bots. 
\begin{eqnarray}
\text{Destination Net Scan Footprint} \\
\text{PDF of First Destination Preference}
\end{eqnarray}

\subsection{Attack Patterns}
\label{sec:attackPatterns}

The general concept of many data mining tasks, like common pattern detection and
clustering, involves the following procedure with three steps 
\cite{jain1988algorithms}:

\begin{enumerate}
\item feature selection and/or extraction, pattern representation
\item definition of a pattern proximity measure appropriate to the data domain
\item grouping similar patterns
\end{enumerate}

The first step includes the extraction of certain features characterizing the
relevant aspects of the data set and representing them with an adequate means,
for example an array of values. An effective measure to describe the similarity
of two data series is done by a similarity distance such as Mahalanobis,
Pearson, Spearman \etc. The grouping or clustering of patterns is done
by clustering algorithms like the K-Means-Algorithm. 

Unfortunately, the discipline of pattern detection does not offer a straight
forward method for all data types. Not every algorithm can handle all cluster
shapes or sizes and runtime or output quality might differ severely on
different data dimensionality and types. Different clustering algorithms produce
different partitions of data, and even the same clustering algorithm is
dependant on its initializations and configurable parameters. Indeed, the real
skill in pattern detection is the choice of a proper clustering algorithm (and
similarity measure) as hundreds of clustering algorithms exist
\cite{jain2010data}. This is the reason why we see so many different approaches
in the field of honeypot attack pattern detection, but also any other clustering
discipline.

The problem of network traffic clustering and anomaly detection is not a new
discipline and has been extensively studied. Approaches commonly utilise
signature based methods in combination with intrusion detection systems,
statistics (e.g. Moving Average Models\cite{Barford:NetworkAnomalies,
Brutlag:AbberantBehaviour}, Principal Component Analysis \cite{
Lakhina:CharacterizationFlow, Lakhina:MiningAnomalies}) or data mining and
unsupervised machine learning (e.g hierarchical clustering
\cite{portnoy2001intrusion}, KNN-clustering \cite{eskin2002geometric}).
However, only some research was done explicitly in the field of honeypot traffic
analysis \cite{Owezarski:ClusterEnsemble}. Therefore, we concentrate on the
publications which effectively deployed and/or analysed honeypot and honeynet
traffic.

The widespread procedure of association rule mining was applied by Pouget
\cite{pouget2004honeypot} to find interesting relationships and patterns
between observed events. With the induction of
 association rules one tries to find sets of
\textit{items}, i.e. events, port numbers, IP-addresses and so on, that
frequently occur together. It originates from customer behaviour analysis, that
tries to recommend products based on sets of collectively bought items. This
means that an association rule $R$ states that if we see specific action $a$ and
$b$, we can be confident, quantified by a percentage, that also action $c$ will
be observed: $a \cap b \Rightarrow c$. The metrics which are applied are
$support$ and $confidence$. Support is the ratio between the number of
transactions that include all items of the rule and all transactions. Confidence
is the ratio between the number of transactions that include all items of the
rule and the transactions that contain the premise. Rules should have a minimal 
support threshold so that only meaningful rules are found.
\begin{small}
\begin{eqnarray}
Supp(R) & = & \frac{\# transactions \,\, incl. \, \{a \cap b \cap c \}}
	{\# transactions} \\
Conf(R) & = & \frac{\# transactions \,\, incl. \, \{a \cap b \cap c\}}
	{\# transactions \,\, incl. \, \{a \cap b \}}
\end{eqnarray}
\end{small}
Pouget applied association rules to each found port sequence group and mined on the
following features: number $T$ of machines in the environment targeted by one
attack source, $n_i$ number of packets sent by attack source to honeypot $i$ and
$N$, which is the total number of packets sent by one attack source to the whole
environment. The resulting rules represented meaningful clusters, which have
been ascribed to attack tools and offer a good alternative to clustering
by port sequences only.

One of the first extensive works on honeypot data was done by Thonnard et al.
\cite{thonnard2008framework}, who group attacks on a honeynet by detecting
common time series patterns. They have used the data from the leurre.com project
and their data aggregation is based on the ordered list of ports targeted by a
source identified by an IP-address. All attackers having the same attack
fingerprint are classified into one set, then the number of unique source counts
per time unit for each class is calculated. Furthermore, only time series' which
have at least one peak of activity with a minimum of $x=10$ sources for a given
time unit are considered. That means, that featureless data (port sequences
which are not mutual) is filtered to assure a certain quality. The symbolic
aggregate approximation (SAX) is used to reduce the dimensionality of the data
and to make a fast similarity distance evaluation possible. SAX approximates
time series data by segmenting into time intervals of equal size and summarizing
each of these intervals by its mean value. Each interval is then mapped to a
finite alphabet symbol. The alphabet is chosen relatively, i.e. symbol $B$
representing the range 20\%-30\% of the maximum unique source count \etc. Time
series can then be interpreted and compared as a string. Thonnard defined
similarity if more then 90\% of the symbols match for a given pattern and used
an alphabet with 8 symbols. If $N$ is the number of elements in time series $T$,
the $dist()$-function returns the inter-symbol distance and $\omega$ is the
number of intervals in the SAX representation, then the \textit{minimum SAX
distance} can be calculated as follows:
\begin{eqnarray}
SAX(W_{T_1}, W_{T_2}) = \sqrt[\sqrt{\frac{N}{\omega}}]
	 {\left(\sum_{i=1}^\omega dist(W_{T_1}(i), W_{T_2}(i))\right)^2}
\end{eqnarray}
Comparing the resemblance of temporal behaviour, Thonnard et al. found only 
three patterns of attacks:
\begin{enumerate}
\item continuous activity
\item sustained bursts
\item ephemeral spikes
\end{enumerate}
Especially the ephemeral spikes can lead to a false similarity measurement, 
because the symbols of the alphabet are chosen relatively due to a standardization process.
This means that temporal patterns with only a few spikes and many zeros or very small
values have a mean value close to zero. SAX calculates a high similarity
degree because all these values are represented by only one symbol. However,
a similarity has not to be the case. Therefore a global and local similarity
measure is necessary. Global similarity $SIM_G$ is defined using the largest 
lower-bounding distance that is theoretically possible, as denoted by the 
abstract variables $W_T$ and $\tilde{W}_T$ which have maximal distance for every
pair of symbols. Local similarity $SIM_L$ compares only values, if one of the
patterns exceeds the upper quantile value $UQ$ at a given time unit, 0.975 was 
used by Thunnard. Again, $W_T$ and $\tilde{W}_T$ are used to calculate the
largest distance between every pair of symbols based on $X_{UQ}$. Both measures
are combined to obtain the total similarity.

\begin{tiny}
\begin{eqnarray}
SIM_G(W_{T_1}, W_{T_2}) &=& 1 - \frac{SAX(W_{T_1}, W_{T_2})}
	{SAX(W_{T}, \tilde{W}_{T})} \\	
X_{UQ} &=& \cup_{k=1}^2 \left(x_i | W_{T_k} (x_i) > UQ, \quad 
	\forall i \in \{1, ..., |W_{T_k}|\} \right) \\	
SIM_L(W_{T_1}, W_{T_2}) &=& 1 - \frac
	{ \sum_{j=1}^{|X_{UQ}|} SAX( W_{T_1}(x_j), W_{T_2}(x_j) ) }
	{ \sum_{j=1}^{|X_{UQ}|} SAX( W_{T}(x_j), \tilde{W}_{T}(x_j) ) } \\
SIM_{total} &=& \frac{SIM_G + SIM_L}{2}
\end{eqnarray}
\end{tiny}
The clustering algorithm applied is a greedy algorithm, which takes a pattern and
combines all other patterns which exceed the similarity threshold to a cluster 
and then excludes the found cluster from the pattern list. Afterwards the next
remaining pattern are analysed. Clique-similar clusters emerge. Despite the 
local decision making scheme, fairly good results are produced and another
advantage is that the number of total cluster has not to be known in beforehand.

\begin{figure}[t!]
	\centering
	\includegraphics[width=0.45\textwidth]{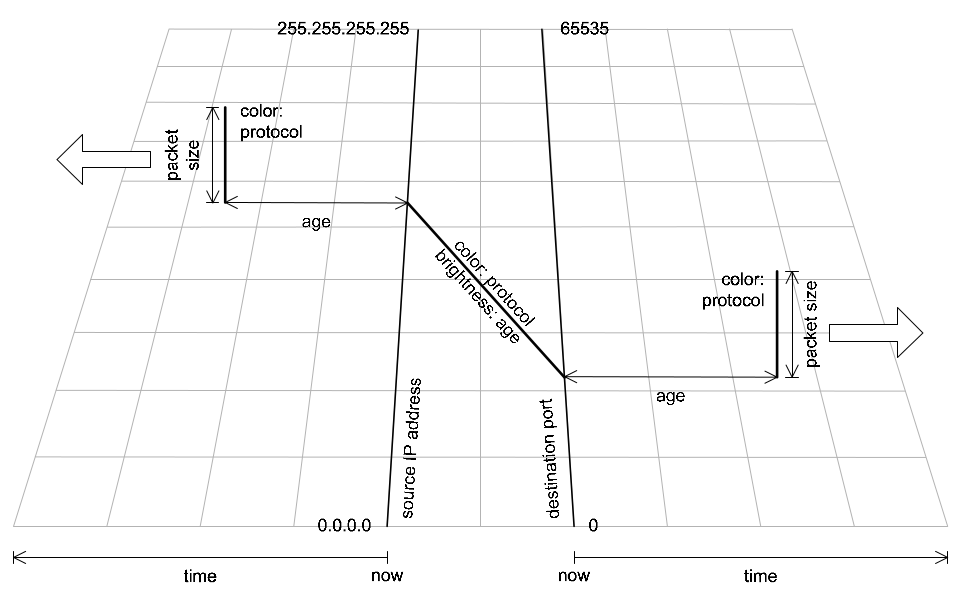}
	\caption[SecVis visualization overview]
		{SecVis visualization overview by Krasser.}
	\label{fig:secVis_structure}
\end{figure}

Krasser \cite{krasser2005real} presented a way to make attack patterns on
honeypots easily detectable to the human eye. It is not based on a statistical
method, but rather presents an intuitive way of visualizing network traffic
information in real-time for monitoring or in playback mode for forensics. It is
a combination of animated scatter plots and parallel coordinate plots.
\autoref{fig:secVis_structure} shows an overview of the visualization. The left
vertical line denotes the source IP-address, 0.0.0.0 is at the bottom,
255.255.255.255 is at the top. The right vertical line denotes the destination
port, port 0 at the bottom, port 65535 at the top. Coloured lines connect those
vertical lines, one line for each packet. UDP packets are visualized by blue
links, TCP packet by green links. Each packets also triggers two bars, whose
height represent the packet size. The bars move away from the vertical lines as
the packets gets older. Krasser successfully demonstrated that this
visualization is useful for visualizing inbound as well as outbound traffic and is
capable of highlighting differences in the traffic bit rates, common attack
ports and sources or patterns like reoccurring worm attacks.

Almotairi \cite{almotairi2008principalComp} has used the principal component
analysis (PCA) to separate latent groups of activities and to find outliers from
cluster groups. The PCA is a multivariate statistical technique which is used to
reduce the dimensionality of a data set into a few linearly uncorrelated
variables, called the principal components. The resulting number of principal
components is less than or equal to the number of original variables $p$ and the
components are defined and ordered in such a way that the first component has
the largest variance. Therefore much of the variance in the original set can be
retained be choosing the first $k$ PCs with $k<p$. PCs are usually selected by
the Kaisers' rule \cite{rencher2003methods}, which suggests the elimination of
all PCs with a Eigenvalue of less than one.
\begin{eqnarray}
C_i &=& \frac{X_i - \bar{X}_i}{\sqrt{s_i}} \\
Z &=& A^TC \\
Z_1 &=& A_1^TC = a_{11}C_1 + a_{12}C_2 + a_{13}C_3 + ... + a_{1p}C_p
\nonumber \\
\vdots \nonumber \\
Z_k &=& A_k^TC = a_{k1}C_1 + a_{k2}C_2 + a_{k3}C_3 + ... + a_{kp}C_p
\nonumber
\end{eqnarray}
To calculate the PCs the p-dimensional vector $X=(X_1, .., X_p)^T$ is
standardized by $C_i$ for all $i=1..p$, where $\bar{X}_i$ is the sample mean and
$s_i$ is the sample variance for $X_i$. Empirical experience shows that PCAs on
honeypot data should be calculated using the correlation matrix $R$ of $C$. The
Eigenvector of R is $A=(A_1, ..., A_k)^T$, the Eigenvalue vector of C is
$l=(l_1, ..., l_p)$ and the first component equals to $Z_1$. The components can
be interpreted by analysing the loading of the components variables, as high
loading indicates significance, which allows the assignment to events like
\textit{targeted attacks against open ports} or \textit{scan activities}.
Eventually, the results can be visualized with scatter plots having a PC on each
axis, which Almotairi used to find interrelations between components and also to
identify extreme activities or outliers. Those can be identified by sight,
however also automatically by applying contours, the population ellipsoids,
around the mean values ($Z_{ik}$ is the score of the $k^{th}$ PC of $i$th
observation, $l_k$ is the $k^{th}$ Eigenvalue). The first and last components
can give interesting insights, as they mark the most and least important properties.
\begin{eqnarray}
\sum_{k=1}^{p} \frac{Z_{ik}^2}{l_k} = \text{\textit{const}}, \quad \quad \quad
	\text{e.g.:} \quad \frac{Z_{i1}^2}{l_1} + \frac{Z_{i2}^2}{l_2} \le 5.8
\end{eqnarray}

One of the latest honeypot attack analysis was published by Owezarski
\cite{Owezarski:ClusterEnsemble}, whose approach uses unsupervised machine
learning methods and applies robust clustering techniques. Eventually, this
method creates signatures automatically based on the clusters. This analysis is
founded on a general algorithm which was applied to recognize anomalies in
production traffic \cite{Owezarski:OriginalAlgo} and was altered to work on
server honeypots traffic. Owezarski aims at increasing the robustness of the
clustering algorithm by the divide and conquer approach sub-space clustering
(SSC) and the notion of clustering ensembles. A clustering ensemble $P$ consists
of a set of $N$ partitions $P_n$ produced for the same data with $n=1,...,N$ by
the same clustering algorithm. Each of these partitions provides different and
independent exposure of data patterns, which then can be combined to a global
cluster including all metrics and facilitating a holistic understanding of the
threat. A shift of the similarity measure from patterns to the clustering
results takes place. This is possible because of the downward closure property
\cite{Agrawal:SubspaceRule}: \textit{If a collection of points is a cluster in a
d-dimensional space, then it is also part of a cluster in any (d-1) projections
of this space.} Moreover, high-dimensional data tends to be sparse, therefore
clustering algorithms create better results in lower dimensions, which also have
a smaller computational cost. The partitions are combined by the Inter
Clustering Result Association (ICRA) method to correlate clusters and Inter
Outlier Association (IOA) method to correlate outliers. The notion of graphs is
used: a vertex is a cluster/outlier from any sub-space and an edge represents a
high similarity. For ICRA, similarity $CS$ is calculated by the ratio of mutual
flows the clusters $C_i$ are based on, a threshold of 0.9 was effective for
Owezarskis data: $CS(C_1, C_2) = \frac{|(C_1 \cap C_2)|}{max(|C_1|, |C_2|)} >
0.9$. IOA links outlier vertices if and only if completely the same flows are
responsible for that outlier. Similar to Thonnard \cite{thonnard2008framework} a
greedy algorithm is used to recognize cluster-cliques because of performance
reasons. The intersection of all flows present within cluster-cliques form the
anomaly of interest. 

Gregio et al. \cite{gregio2007evaluation} have demonstrated the effectiveness of
classical data mining and supervised classification approaches in the honeypot
data log analysis using the well-established k-nearest neighbour (KNN)
algorithm, neural networks and decision trees. The purpose of this analysis is
to differentiate between internet noise like inoffensive or known traffic and
anomalies and real attacks. So their setup does not only include honeypot
traffic which is labelled as an attack but also productive traffic which is
labelled as normal. Both traffic types are used to support a learning process
and to make a classification of traffic possible. KNN uses a simple
classification principle: For each instance of data of unknown class the
distance to data with known classes is calculated and the $k$ nearest neighbours
decide by a majority vote which class is selected for the unknown element
\cite{jain2010data}. Artificial neural networks are computing structures that
are composed of single processing units called neurons. Different forms of
neural networks exist, however a well-understood and commonly used model is the
multilayer perceptron (MLP). MLP utilizes a supervised learning technique called
backpropagation for training the network and consists of multiple layers of
neurons in a directed graph \cite{Haykin1998Book}. A decision tree is a machine
learning algorithm which performs successive partitioning of the original data
set into successively more homogeneous subgroups. Each node in a decision tree
resembles a partition decision. Hence, the higher a decision tree is the higher
the detail level of clustering \cite{Quinlan1993Book}. For the most simple
classification of traffic, either suspicious or normal, the KNN algorithm was
too slow to create reasonable results, neuronal networks performed good and
produced good results, however with a quite high share of false negative.
Decision trees preformed best and produced correct results with few false
positives and a moderate number of false negatives. 

Honeycomb \cite{Kreibich2004Honeycomb} is one of the most famous HoneyD plugins
and scans incoming traffic to detect repeating patterns in packet payloads using
the longest common substring (LCS) algorithm. This implementation is based on
suffix trees, which are used as building blocks for a variety of string
algorithms. Using suffix trees, the longest common substring of two strings is
straightforward to find in linear time. Suffix trees can be generated for
example with the Ukkonen's algorithm \cite{ukkonen1995suffix}. Honeycomb
performs a \textit{protocol analysis} which orders traffic
by network and transport layer header information (IP-address or Ports).
After that LCS is applied in two different ways: \textit{Vertical pattern
detection} concatenates for two connections the incoming messages into one
string, respectively, and compares then the resulting string. \textit{Horizontal
pattern detection} compares for two connections two messages at the same depth
in the dialogue, that means that LCS is applied to the n-th messages.

Sampling is the process of selecting a subset of individuals from within a 
statistical population to estimate the overall characteristics. Connection sampling 
can greatly benefit the analysis by reducing 
bandwidth and memory requirements, which ultimately improves the scalability.
Yegneswaran et. al \cite{yegneswaran2004design} have shown, that subset
properties from honeypot data are able do describe trends and patterns of the
overall data, such as finding \"heavy hitters\". The data set was reduced by two
methods. The first method chooses randomly flows and counts the related packets.
The second method considers only a subnet of the observed network.

\subsection{Attack Root Cause Identification}

Attack root causes can be defined as the most basic cause that can be reasonably
identified as the origin of a attack. The root cause can be associated to a
specific attack tool, or one of its variants or configurations. One of the main
tasks of honeypot data analysis is the assignment of clusters and recognized
patterns to root causes. However, this is not necessarily an one-to-one
relationship, as it is difficult to guarantee that a found cluster is caused by
only one attack tool and that one attack tool does not cause two clusters, for
example by different attack configurations. Ultimately, a cluster should always
remain explicable in its formation \cite{pouget2004honeypot}.

Before a cluster can be assigned to an attack root cause, it is necessary to
validate the cluster coherency, that is if we found good (meaningful) clusters.
Undoubted, different attacks can create the same number of packets on the same
ports, therefore a pure statistical analysis of transactional data might be not
enough. One possible way of determining the coherence is by considering the
packet data content. The payload of all packets sent from one attack source can
be transformed into strings and concatenated. This creates an attack
fingerprint, which then can be used to check the cluster coherency by comparing
the fingerprints by a simple string distance measure like the Levenshtein
distance. Pouget \cite{pouget2004honeypot} used this method to prove that his
clustering technique is meaningful, as the fingerprints $i$ from a Cluster $C$
mostly have a distance $d_i=0$ to their cluster partners, resulting in a very
low average distance $D_C$ for a cluster, which in turn means that a mutual
payload exists.
\begin{eqnarray}
d_i &=& \sum_{j \in C} \frac{D(i,j)}{n-1} \\
D_C &=& \sum_{i \in C} \frac{d_i}{n} = \sum_{i,j \in C, i<j} 
	\frac{2D(i,j)}{(n-1)n}
\end{eqnarray}

Polymorphic attacks are attacks that are able to change their appearance with
every instance. Thus, polymorphic worms pose a big challenge to the honeypot
pattern detection, and more specifically to root cause identification, as worms
might change the attack vector for exploiting the vulnerability or the attack
body might change by garbage insertions, encryption, instruction reordering and
so on. Therefore (sub-) string based methods like LCS are insufficient.
Different approaches exist, however their research is based on the mutual
premise that despite the polymorphism the worms must have some invariant
substrings. Indeed, such invariants exist \cite{newsome2005polygraph} and the 
meaningful strings which can be used for classification have to be found.

Tang and Chen \cite{tang2005doubleHP} proposed the design of a 
\textit{double-honeypot} as a counter measure to worms. The novelty of 
this system is the ability to distinguish worm activities from non-attacking 
behaviour on honeypots, as for example misconfigurations. This system is 
composed of two independent honeypot arrays, the inbound array consisting of 
high-interaction honeypots which allows compromise and an outbound array 
consisting  low-interaction honeypots. If a compromised inbound honeypot tries 
to find and infect other victims, all outgoing traffic initiated by the honeypot 
will be redirected by a network translator to the outbound array. If one of the
outbound honeypots sees network traffic, a compromise on the inbound honeypot
took definitely place. Expanding this work, 
Mohammed, Hashim et al. \cite{mohammed2010polymorphic} proposed a \textit
{double-honeynet} to solve some of the limitations of the double-honeypot. The
double-honeynet is a combination of two honeynets consisting only of server
high-interaction honeypots. One honeynet is destined to receive incoming
connection requests Once a honeypot from the first honeynet $H_1$, is
compromised, a worm will attempt to make outbound connections. An internal
translator intercepts all outbound traffic and redirects it to the second
honeynet $H_2$, which also allows the infection. This procedure repeats, causing
the worm to spread back and forth across the honeynets and to manifest different
instances. Those instance are collected centrally and a signature generation
algorithm is initiated: First a substrings extraction process takes place which
creates the set of all possible substrings and determines their
frequency. Then the PCA is applied on the frequency count to reduce the
dimension and get the most significant strings, which then can be used to create
a signature and to define a root cause.

Another way of assigning the attacks root cause is by determining which
attack tool was used to convey the attack. This can be done by
analysing port sequences or the TCP Initial Sequence Number (ISN).
\begin{eqnarray}
\text{ISN} \\
\text{Port Sequence}
\end{eqnarray}

Some attack tools use always the same ISN or a bad random number generator with
a low entropy \cite{francois2008graphs}, which makes it possible to assign some
ISN to specific attack tools. Moreover, honeypots receive backscatter packets,
which are SYN-ACK replies to spoofed SYN packets and therefore possess the ISN+1.
Again, these are typical ISNs, which could be linked to specific tools.

%
%
The analysis by port sequences was introduced by Pouget at al.
\cite{pouget2004honeypot} in order to show that frequent/repetitive attacks on
honeypots create large amount of data, which might lead to misleading results if
general statistics are applied. However, a closer look on port sequences can
reveal some hidden phenomena. A port sequence is a time ordered sequence of
ports without duplicates \cite{leurre2008long} that represents the order in
which the attack sources (IP-address with a timeout of 1 day for Pouget) sent
packets to specific ports, for example: Attacker sents TCP requests to port 135,
again on 135 and then on port 4444 creates $\{135T, 4444T\}$. Port Sequences can
be created for observations on a single honeypot or for observations on several
honeypots. Preliminary results from Pouget showed that each sequence is often
limited to only one port, and that a port sequence represented as a \textit{set}
is almost uniquely identified by this set, however because of some rare cases
the ordered sequence was preferred. Pouget observed more than twice as many port
sequences as unique targeted ports. The distribution is similar to other
metrics, as the top 8 sequences already characterize the activity of about 75\%
of the attacks \cite{dacier2004attack}. These results motivated a further in-
depth investigation.

Another possibility to deduce an attack to a specific attack tool was discovered
by Kohlrausch \cite{kohlrausch2009experiences}. During a buffer overflow the
instruction counter (EIP) is overwritten with a new return address at which the
shell code can be assumed. Moreover, honeypots like Argos track the preceding
value of the instruction counter (faulty EIP) which is the last legitimate
instruction before the exploit had taken over the control. Evaluations show,
that the values of the EIP and faulty EIP are characteristic for an exploit tool
and operating system pair. However, this analysis might be strongly biased by 
active address space randomization algorithms. 
\begin{eqnarray}
\text{EIP and faulty EIP} 
\end{eqnarray}

Another major problem in identifying a single root cause for attacks on
honeypots were discovered by Alata and Pouget \cite{alata2007lessons,
pouget2008understanding}. Pouget made the observation that different sets of
compromised machines are used to carry out the various stages of planned
attacks. That means, that a single attacker causes different attack patterns
from different machines on the honeypot. In addition, Alata observed this
phenomenon also on a SSH high-interaction honeypot. Two groups of attacking
machines have been spotted: The first group is composed specifically to scan
hosts and perform dictionary-attacks. If they are successfully, usually a day
later a machine from the second group appears. This group has no intersection
with the first group in terms of IP-address and geographical lookups even reveal
different countries. After a login the second machine tries to run own services
or get root access. Interestingly, comparing the attack sources between low- and
high-interaction honeypot data sets demonstrates that mutual IP-addresses are
from the scanning group only, the second intrusion group also never appears on 
low-interaction honeypots.

In general, attack root cause identification requires a good knowledge of 
black-hat tools or recent participation on security pages and mailing pages, which
usually inform about Common Vulnerabilities and Exposures (CVE). That is why it
can be really difficult to assign a honeypot attack cluster to a known attack
tool. Moreover, because of the increasing complexity of worms it is necessary to
rather perform payload analysis (byte sequences, shellcode commands etc.) 
\cite{newsome2005polygraph} than pure statistical evaluations and the mere
detection of a exploit. Honeypots are still used to collect worms, however the
signature generation for worms evolved into its own broad field of studies and
are rather the domain of intrusion detection systems \cite{li2008survey}.

\subsection{Attack Risk Assessment}
\label{sec:attackRiskAssessment}

The risk estimation is rather done on high interaction honeypots, as it can be
assessed based on the severity of the vulnerability and the analysis of the
exploit. However, risk estimation can also be done for low-interaction honeypots
based on the scope of the attack, which can be measured by three features
describing the amount of communication: The number of packets of the attack, the
amounts of bytes exchanged in the attack and the communication duration. 

$$ risk = \log(nPackets)+\log(nBytes)+\log(duration+1) $$ 

Ozewaski \cite{Owezarski:ClusterEnsemble} extended this risk estimation by
multiplying the risk value by the number of sub-spaces the attack is found in,
which is an indication for how many network-features the attack affects.

\begin{eqnarray}
risk_{subspace}  & = & C * \log(nPackets)+\log(nBytes)
\nonumber \\
& & +\log(duration+1) \nonumber
\end{eqnarray}

SweetBait \cite{portokalidis2007sweetbait} is designed to be an automated
response system that protects from random IP scanning worms using 
low-interaction (honeyD) and high-interaction (Argos) honeypots. Honeypots are 
used to create signatures, which then are sent to IDS/IPS sensors in order to
determine the virulence of worms on production systems. The expected virulence
of worms is based on their aggressiveness, which is quantified by the
exponentially weighted moving average (EWMA) of the number of alerts generated
by each signature on each period: $m' = w \times a + (1 - w) \times m$, where
$m'$ is the new value, $m$ the previous value and the weight $0 < w \le 1$
configures the computation to follow more or less aggressively the recent
changes in activity levels, whereby values below $0.5$ were not useful.
This EWMA value is then used to predict the virulence by adjusting the value by
port and protocol bias value, which is useful for especially active ports like
145: $A = m \times port\,bias \times protocol\,bias$. The EWMA can be used not
only to predict signatures observed by IDS, but also any temporal pattern on 
honeypots.

\subsection{Exploit Detection}
\label{sec:ExploitDetection}

As high interaction honeypots are actively exploited, they also consider the
vulnerabilities which were exploited in their analysis. Two main procedures to
detect exploitations and to find vulnerabilities are either the data-driven
technique (example: Argos) or the operating system state monitoring (example:
Capture-HPC) \cite{kohlrausch2009experiences}. The former detects exploitations
by dynamic taint analysis, which is based on the idea that all data from the
internet is potentially malicious and therefore is marked as tainted. The 
data-flow of tainted data is monitored. The exploitation of a honeypot is then
specified as the direct execution of tainted shell-code. Dynamic taint analysis
is very accurate and reliable to detect attacks utilizing buffer overflows. The
latter inspects the states of operating systems and tries to spot illegitimate
actions in the file system or process management. An exploitation has occurred
if modifications (sometimes even read operations) are done to this specific
locations, no active execution of those modified files is required by
definition, however this usually happens implicitly.  Modifications can be
recognized comparing the files with a backup, by comparing hash-values or by
controlling the kernel-log for sensitive calls.

Argos memory tainting technique was extended by SweetBait
\cite{portokalidis2007sweetbait}, which inserted its own shellcode into the
code that is under attack, which makes the gathering of more information about
the memory and process states possible.


\subsection{Overview Honeypot Data Analysis}

\autoref{tab:honeypotMetrics} gives an overview over the various honeypot 
metrics used in research projects.
Interestingly, one finding of this overview is that most of the researchers tend
to pose the first three presented questions, which refer to the attack source,
attack target and the frequency. Furthermore, a common consensus exists in
identifying sources or targets and in describing the frequency, as many of the
metrics and analysis methods are reused throughout the publications. The reason
behind this circumstance is that \textit{direct (apparent) information} is
evaluated and in cases like the country-mapping extended by simple lookups.
Direct information describes the observations and is recorded in honeypot logs
during common operation: Usual honeypot logfiles contain the source, the target
and the timestamp of an attack based on the IP information. It is important to
note, especially for the IP, that a communication without an IP- address for the
source and target would be not possible and each event has a timestamp.
Therefore, it is straightforward and natural to pose analysis questions based on
these features.

However, this situation is different for the remaining questions, because they
attempt to \textit{derive information}. Derived information explains, assesses
or localizes the cause of the observations which is fundamentally more complex
than mere description. Since the analysis is more complex, such research
appeared later than simple descriptive analysis and less overlaps between
methods exist. This is especially true for the pattern-detection, which can be
done by many different similarity measures and clustering algorithms (as
explained in \autoref{sec:attackPatterns}). The conduct of such analysis is
growing into an interdisciplinary approach, because in order to derive
information, basic statistics usually do not suffice any more: sophisticated
honeypot networks and methods from other fields like association rule mining,
neuronal networks, memory tainting in virtual machines to name a few, have
become necessary. In general, the bond between honeypots and other 
research fields has intensified during the last years.

\newcommand{\ficolumn}{4cm}
\newcommand{\fcolumn}{0.25\textwidth}
\newcommand{\scolumn}{0.35\textwidth}
\newcommand{\tcolumn}{0.30\textwidth}
\begin{table*}
\center
	\caption{Metrics used in honeypot data analysis and related publications.}
	\label{tab:honeypotMetrics}
\begin{tabular}{p{\fcolumn}p{\scolumn}p{\tcolumn}}
	\toprule
		Problem Statement & Analysis & Examples \\ 
		\midrule
		
		\multirow{7}{\ficolumn}{Do common attack origins exist?} & 
			
			IP-Address or IP-Prefix & \cite{leurre2008long, pouget2004honeypot,
				francois2008graphs, kyotoDataset, yegneswaran2005using,
				kaaniche2007empirical} \\
			
			\cmidrule(l){2-3} &

			Autonomous System Number & --- \\
			
			\cmidrule(l){2-3} &

			Domain Name, URL, URL-Type  & \cite{ikinci2008monkey,
				mcgrew2006experiences, kyotoDataset, trivedi2007analyzing} \\
			
			\cmidrule(l){2-3} &
			
			Country & \cite{leurre2008long, pouget2004honeypot, 
				mcgrew2006experiences, kyotoDataset, trivedi2007analyzing} \\
			
			\cmidrule(l){2-3} &
			
			UserID / Email & \cite{trivedi2007analyzing, alata2007lessons} \\
			
			\cmidrule(l){2-3} &

			(Worm-) Signature & \cite{chun2003netbait, kyotoDataset,
				yegneswaran2005using,  portokalidis2007sweetbait} \\

			\cmidrule(l){2-3} &

			User Agent & \cite{valli2010voip} \\

			\cmidrule(l){2-3} &
			
			Operating System & \cite{leurre2008long, pouget2004honeypot} \\

		\midrule

		\multirow{6}{\ficolumn}{What is the target of the attack?} & 

			IP & \cite{kaaniche2007empirical, Owezarski:ClusterEnsemble,
				mohammed2010polymorphic, yegneswaran2004design} \\

			\cmidrule(l){2-3} &

			Port and Transport Protocol & \cite{leurre2008long, 
				pouget2004honeypot, 
				francois2008graphs, almotairi2008principalComp, 
				thonnard2008framework, gregio2007evaluation} \\

			\cmidrule(l){2-3} &

			Service & \cite{pouget2004honeypot, mcgrew2006experiences, 
				alata2007lessons}\\

			\cmidrule(l){2-3} &

			Software Client, Plugins & \cite{ikinci2008monkey, Paper:YALIH}\\

			\cmidrule(l){2-3} &

			Vulnerability & \cite{Argos:2006Qemu, alata2007lessons} \\

			\cmidrule(l){2-3} &

			OS & \cite{leurre2008long, kyotoDataset} \\

		\midrule
		
		\multirow{21}{\ficolumn}{What are the attack frequencies?} & 
		
			Time until First Incoming Connection & 
				\cite{mcgrew2006experiences, alata2007lessons} \\

			\cmidrule(l){2-3} &

			Number of Incoming Connections per Time Unit & 
				\cite{kaaniche2007empirical} \\ 
			
			\cmidrule(l){2-3} &

			Number of Sources per Time Unit & 
				\cite{pouget2005advantages:seq, 
				francois2008graphs, yegneswaran2005using, 
				thonnard2008framework} \\
			
			\cmidrule(l){2-3} &
		
			Number of new Sources per Time Unit (CDF) & 
				\cite{yegneswaran2005using, leurre2008long} \\
			
			\cmidrule(l){2-3} &

			Interarrival Time (Distribution for equally-sized Source Intervals)&
				\cite{yegneswaran2005using, kaaniche2007empirical,
				mohammed2010polymorphic} \\

			\cmidrule(l){2-3} &

			Number of Sources versus the number of attacks per Source & 
				\cite{kaaniche2007empirical} \\
			
			\cmidrule(l){2-3} &

			Received Packets per Time Unit & \cite{francois2008graphs, 
				gregio2007evaluation, mohammed2010polymorphic, yegneswaran2004design}\\

			\cmidrule(l){2-3} &

			Received Data (kB) per Time Unit & \cite{francois2008graphs,
				gregio2007evaluation, mohammed2010polymorphic, yegneswaran2004design} \\

			\cmidrule(l){2-3} &

			Messages/Emails Received per Time Unit & 
				\cite{trivedi2007analyzing} \\

			\cmidrule(l){2-3} &

			URLs / Attachments Received per Time Unit &
				\cite{trivedi2007analyzing} \\ 

			\cmidrule(l){2-3} &

			Received Data (kB) per Message & \cite{trivedi2007analyzing,
				krasser2005real} \\

			\cmidrule(l){2-3} &

			Exploitations per Time Unit & \cite{kohlrausch2009experiences,
				alata2007lessons}\\

			\cmidrule(l){2-3} &

			Sessions per Time Unit & \cite{kaaniche2007empirical, 
				gregio2007evaluation} \\

			\cmidrule(l){2-3} &

			Session Duration & \cite{kaaniche2007empirical, 
				gregio2007evaluation} \\

			\cmidrule(l){2-3} &

			Time between Sessions & 
				\cite{mcgrew2006experiences} \\

			\cmidrule(l){2-3} &

			Source Lifetime & \cite{yegneswaran2005using} \\
			
			\cmidrule(l){2-3} &

			(Un-)Known Attack Sessions per Time Unit & \cite{chun2003netbait, 
				kyotoDataset, yegneswaran2005using} \\

			\cmidrule(l){2-3} &

			Number of Basic Flows per Time Unit & 
				\cite{almotairi2008principalComp, Owezarski:ClusterEnsemble} \\

			\cmidrule(l){2-3} &

			Number of Activity Flows per Time Unit & 
				\cite{almotairi2008principalComp} \\








		\midrule

		\multirow{4}{\ficolumn}{How to detect changes in attacks?} & 
			
			Ratios between different Time Units & 
				\cite{thonnard2008framework} \\

			\cmidrule(l){2-3} &

			Sliding Window and Locality Statistics & 
				\cite{francois2008graphs} \\

			\cmidrule(l){2-3} &

			Deviation $\beta$ value & \cite{yegneswaran2005using} \\

			\cmidrule(l){2-3} &

			Linear Regression & \cite{kaaniche2007empirical}\\

		\midrule

\end{tabular}
\end{table*}
\begin{table*}
\center
\begin{tabular}{p{\fcolumn}p{\scolumn}p{\tcolumn}}
		\midrule
		\multirow{8}{\ficolumn}{How to compare propagation?} &

			Propagation Graph & \cite{kaaniche2007empirical} \\

			\cmidrule(l){2-3} &

			Attack Graph & \cite{vasilomanolakis2015honeypot}\\

			\cmidrule(l){2-3} &

			\#Unique Mutual Attackers per \#Targets per Time Slot & \cite{vasilomanolakis2015honeypot} \\

			\cmidrule(l){2-3} &

			Phase Plots & \cite{yegneswaran2005using} \\

			\cmidrule(l){2-3} &

			Association Rule Mining & \cite{pouget2004honeypot} \\

			\cmidrule(l){2-3} &

 			Destination Net Footprint Scan & 
 				\cite{yegneswaran2005using} \\

 			\cmidrule(l){2-3} &

			Ratio of Attackers per Number of Target Sensors & 
				\cite{vasilomanolakis2015honeypot} \\

			\cmidrule(l){2-3} &

			PDF of First Destination Preference & 
 				\cite{yegneswaran2005using} \\

		\midrule

		\multirow{6}{\ficolumn}{How to detect attack patterns?} &

			SecViz Visualization & \cite{krasser2005real} \\

			\cmidrule(l){2-3} &

			Symbolic Aggregate approXimation (SAX) & 
				\cite{thonnard2008framework} \\

			\cmidrule(l){2-3} &

			Principal Component Analysis (PCA) & 
				\cite{almotairi2008principalComp, mohammed2010polymorphic} \\

			\cmidrule(l){2-3} &

			Sub-Space Clustering (SSC) & \cite{Owezarski:ClusterEnsemble} \\

			\cmidrule(l){2-3} &

			Multilayer Perceptron (MLP) & \cite{gregio2007evaluation} \\
			
			\cmidrule(l){2-3} &

			Longest Common Subsequence (LCS) & \cite{Kreibich2004Honeycomb,
			mohammed2010polymorphic} \\

		\midrule

		\multirow{5}{\ficolumn}{How to identify a root causes?} &

			Cluster Coherency & \cite{pouget2004honeypot} \\

			\cmidrule(l){2-3} &

			Double-Honeynet & \cite{mohammed2010polymorphic} \\

			\cmidrule(l){2-3} &

			ISN & \cite{francois2008graphs} \\

			\cmidrule(l){2-3} &

			Port Sequence & \cite{pouget2004honeypot} \\

			\cmidrule(l){2-3} &

			Faulty EIP & \cite{kohlrausch2009experiences} \\

		\midrule

		\multirow{2}{\ficolumn}{How to assess the risk?} &

			Communication Scope & \cite{Owezarski:ClusterEnsemble} \\

			\cmidrule(l){2-3} &

			EWMA & \cite{portokalidis2007sweetbait} \\

		\midrule

		\multirow{2}{\ficolumn}{How to recognize exploits?} & 
			
			Data-Driven Technique & \cite{Argos:2006Qemu} \\

			\cmidrule(l){2-3} &

			Memory Tainting & \cite{CaptureHPCHP} \\

		\bottomrule
\end{tabular}
\end{table*}

\section{Legal and Ethical Concerns}
\label{sec:honeypotLegality}

The deployment of honeypots involves the discussion about legal and ethical
liability. It is important to announce beforehand, that the legal situation is
country-dependent, furthermore, it is often difficult to decide which laws apply
if attacker and victim are situated in different countries. However, this
chapter aims to highlight possible pitfalls and general reasoning which has to
be considered.

Lance Spitzner \cite{HoneypotLegalIssues} formulated the first two problems,
which are entrapment and privacy. 

\paragraph{Entrapment Challenges}
Entrapment can be defined as the persuasion of an entity to commit 
a crime although no previous intent to commit
such a crime existed. It can be argued that this does not apply to honeypots, as
they do not actively persuade anyone. Server honeypots mimic production systems
or services and wait for incoming connections, they are virtually invisible to
other network participants unless they decide to approach those system, for
example due to an IP-address range scan. Client honeypots make the first
request, however, they do not persuade to commit a crime, as the exploitation is
already prepared server-sided, which means that the intent of a crime already
existed. On the other hand, a honeypot remains a system consciously set up for
attacks and sometimes even with known security vulnerabilities which makes an
attack possible in the first place.

\paragraph{Privacy Challenges}
The privacy issue elaborates the data collection. The question arises whether a
honeypot is allowed to collect information about the attackers without their
knowledge or permission and thus violate their privacy? In order to answer that
question one has to consider which type of data is saved by the honeypot and how
the data is used by the honeypot systems and administrators.

Data is usually divided into two general categories, meta-information and
content. Meta-information stores information about content and connections. For
the IP this is operational/transactional data like the IP-address, IP-header,
session-cookies, timestamp of the communication and so forth. This type of
information is rather collected by low-interaction honeypots, however, some
content might be stored depending on the level of emulation. Content data is the
data which a sender actually intends to transmit, like email-text, keystrokes
and files. High-interaction honeypots focus on saving large amounts of content
data, although meta-information is stored additionally. Furthermore, it is
important to highlight the purpose of a honeypot and how the meta-information or
content is used. As already stated, production honeypots can  be used to defend
production systems. Although research honeypots do not protect an organization
directly, they help to understand threats, develop  countermeasures and to fix
exploitable bugs, therefore they contribute indirectly to the safety. 
Content data has more privacy issues then transactional data, as it
also underlies the copyright of the author. However, one has to consider that on
the one hand the attacker does not have any authorized means to save data on
those honeypots (like a legitimate account) and on the other hand does decide
voluntarily to transmit the content.

Common practice in the internet shows that the processing of transactional data 
is done frequently and is not persecuted by many countries. However, some countries
such as Germany rate IP-addresses with an affiliated timestamp as personal data
since backtracking of individuals is possible with the help of ISPs. That is why a storage
of non-anonymized IP-addresses is only allowed for 7 days in order to ensure undisturbed 
operation of services \cite{TKGGermany, BGHAZ}. Such local regulations have to be 
considered although they might interfere with the scientific freedom.

As long as tools attempt to secure one's own systems and their usage-emphasis
lies in the improvement of the protection, the legal risks \textit{seem} to be
negligible. One should always keep in mind that honeypots do not only
communicate with criminal, malicious entities but also with their victims.
Hijacked systems are used to propagate attacks, backscatter packets arrive 
a result of spoofed IP-addresses. Because of that honeypot log records should
always be handled with care. As far as is known, there was up to date no trial
concerning honeypots explicitly. However, this holds only as long as the
honeypot does not cause any harm, which leads us to the next problem statement.

\paragraph{Liability Challenges}
The third concern is the liability in case of harming other systems. As
honeypots offer known vulnerabilities to attackers, they can be used to harm
other systems. Low-interaction honeypots emulate protocols and therefore they
can be subject of a IP spoofing and amplification attack. High-interaction
honeypot might allow arbitrary code execution on the machine and therefore cause
even much higher damage to other systems. This means, that the original attacker
is not visible to the victim, which in turn would sue the operator of honeypots
they appear to be the real attacker. The argument here is, that if an
administrator had taken proper precautions to keep the (honeypot) systems
secure, the attacker would not have been possible, therefore the administrator
is jointly responsible for any damage which has occurred. The higher the
interaction level of the honeypot the higher the risk of a harmful utilization. 
Therefore verification checks should be done more often on high-interaction
honeypots. It is recommended to reset virtual honeypots
as often as possible and to use a reverse firewall to limit the amount of
malicious traffic that can leave the honeypot, as it shows consciousness and the
attempt to minimize possible damage, which might limit the legal liability.
Another possibilities are containment systems, which restrict the rate at which
a computer is allowed to make  connections to other machines, for example Dantus
feedback control system \cite{Dantu2007FWC}.

Not only the deployment has to be considered, but also the development and
publication of honeypot software, as many countries including Germany have a
valid \textit{hacker paragraph}, compare the German criminal law code §202c
\cite{HackerParagraph}. This paragraph makes the publication of software whose
purpose is to spy out or intercept data and facilitating hacking attacks a
punishable offence, as it is seen as a preparation of a crime. It is not
completely clarified which software is affected by this law. However, a honeypot
is collecting similar information, in the same technical manner, as many other
security tools like IDS sensors or even system logs, and this category of
software is not banned.

\balance

\bibliographystyle{IEEEtran}
\bibliography{bib/bibliography}

\begin{thebibliography}{100}
\providecommand{\url}[1]{#1}
\csname url@samestyle\endcsname
\providecommand{\newblock}{\relax}
\providecommand{\bibinfo}[2]{#2}
\providecommand{\BIBentrySTDinterwordspacing}{\spaceskip=0pt\relax}
\providecommand{\BIBentryALTinterwordstretchfactor}{4}
\providecommand{\BIBentryALTinterwordspacing}{\spaceskip=\fontdimen2\font plus
\BIBentryALTinterwordstretchfactor\fontdimen3\font minus
  \fontdimen4\font\relax}
\providecommand{\BIBforeignlanguage}[2]{{%
\expandafter\ifx\csname l@#1\endcsname\relax
\typeout{** WARNING: IEEEtran.bst: No hyphenation pattern has been}%
\typeout{** loaded for the language `#1'. Using the pattern for}%
\typeout{** the default language instead.}%
\else
\language=\csname l@#1\endcsname
\fi
#2}}
\providecommand{\BIBdecl}{\relax}
\BIBdecl

\bibitem{yegneswaran2005using}
V.~Yegneswaran, P.~Barford, and V.~Paxson, ``Using honeynets for internet
  situational awareness,'' in \emph{Proceedings of the Fourth Workshop on Hot
  Topics in Networks (HotNets IV)}.\hskip 1em plus 0.5em minus 0.4em\relax
  Citeseer, 2005, pp. 17--22.

\bibitem{leurre2008long}
C.~Leita, V.~Pham, O.~Thonnard, E.~Ramirez-Silva, F.~Pouget, E.~Kirda, and
  M.~Dacier, ``The leurre. com project: collecting internet threats information
  using a worldwide distributed honeynet,'' in \emph{Information Security
  Threats Data Collection and Sharing, 2008. WISTDCS'08. WOMBAT Workshop
  on}.\hskip 1em plus 0.5em minus 0.4em\relax IEEE, 2008, pp. 40--57.

\bibitem{symantecReport}
{Symantec}, ``Internet security threat report 2015,''
  \url{https://www4.symantec.com/mktginfo/whitepaper/ISTR/21347932_GA-internet-security-threat-report-volume-20-2015-social_v2.pdf},
  last accessed Frebruary 2016.

\bibitem{AtlasAttacks}
{ATLAS}, ``Summary report - global attacks,''
  \url{https://atlas.arbor.net/summary/attacks}, last accessed July 2016.

\bibitem{vasilomanolakis2015taxonomy}
E.~Vasilomanolakis, S.~Karuppayah, M.~M{\"u}hlh{\"a}user, and M.~Fischer,
  ``Taxonomy and survey of collaborative intrusion detection,'' \emph{ACM
  Computing Surveys (CSUR)}, vol.~47, no.~4, p.~55, 2015.

\bibitem{Spitzner:TrappingHackers:DefHoneypot}
\BIBentryALTinterwordspacing
L.~Spitzner, ``The honeynet project: Trapping the hackers,'' \emph{IEEE
  Security and Privacy}, vol.~1, no.~2, pp. 15--23, Mar. 2003. [Online].
  Available: \url{http://dx.doi.org/10.1109/MSECP.2003.1193207}
\BIBentrySTDinterwordspacing

\bibitem{vasilomanolakis2015honeypot}
E.~Vasilomanolakis, S.~Karuppayah, P.~Kikiras, and M.~M{\"u}hlh{\"a}user, ``A
  honeypot-driven cyber incident monitor: lessons learned and steps ahead,'' in
  \emph{Proceedings of the 8th International Conference on Security of
  Information and Networks}.\hskip 1em plus 0.5em minus 0.4em\relax ACM, 2015,
  pp. 158--164.

\bibitem{pouget2003white}
F.~Pouget and M.~Dacier, ``White paper: Honeypot, honeynet: A comparative
  survey,'' Technical Report RR-03-082, Institut Eurecom, Tech. Rep., 2003.

\bibitem{ikinci2008monkey}
A.~Ikinci, T.~Holz, and F.~C. Freiling, ``Monkey-spider: Detecting malicious
  websites with low-interaction honeyclients.'' in \emph{Sicherheit}, vol.~8,
  2008, pp. 407--421.

\bibitem{AccPointHP2009Survey}
N.~Al-Gharabally, N.~El-Sayed, S.~Al-Mulla, and I.~Ahmad, ``Wireless honeypots:
  survey and assessment,'' in \emph{Proceedings of the 2009 conference on
  Information Science, Technology and Applications}.\hskip 1em plus 0.5em minus
  0.4em\relax ACM, 2009, pp. 45--52.

\bibitem{almutairi2012survey}
{Almutairi, Abdulrazzaq and Parish, David and Phan, Raphael}, ``Survey of high
  interaction honeypot tools: Merits and shortcomings,'' 2012.

\bibitem{Kambow2014:IndianSurvey3}
N.~Kambow and L.~K. Passi, ``Honeypots: The need of network security,''
  \emph{International Journal of Computer Science and Information Technologies,
  Vol. 5}, 2014.

\bibitem{mairh2011honeypot}
A.~Mairh, D.~Barik, K.~Verma, and D.~Jena, ``Honeypot in network security: a
  survey,'' in \emph{Proceedings of the 2011 International Conference on
  Communication, Computing \& Security}.\hskip 1em plus 0.5em minus 0.4em\relax
  ACM, 2011, pp. 600--605.

\bibitem{bringer2012survey}
M.~L. Bringer, C.~A. Chelmecki, and H.~Fujinoki, ``A survey: Recent advances
  and future trends in honeypot research,'' \emph{International Journal},
  vol.~4, 2012.

\bibitem{firstHPsansDef}
{Loras R. Even}, ``Honey pot systems explained,''
  \url{https://www.sans.org/security-resources/idfaq/honeypot3.php}, July 2000,
  last accessed January 2016.

\bibitem{SpitznerHoneypotPartOne}
{Lance Spitzner, Marty Roesch}, ``The value of honeypots, part one: Definitions
  and values of honeypots, october 10, 2001,''
  \url{http://www.symantec.com/connect/articles/value-honeypots-part-one-definitions-and-values-honeypots},
  last accessed on January 2016.

\bibitem{Stoll:CuckoosEgg}
C.~Stoll, \emph{Cuckoo's Egg}.\hskip 1em plus 0.5em minus 0.4em\relax Pocket,
  1990.

\bibitem{Cohen:FirstHoneypot}
F.~Cohen, ``A note on the role of deception in information protection,''
  \emph{Computers and Security}, 1999.

\bibitem{Cheswick:Berferd}
B.~Cheswick, ``An evening with berferd in which a cracker is lured, endured,
  and studied,'' \emph{AT\&T Bell Laboratories}, 1991.

\bibitem{pathan2014:wormHistory}
A.-S.~K. Pathan, \emph{The State of the Art in Intrusion Prevention and
  Detection}.\hskip 1em plus 0.5em minus 0.4em\relax CRC Press, 2014.

\bibitem{Homepage:SANS}
{SANS Institute (SysAdmin, Networking and Security)}, ``Cooperative research
  and education organization,'' \url{https://www.sans.org/}, last accessed on
  January 2016.

\bibitem{Cert:FirstUnknownMW}
{CERT/CC, Computer Emergency Response Team}, ``Exploitation of vulnerability in
  cde subprocess control service,''
  \url{https://www.cert.org/historical/advisories/CA-2002-01.cfm}, last
  accessed on January 2016.

\bibitem{Schneier:SecurityDefinition}
B.~Schneier, \emph{Secrets \& Lies: Digital Security in a Networked World},
  1st~ed.\hskip 1em plus 0.5em minus 0.4em\relax New York, NY, USA: John Wiley
  \& Sons, Inc., 2000.

\bibitem{Mokube:Challenges:HighLowDef}
\BIBentryALTinterwordspacing
I.~Mokube and M.~Adams, ``Honeypots: Concepts, approaches, and challenges,'' in
  \emph{Proceedings of the 45th Annual Southeast Regional Conference}, ser.
  ACM-SE 45.\hskip 1em plus 0.5em minus 0.4em\relax New York, NY, USA: ACM,
  2007, pp. 321--326. [Online]. Available:
  \url{http://doi.acm.org/10.1145/1233341.1233399}
\BIBentrySTDinterwordspacing

\bibitem{HoneyC:HPClient}
C.~Seifert, I.~Welch, P.~Komisarczuk \emph{et~al.}, ``Honeyc-the
  low-interaction client honeypot,'' \emph{Proceedings of the 2007 NZCSRCS,
  Waikato University, Hamilton, New Zealand}, 2007.

\bibitem{SpitznerLowHighOnly}
{Lance Spitzner}, ``Seclists email lists, honeypot tools categorization,
  2004.'' \url{http://seclists.org/honeypots/2004/q3/5}, last accessed on
  January 2016.

\bibitem{HoneyDBackground}
{Niels Provos}, ``Honeyd - honeypot background, 2004,''
  \url{http://www.honeyd.org/background.php}, last accessed January 2016.

\bibitem{Provos2004HoneyD}
N.~Provos \emph{et~al.}, ``A virtual honeypot framework.'' in \emph{USENIX
  Security Symposium}, vol. 173, 2004.

\bibitem{defAttackComputer}
ISO/IEC, ``Information technology - security techniques - information security
  management systems - overview and vocabulary,'' \emph{INTERNATIONAL STANDARD,
  ISO/IEC 27000}, 2009.

\bibitem{kaurcomparison}
T.~Kaur, V.~Malhotra, and D.~Singh, ``Comparison of network security
  tools-firewall, intrusion detection system and honeypot,'' \emph{nternational
  Journal of Enhanced Research in Science Technology \& Engineering, ISSN}, pp.
  2319--7463, 2004.

\bibitem{FirstHoneypot:DTK}
{All.Net \& Affiliated Companies}, ``The deception toolkit home page,''
  \url{http://www.all.net/dtk/}, last accessed on January 2016.

\bibitem{Zhang2003HP:bof}
F.~Zhang, S.~Zhou, Z.~Qin, and J.~Liu, ``Honeypot: a supplemented active
  defense system for network security,'' in \emph{Parallel and Distributed
  Computing, Applications and Technologies, 2003. PDCAT'2003. Proceedings of
  the Fourth International Conference on}.\hskip 1em plus 0.5em minus
  0.4em\relax IEEE, 2003, pp. 231--235.

\bibitem{honeypotSoftware:Cybercop}
{Networks Associates Technology, Inc.}, ``Cybercop sting - getting started
  guide,'' \url{http://www.scn.rain.com/~neighorn/PDF/Cstguide.pdf}, last
  accessed on December 2015.

\bibitem{HoneyBotPage}
{Atomic Software Solutions}, ``Honeybot - a medium interaction honeypot for
  windows,'' \url{http://www.atomicsoftwaresolutions.com/}, last accessed on
  December 2015.

\bibitem{HOASAnnouncement}
{Brazilian Distributed Honeypots Project}, ``Announce: Hoacd 1.0 (bootable
  openbsd + honeyd cd),'' \url{https://lwn.net/Articles/90941/}, last accessed
  on December 2015.

\bibitem{HoneyPerl}
------, ``Honeyperl honeypot,''
  \url{http://sourceforge.net/projects/honeyperl/files/honeyperl/}, last
  accessed on December 2015.

\bibitem{ImpostHP}
{ziplock / Sickbeatz, b4b0 security}, ``Impost,''
  \url{http://impost.sourceforge.net/}, last accessed on December 2015.

\bibitem{KFSensorHP}
{Keyfocus}, ``Kfsensor,'' \url{http://www.keyfocus.net/kfsensor/}, last
  accessed on January 2016.

\bibitem{LaBreaHP}
{Tom Liston}, ``Labrea: Sticky honeypot and ids,''
  \url{http://labrea.sourceforge.net/labrea-info.html}, last accessed on
  January 2016.

\bibitem{chun2003netbait}
B.~N. Chun, J.~Lee, and H.~Weatherspoon, ``Netbait: a distributed worm
  detection service,'' \emph{Intel Research Berkeley Technical Report
  IRB-TR-03}, vol.~33, 2003.

\bibitem{NetBaitService}
{NetBait Inc.}, ``Netbait service faq,''
  \url{http://netbaitinc.com/support/supp_faq.shtml}, last accessed on December
  2015.

\bibitem{PolymorphicWorms:HoneypotOverviewChapter}
A.-S. K.~P. Mohssen~Mohammed, \emph{Automatic Defense Against Zero-day
  Polymorphic Worms in Communication Networks}.\hskip 1em plus 0.5em minus
  0.4em\relax Auerbach Publications, 2013.

\bibitem{SingleHoneypotHP}
{Luis Wong}, ``single-honeypot,''
  \url{http://single-honeypot.sourceforge.net/}, last accessed on January 2016.

\bibitem{SpecterHP}
{NETSEC Inc.}, ``Specter,'' \url{http://www.specter.com/default50.htm}, last
  accessed on January 2016.

\bibitem{SymantecHP}
{Symantec Inc.}, ``Symantec decoy server,''
  \url{https://www.symantec.com/about/news/release/article.jsp?prid=20030623_01},
  last accessed on January 2016.

\bibitem{TinyHoneypotGH}
{George Bakos}, ``Tiny honeypot,'' \url{https://github.com/tiny-honeypot/thp},
  last accessed on December 2015.

\bibitem{NepenthesHP}
{Carnivore Project}, ``Nepenthes - finest collection -,''
  \url{http://nepenthes.carnivore.it/documentation:readme}, last accessed
  December 2015.

\bibitem{baecher2006nepenthes}
P.~Baecher, M.~Koetter, T.~Holz, M.~Dornseif, and F.~Freiling, ``The nepenthes
  platform: An efficient approach to collect malware,'' in \emph{Recent
  Advances in Intrusion Detection}.\hskip 1em plus 0.5em minus 0.4em\relax
  Springer, 2006, pp. 165--184.

\bibitem{DionaeaHP}
{Carnivore Project}, ``Dionaea honeypot,'' \url{http://dionaea.carnivore.it/},
  last accessed December 2015.

\bibitem{libemuHP}
------, ``libemu – x86 shellcode emulation,''
  \url{http://libemu.carnivore.it/}, last accessed on December 2015.

\bibitem{HoneytrapGit}
{Tillmann Werner}, ``Honeytrap – a dynamic meta-honeypot daemon,''
  \url{http://src.carnivore.it/honeytrap/},
  \url{http://honeytrap.carnivore.it/details.html}, last accessed January 2016.

\bibitem{bacher2005know}
{Bacher, Paul and Holz, Thorsten and Kotter, Markus and Wicherski, Georg},
  ``Know your enemy: Tracking botnets,''
  \url{https://www.honeynet.org/papers/bots/}, 2005, last accessed December
  2015.

\bibitem{wicherski2010placing}
G.~Wicherski, ``Placing a low-interaction honeypot in-the-wild: A review of
  mwcollectd,'' \emph{Network Security}, vol. 2010, no.~3, pp. 7--8, 2010.

\bibitem{RFC-5321}
J.~Klensin, ``{Simple Mail Transfer Protocol},'' IETF, RFC 5321, October 2008.

\bibitem{grimes2005honeypots}
R.~A. Grimes, \emph{Honeypots for Windows}.\hskip 1em plus 0.5em minus
  0.4em\relax Apress, 2005.

\bibitem{SpamDBSD}
{OpenBSD}, ``spamd - spam deferral daemon,''
  \url{http://www.openbsd.org/cgi-bin/man.cgi/OpenBSD-current/man8/spamd.8},
  last accessed on December 2015.

\bibitem{HoneypotNETOverview}
{Honeypots.NET}, ``Honeypot software, honeypot products, deception software,''
  \url{http://www.honeypots.net/honeypots/products}, last accessed on January
  2016.

\bibitem{SMTPotMirror}
{Karl A. Krueger}, ``Smtpot - a simple smtp honeypot,''
  \url{http://mirror.unpad.ac.id/orari/library/library-sw-hw/linux-1/HONEYPOTS/smtpot/},
  last accessed on January 2016.

\bibitem{SpamHole}
{Druid Cau}, ``Spamhole,'' \url{http://sourceforge.net/projects/spamhole/},
  last accessed on January 2016.

\bibitem{Spampot}
{Neale Pikett}, ``Spampot,''
  \url{https://mail.python.org/pipermail/spambayes/2003-January/002883.html},
  last accessed on December 2015.

\bibitem{y-ssh-95}
T.~Yl{\"o}nen, ``Announcement: Ssh (secure shell) remote login program,''
  Newsgroups: comp.security.unix, talk.politics.crypto.
  \url{https://groups.google.com/group/comp.security.unix/msg/67079d812a19f499},
  1995.

\bibitem{RFC-4251}
T.~Ylonen and C.~Lonvick, ``{The Secure Shell (SSH) Protocol Architecture},''
  IETF, RFC 4251, January 2006.

\bibitem{KojoneySourceforge}
{Jose Antonio Coret}, ``Kojoney - a honeypot for the ssh service,''
  \url{http://kojoney.sourceforge.net/}, last accessed on January 2016.

\bibitem{Kojoney2Github}
{University of Pennsylvania's School of Arts \& Sciences, Justin Klein Keane},
  ``Kojoey2 ssh medium-interaction honeypot,''
  \url{https://github.com/madirish/kojoney2}, last accessed on December 2015.

\bibitem{KippoGithub}
{Upi Tamminen (desaster)}, ``Kippo, ssh medium-interaction honeypot,''
  \url{https://github.com/desaster/kippo/}, last accessed on December 2015.

\bibitem{CowrieGithub}
{Github Michel Oosterhof}, ``Cowrie - medium-interaction honeypot,''
  \url{https://github.com/micheloosterhof/cowrie}, last accessed on December
  2015.

\bibitem{GlastopfHomepage}
{Glastopf Web Honeypot Project, Lukas Rist}, ``Glastopf honeypot,''
  \url{glastopf.org/}, last accessed on January 2016.

\bibitem{HoneyWebHTTP}
{Kevin Timm}, ``Honeyweb - http honeypot,''
  \url{http://www.citi.umich.edu/u/provos/honeyd/contrib/ktimm/}, last accessed
  on December 2015.

\bibitem{elastichoneyHP}
{Jordan Wright.}, ``elastichoney,''
  \url{https://github.com/jordan-wright/elastichoney}, last accessed February
  2016.

\bibitem{GoogleHackHoneypot}
{Ryan McGeehan Greg Smith Brian Engert Kevin Reedy Kevin Benes}, ``Google hack
  honeypot,'' \url{http://ghh.sourceforge.net/}, last accessed on January 2016.

\bibitem{oms-whapu-16}
J.~Onaolapo, E.~Mariconti, and G.~Stringhini, ``{What Happens After You Are
  Pwnd: Understanding The Use Of Leaked Account Credentials In The Wild},'' in
  \emph{Proc. of the ACM Internet Measurement Conference (IMC)}.\hskip 1em plus
  0.5em minus 0.4em\relax New York, NY, USA: ACM, 2016.

\bibitem{gruber2015global}
M.~Gruber, D.~Hoffstadt, A.~Aziz, F.~Fankhauser, C.~Schanes, E.~Rathgeb, and
  T.~Grechenig, ``Global voip security threats-large scale validation based on
  independent honeynets,'' in \emph{IFIP Networking Conference (IFIP
  Networking), 2015}.\hskip 1em plus 0.5em minus 0.4em\relax IEEE, 2015, pp.
  1--9.

\bibitem{carmo2011artemisa}
R.~D. Carmo, M.~Nassar, and O.~Festor, ``Artemisa: An open-source honeypot
  back-end to support security in voip domains,'' in \emph{Integrated Network
  Management (IM), 2011 IFIP/IEEE International Symposium on}.\hskip 1em plus
  0.5em minus 0.4em\relax IEEE, 2011, pp. 361--368.

\bibitem{ph-vhfbt-08}
N.~Provos and T.~Holz, \emph{{Virtual Honeypots. From Botnet Tracking to
  Intrusion Detection}}, 2nd~ed.\hskip 1em plus 0.5em minus 0.4em\relax Upper
  Saddle River, NJ: Addison--Wesley, 2008.

\bibitem{k-mmhtu-04}
B.~Krishnamurthy, ``{Mohonk: MObile Honeypots to Trace Unwanted Traffic
  Early},'' in \emph{Proc. of the ACM SIGCOMM Workshop on Network
  Troubleshooting (NetT)}.\hskip 1em plus 0.5em minus 0.4em\relax New York, NY,
  USA: ACM, 2004, pp. 277--282.

\bibitem{wtkss-fimh-12}
M.~W{\"a}hlisch, S.~Trapp, C.~Keil, J.~Sch{\"o}nfelder, T.~C. Schmidt, and
  J.~Schiller, ``{First Insights from a Mobile Honeypot},'' \emph{ACM Computer
  Communication Review (CCR)}, vol.~42, no.~4, pp. 305--306, August 2012.

\bibitem{wvkss-diomh-13}
\BIBentryALTinterwordspacing
M.~W{\"a}hlisch, A.~Vorbach, C.~Keil, J.~Sch{\"o}nfelder, T.~C. Schmidt, and
  J.~H. Schiller, ``Design, implementation, and operation of a mobile
  honeypot,'' Open Archive: arXiv.org, Technical Report arXiv:1205.4778, 2013.
  [Online]. Available: \url{http://arxiv.org/abs/1301.7257}
\BIBentrySTDinterwordspacing

\bibitem{honeynetproject}
\BIBentryALTinterwordspacing
honeynet Project, ``{The Honeynet Project Chinese Chapter Status Report (Period
  Apr 2007 to Dec 2008)},'' 2009. [Online]. Available:
  \url{http://www.honeynet.org/node/336}
\BIBentrySTDinterwordspacing

\bibitem{Github:FakeAP}
{2005 Black Alchemy Enterprises, DEFCON X Proof of Concept}, ``Fakeap,''
  \url{https://github.com/tbennett6421/Black-Alchemy---FakeAP}, last accessed
  on December 2015.

\bibitem{cs-hfivh-10}
T.~OConnor and B.~Sangster, ``{honeyM: A Framework for Implementing Virtual
  Honeyclients for Mobile Devices},'' in \emph{Proc. of the third ACM
  WiSec}.\hskip 1em plus 0.5em minus 0.4em\relax New York, NY, USA: ACM, 2010,
  pp. 129--138.

\bibitem{bluepotGithub}
{Andrew Michael Smith}, ``bluepot,''
  \url{https://github.com/andrewmichaelsmith/bluepot}, last accessed February
  2016.

\bibitem{s-hwhma-07}
\BIBentryALTinterwordspacing
R.~Siles, ``{HoneySpot: The Wireless Honeypot. Monitoring the Attacker's
  Activities in Wireless Networks. A design and architectural overview},'' The
  Spanish Honeynet Project, Research Project, December 2007. [Online].
  Available:
  \url{http://honeynet.org.es/papers/honeyspot/HoneySpot_20071217.pdf}
\BIBentrySTDinterwordspacing

\bibitem{mulliner2011poster}
C.~Mulliner, S.~Liebergeld, and M.~Lange, ``Poster: Honeydroid-creating a
  smartphone honeypot,'' in \emph{IEEE Symposium on Security and Privacy},
  2011.

\bibitem{RFC-7228}
C.~Bormann, M.~Ersue, and A.~Keranen, ``{Terminology for Constrained-Node
  Networks},'' IETF, RFC 7228, May 2014.

\bibitem{IoTPOTPaper}
\BIBentryALTinterwordspacing
Y.~M.~P. Pa, S.~Suzuki, K.~Yoshioka, T.~Matsumoto, T.~Kasama, and C.~Rossow,
  ``Iotpot: Analysing the rise of iot compromises,'' in \emph{9th USENIX
  Workshop on Offensive Technologies (WOOT 15)}.\hskip 1em plus 0.5em minus
  0.4em\relax Washington, D.C.: USENIX Association, Aug. 2015. [Online].
  Available:
  \url{https://www.usenix.org/conference/woot15/workshop-program/presentation/pa}
\BIBentrySTDinterwordspacing

\bibitem{cameraHP}
{Alex Bredo, EDAG Production Solutions, Github.}, ``honeypot-camera,''
  \url{https://github.com/alexbredo/honeypot-camera}, last accessed February
  2016.

\bibitem{ShockpotGithub}
{ThreatStream}, ``Shockpot,'' \url{https://github.com/threatstream/shockpot},
  last accessed February 2016.

\bibitem{ShellshockCVE}
{Common Vulnerabilities and Exposures (CVE), National Vulnerability Database
  (NVD)}, ``Cve-2014-6271,''
  \url{https://cve.mitre.org/cgi-bin/cvename.cgi?name=CVE-2014-6271}, last
  accessed February 2016.

\bibitem{conpotHoneypot}
{MushMush Foundation at Github}, ``Conpot,''
  \url{https://github.com/mushorg/conpot}, last accessed February 2016.

\bibitem{HoneyPointServer}
{MicroSolved, Inc.}, ``Honeypoint security server,''
  \url{http://microsolved.com/HoneyPoint-server.html}, last accessed on
  December 2015.

\bibitem{HoneySinkHP}
{Google Summer of Code 2011}, ``Network sinkhole, honeysink,''
  \url{https://redmine.honeynet.org/projects/sinkhole}, last accessed January
  2016.

\bibitem{Argos:2006Qemu}
\BIBentryALTinterwordspacing
G.~Portokalidis, A.~Slowinska, and H.~Bos, ``Argos: An emulator for
  fingerprinting zero-day attacks for advertised honeypots with automatic
  signature generation,'' in \emph{Proceedings of the 1st ACM SIGOPS/EuroSys
  European Conference on Computer Systems 2006}, ser. EuroSys '06.\hskip 1em
  plus 0.5em minus 0.4em\relax New York, NY, USA: ACM, 2006, pp. 15--27.
  [Online]. Available: \url{http://doi.acm.org/10.1145/1217935.1217938}
\BIBentrySTDinterwordspacing

\bibitem{HoneywallCD}
{Honeynet Project}, ``Honeywall cdrom project documentation,''
  \url{https://www.honeynet.org/project/HoneywallCDROM}, last accessed on
  December 2015.

\bibitem{genericPHPhoneypot}
M.~Mueter, F.~Freiling, T.~Holz, and J.~Matthews, ``A generic toolkit for
  converting web applications into high-interaction honeypots,''
  \emph{University of Mannheim}, vol. 280, 2008.

\bibitem{homepagePHPhoneypot}
{Mueter, Michael and Freiling, Felix and Holz, Thorsten and Matthews, Jeanna},
  ``High interaction honeypot analysis tool,''
  \url{http://sourceforge.net/projects/hihat/}, last accessed on December 2015.

\bibitem{honeybowSensor}
{Chinese Honeynet project}, ``Honeybow sensor,''
  \url{http://sourceforge.net/projects/honeybow/}, last accessed on December
  2015.

\bibitem{SebekHP}
{Honeynet Project}, ``Sebek - a data capture tool,''
  \url{https://projects.honeynet.org/sebek/}, last accessed December 2015.

\bibitem{SebekArticle}
T.~H. Project, ``Sebek a kernel based data capture tool,'' \emph{Know Your
  Enemy Journal}, November 2003.

\bibitem{CanarytokensGithub}
{Thinkst Applied Research}, ``Github canarytokens-docker,''
  \url{https://github.com/thinkst/canarytokens-docker}, last accessed July
  2016.

\bibitem{HoneytokensSpitzner}
{Lance Spitzner, July 2003}, ``Honeytokens: The other honeypot,''
  \url{http://www.symantec.com/connect/articles/honeytokens-other-honeypot},
  last accessed July 2016.

\bibitem{CanarytokensService}
{Thinkst Applied Research}, ``Service canarytokens,''
  \url{http://canarytokens.org/generate}, last accessed July 2016.

\bibitem{phoneycPaper}
J.~Nazario, ``Phoneyc: A virtual client honeypot,'' in \emph{Proceedings of the
  2nd USENIX conference on Large-scale exploits and emergent threats: botnets,
  spyware, worms, and more}.\hskip 1em plus 0.5em minus 0.4em\relax USENIX
  Association, 2009, pp. 6--6.

\bibitem{phoneycHomepage}
{J Nazario}, ``phoneyc python honeyclient,''
  \url{https://github.com/honeynet/phoneyc}, last accessed December 2015.

\bibitem{HoneyCHome}
{The Honeynet Project}, ``Honeyc,'' \url{https://projects.honeynet.org/honeyc},
  last accessed January 2016.

\bibitem{ThugGithub}
{Angelo Dell'Aera}, ``Thug honeypot,'' \url{https://github.com/buffer/thug/},
  last accessed on January 2016.

\bibitem{Paper:YALIH}
\BIBentryALTinterwordspacing
M.~Mansoori, I.~Welch, and Q.~Fu, ``Yalih, yet another low interaction
  honeyclient,'' in \emph{Proceedings of the Twelfth Australasian Information
  Security Conference - Volume 149}, ser. AISC '14.\hskip 1em plus 0.5em minus
  0.4em\relax Darlinghurst, Australia, Australia: Australian Computer Society,
  Inc., 2014, pp. 7--15. [Online]. Available:
  \url{http://dl.acm.org/citation.cfm?id=2667510.2667512}
\BIBentrySTDinterwordspacing

\bibitem{YalihGithub}
{Victoria University of Wellington (Masood Mansoori) Singapore Polytechnic (Lai
  Qi Wei, Ritchie Lam Qiaowei)}, ``Yalih - yet another low interaction
  honeyclient,'' \url{https://github.com/Masood-M/YALIH}, last accessed on
  January 2016.

\bibitem{SpyByeHP}
{Niels Provos}, ``Spybye - finding malware,''
  \url{http://www.monkey.org/~provos/spybye/}, last accessed January 2016.

\bibitem{MonkeySpiderHP}
{Ali Ikinci}, ``The monkey-spider project,''
  \url{http://monkeyspider.sourceforge.net/}, last accessed January 2016.

\bibitem{dewald2010adsandbox}
A.~Dewald, T.~Holz, and F.~C. Freiling, ``Adsandbox: Sandboxing javascript to
  fight malicious websites,'' in \emph{Proceedings of the 2010 ACM Symposium on
  Applied Computing}.\hskip 1em plus 0.5em minus 0.4em\relax ACM, 2010, pp.
  1859--1864.

\bibitem{CaptureBat}
{New Zealand Honeynet Project and Victoria University of Wellington, Christian
  Seifert}, ``Capture-bat download page,''
  \url{https://www.honeynet.org/node/315} (reupload), last accessed on December
  2015.

\bibitem{HoneyClientGit}
{MITRE}, ``Honeyclient project,''
  \url{https://github.com/dkindlund/honeyclient}, last accessed December 2015.

\bibitem{wang2005Presi}
K.~Wang, ``Using honeyclients to detect new attacks,'' in \emph{RECON
  Conference (June 2005)}, 2005.

\bibitem{CaptureHPCHP}
{honeynet project, Christian Seifert and Ramon Steenson of the New Zealand
  Chapter}, ``Capture-hpc client honeypot,''
  \url{https://projects.honeynet.org/capture-hpc}, last accessed December 2015.

\bibitem{HoneyMonkeyHP}
{Microsoft Research Team}, ``Strider honeymonkey exploit detection,''
  \url{http://research.microsoft.com/en-us/um/redmond/projects/strider/honeymonkey/},
  last accessed January 2016.

\bibitem{wang2006automated}
Y.-M. Wang, D.~Beck, X.~Jiang, R.~Roussev, C.~Verbowski, S.~Chen, and S.~King,
  ``Automated web patrol with strider honeymonkeys,'' in \emph{Proceedings of
  the 2006 Network and Distributed System Security Symposium}, 2006, pp.
  35--49.

\bibitem{TrigonaHP}
{Australian Honeynet Project}, ``Trigona,''
  \url{http://honeynet.org.au/?q=node/63}, last accessed on December 2015 using
  the web archive.

\bibitem{HoneyspiderNetwork}
{NASK/CERT Polska, National Cyber Security Centre (Netherlands)}, ``Honeyspider
  network,'' \url{http://www.honeyspider.net/}, last accessed on December 2015.

\bibitem{ShielaHP}
{Joan Robert Rocaspana at Vrije Universiteit Amsterdam, Herbert Bos}, ``Shelia:
  a client-side honeypot for attack detection,''
  \url{https://www.cs.vu.nl/~herbertb/misc/shelia/}, last accessed on January
  2016.

\bibitem{UW2006spycrawler}
A.~Moshchuk, T.~Bragin, S.~D. Gribble, and H.~M. Levy, ``A crawler-based study
  of spyware in the web.'' in \emph{NDSS}, 2006.

\bibitem{WEFhomepage}
{Thomas Müller Benjamin Mack Mehmet Arziman, Hochschule der Medien (HdM),
  Stuttgart}, ``Web exploit finder,''
  \url{http://www.xnos.org/security/web-exploit-finder.html}, last accessed
  January 2016.

\bibitem{xie2007honeyim}
M.~Xie, Z.~Wu, and H.~Wang, ``Honeyim: Fast detection and suppression of
  instant messaging malware in enterprise-like networks,'' in \emph{Computer
  Security Applications Conference, 2007. ACSAC 2007. Twenty-Third
  Annual}.\hskip 1em plus 0.5em minus 0.4em\relax IEEE, 2007, pp. 64--73.

\bibitem{pwnypotGit}
{Shahriyar Jalayeri, TOBIAS JARMUZEK}, ``Pwnypot, high interaction client
  honeypot,'' \url{https://github.com/shjalayeri/pwnypot}, last accessed
  January 2016.

\bibitem{baitNswitch2003}
{Team Violating.}, ``Bait and switch honeypot,''
  \url{http://baitnswitch.sourceforge.net/}, last accessed on December 2015.

\bibitem{HoneynetSecurityConsole}
{Activeworx, Inc.}, ``Honeynet security console,''
  \url{http://seclists.org/focus-ids/2004/May/74}, last accessed on December
  2015.

\bibitem{Honeyweb2009}
{Google Summer of Code GSOC 2009}, ``Honeyweb,''
  \url{https://code.google.com/p/gsoc-honeyweb/}, last accessed on January
  2016.

\bibitem{honeysnapHP}
{The Honeynet Project}, ``Honeysnap,''
  \url{https://projects.honeynet.org/honeysnap/}, last accessed on January
  2016.

\bibitem{PEHunterHP}
{Carnivore Project (Tillmann Werner)}, ``Pe hunter,''
  \url{http://src.carnivore.it/pehunter/}.

\bibitem{HoneyMoleHP}
{Honeynet Project, Portugese Chapter}, ``Honeymole,''
  \url{http://www.honeynet.org.pt/index.php/HoneyMole}, last accessed on
  January 2016 using the web archive.

\bibitem{dionaeaDetection}
{C(yber) Brian}, ``Dionaea honeypot obfuscation,''
  \url{https://www.cyberbrian.net/2014/09/dionaea-honeypot-obfuscation/}, last
  accessed January 2016.

\bibitem{kippoDetection}
{Andrew Morris}, ``Detecting kippo ssh honeypots, bypassing patches, and all
  that jazz.'' \url{http://morris.guru/detecting-kippo-ssh-honeypots/}, last
  accessed December 2015.

\bibitem{fu2006recognizing}
X.~Fu, W.~Yu, D.~Cheng, X.~Tan, K.~Streff, and S.~Graham, ``On recognizing
  virtual honeypots and countermeasures,'' in \emph{Dependable, Autonomic and
  Secure Computing, 2nd IEEE International Symposium on}.\hskip 1em plus 0.5em
  minus 0.4em\relax IEEE, 2006, pp. 211--218.

\bibitem{holz2005detecting}
T.~Holz and F.~Raynal, ``Detecting honeypots and other suspicious
  environments,'' in \emph{Information Assurance Workshop, 2005. IAW'05.
  Proceedings from the Sixth Annual IEEE SMC}.\hskip 1em plus 0.5em minus
  0.4em\relax IEEE, 2005, pp. 29--36.

\bibitem{mukkamala2007detection}
S.~Mukkamala, K.~Yendrapalli, R.~Basnet, M.~Shankarapani, and A.~Sung,
  ``Detection of virtual environments and low interaction honeypots,'' in
  \emph{Information Assurance and Security Workshop, 2007. IAW'07. IEEE
  SMC}.\hskip 1em plus 0.5em minus 0.4em\relax IEEE, 2007, pp. 92--98.

\bibitem{zou2006honeypot}
C.~C. Zou and R.~Cunningham, ``Honeypot-aware advanced botnet construction and
  maintenance,'' in \emph{Dependable Systems and Networks, 2006. DSN 2006.
  International Conference on}.\hskip 1em plus 0.5em minus 0.4em\relax IEEE,
  2006, pp. 199--208.

\bibitem{wang2010honeypot}
P.~Wang, L.~Wu, R.~Cunningham, and C.~C. Zou, ``Honeypot detection in advanced
  botnet attacks,'' \emph{International Journal of Information and Computer
  Security}, vol.~4, no.~1, pp. 30--51, 2010.

\bibitem{phrackHoneyDetection}
{Joseph Corey}, ``Local honeypot identification, phrack (unofficial), volume
  0x0b, issue 0x3e,''
  \url{http://repo.hackerzvoice.net/depot_ouah/p62-0x07.txt}, last accessed
  January 2016.

\bibitem{redPillDetection}
{Joanna Rutkowska}, ``Red pill... or how to detect vmm using (almost) one cpu
  instruction,''
  \url{http://repo.hackerzvoice.net/depot_ouah/Red_%20Pill.html}, last accessed
  January 2016.

\bibitem{honeynet2004knowUrEnemy}
H.~Project, \emph{Know Your Enemy: Learning about Security Threats}.\hskip 1em
  plus 0.5em minus 0.4em\relax Addison-Wesley, 2004.

\bibitem{NoahProject}
{Foundation for Research and Technology Hellas (FORTH) and Partners: DFN-Cert
  Alcatel ETHZ TERENA Virtual Trip VU)}, ``European network of affined
  honeypots, noah-project,''
  \url{https://www.fp6-noah.org/publications/index.html}, last accessed January
  2016.

\bibitem{allianceMwcollectHP}
{Based on work of Paul Bächer, Markus Kötter and Georg Wicherski.},
  ``mwcollect alliance collaborative malware collection and sensing,''
  \url{https://alliance.mwcollect.org/}, last accessed January 2016.

\bibitem{wicherski2006medium}
G.~Wicherski, ``Medium interaction honeypots,'' \emph{German Honeynet Project},
  2006.

\bibitem{sicherheitstacho}
{Telekom , DTAG}, ``Frühwarnsystem, sicherheitstacho,''
  \url{http://www.sicherheitstacho.eu/}, last accessed January 2016.

\bibitem{pouget2005advantages:seq}
F.~Pouget, M.~Dacier, and V.~Pham, ``Leurre.com: On the advantages of deploying
  a large scale distributed honeypot platform,'' in \emph{Proceedings of the
  E-Crime and Computer Evidence Conference}, 2005.

\bibitem{pouget2004honeypot}
F.~Pouget, M.~Dacier \emph{et~al.}, ``Honeypot-based forensics,'' in
  \emph{AusCERT Asia Pacific Information Technology Security Conference}, 2004.

\bibitem{mcgrew2006experiences}
R.~McGrew, ``Experiences with honeypot systems: Development, deployment, and
  analysis,'' in \emph{System Sciences, 2006. HICSS'06. Proceedings of the 39th
  Annual Hawaii International Conference on}, vol.~9.\hskip 1em plus 0.5em
  minus 0.4em\relax IEEE, 2006, pp. 220a--220a.

\bibitem{francois2008graphs}
J.~Francois, O.~Festor \emph{et~al.}, ``Activity monitoring for large honeynets
  and network telescopes,'' \emph{International Journal on Advances in Systems
  and Measurements}, vol.~1, no.~1, pp. 1--13, 2008.

\bibitem{kyotoDataset}
J.~Song, H.~Takakura, Y.~Okabe, M.~Eto, D.~Inoue, and K.~Nakao, ``Statistical
  analysis of honeypot data and building of kyoto 2006+ dataset for nids
  evaluation,'' in \emph{Proceedings of the First Workshop on Building Analysis
  Datasets and Gathering Experience Returns for Security}.\hskip 1em plus 0.5em
  minus 0.4em\relax ACM, 2011, pp. 29--36.

\bibitem{trivedi2007analyzing}
A.~J. Trivedi, P.~Q. Judge, and S.~Krasser, ``Analyzing network and content
  characteristics of spim using honeypots,'' in \emph{Proceedings of the 3rd
  USENIX SRUTI}, 2007.

\bibitem{valli2010voip}
C.~Valli, ``An analysis of malfeasant activity directed at a voip honeypot,''
  \emph{Proceedings of the 8th Australian Digital Forensics Conference}, 2010.

\bibitem{yegneswaran2004design}
V.~Yegneswaran, P.~Barford, and D.~Plonka, ``On the design and use of internet
  sinks for network abuse monitoring,'' in \emph{Recent Advances in Intrusion
  Detection}.\hskip 1em plus 0.5em minus 0.4em\relax Springer, 2004, pp.
  146--165.

\bibitem{kaaniche2007empirical}
M.~Kaaniche, Y.~Deswarte, E.~Alata, M.~Dacier, and V.~Nicomette, ``Empirical
  analysis and statistical modeling of attack processes based on honeypots,''
  \emph{arXiv preprint arXiv:0704.0861}, 2007.

\bibitem{almotairi2008principalComp}
S.~Almotairi, A.~Clark, G.~Mohay, and J.~Zimmermann, ``Characterization of
  attackers' activities in honeypot traffic using principal component
  analysis,'' in \emph{Network and Parallel Computing, 2008. NPC 2008. IFIP
  International Conference on}.\hskip 1em plus 0.5em minus 0.4em\relax IEEE,
  2008, pp. 147--154.

\bibitem{thonnard2008framework}
O.~Thonnard and M.~Dacier, ``A framework for attack patterns' discovery in
  honeynet data,'' \emph{digital investigation}, vol.~5, pp. S128--S139, 2008.

\bibitem{durumeric2013zmap}
Z.~Durumeric, E.~Wustrow, and J.~A. Halderman, ``Zmap: Fast internet-wide
  scanning and its security applications.'' in \emph{Usenix Security}, 2013,
  pp. 605--620.

\bibitem{jain1988algorithms}
A.~K. Jain and R.~C. Dubes, \emph{Algorithms for clustering data}.\hskip 1em
  plus 0.5em minus 0.4em\relax Prentice-Hall, Inc., 1988.

\bibitem{jain2010data}
A.~K. Jain, ``Data clustering: 50 years beyond k-means,'' \emph{Pattern
  recognition letters}, vol.~31, no.~8, pp. 651--666, 2010.

\bibitem{Barford:NetworkAnomalies}
\BIBentryALTinterwordspacing
P.~Barford, J.~Kline, D.~Plonka, and A.~Ron, ``A signal analysis of network
  traffic anomalies,'' in \emph{Proceedings of the 2Nd ACM SIGCOMM Workshop on
  Internet Measurment}, ser. IMW '02.\hskip 1em plus 0.5em minus 0.4em\relax
  New York, NY, USA: ACM, 2002, pp. 71--82. [Online]. Available:
  \url{http://doi.acm.org/10.1145/637201.637210}
\BIBentrySTDinterwordspacing

\bibitem{Brutlag:AbberantBehaviour}
\BIBentryALTinterwordspacing
J.~D. Brutlag, ``Aberrant behavior detection in time series for network
  monitoring,'' in \emph{Proceedings of the 14th USENIX Conference on System
  Administration}, ser. LISA '00.\hskip 1em plus 0.5em minus 0.4em\relax
  Berkeley, CA, USA: USENIX Association, 2000, pp. 139--146. [Online].
  Available: \url{http://dl.acm.org/citation.cfm?id=1045502.1045530}
\BIBentrySTDinterwordspacing

\bibitem{Lakhina:CharacterizationFlow}
\BIBentryALTinterwordspacing
A.~Lakhina, M.~Crovella, and C.~Diot, ``Characterization of network-wide
  anomalies in traffic flows,'' in \emph{Proceedings of the 4th ACM SIGCOMM
  Conference on Internet Measurement}, ser. IMC '04.\hskip 1em plus 0.5em minus
  0.4em\relax New York, NY, USA: ACM, 2004, pp. 201--206. [Online]. Available:
  \url{http://doi.acm.org/10.1145/1028788.1028813}
\BIBentrySTDinterwordspacing

\bibitem{Lakhina:MiningAnomalies}
\BIBentryALTinterwordspacing
------, ``Mining anomalies using traffic feature distributions,'' \emph{SIGCOMM
  Comput. Commun. Rev.}, vol.~35, no.~4, pp. 217--228, Aug. 2005. [Online].
  Available: \url{http://doi.acm.org/10.1145/1090191.1080118}
\BIBentrySTDinterwordspacing

\bibitem{portnoy2001intrusion}
L.~Portnoy, E.~Eskin, and S.~Stolfo, ``Intrusion detection with unlabeled data
  using clustering,'' in \emph{In Proceedings of ACM CSS Workshop on Data
  Mining Applied to Security (DMSA-2001}.\hskip 1em plus 0.5em minus
  0.4em\relax Citeseer, 2001.

\bibitem{eskin2002geometric}
E.~Eskin, A.~Arnold, M.~Prerau, L.~Portnoy, and S.~Stolfo, ``A geometric
  framework for unsupervised anomaly detection,'' in \emph{Applications of data
  mining in computer security}.\hskip 1em plus 0.5em minus 0.4em\relax
  Springer, 2002, pp. 77--101.

\bibitem{Owezarski:ClusterEnsemble}
\BIBentryALTinterwordspacing
P.~Owezarski, ``A near real-time algorithm for autonomous identification and
  characterization of honeypot attacks,'' in \emph{Proceedings of the 10th ACM
  Symposium on Information, Computer and Communications Security}, ser. ASIA
  CCS '15.\hskip 1em plus 0.5em minus 0.4em\relax New York, NY, USA: ACM, 2015,
  pp. 531--542. [Online]. Available:
  \url{http://doi.acm.org/10.1145/2714576.2714580}
\BIBentrySTDinterwordspacing

\bibitem{krasser2005real}
S.~Krasser, G.~Conti, J.~Grizzard, J.~Gribschaw, and H.~Owen, ``Real-time and
  forensic network data analysis using animated and coordinated
  visualization,'' in \emph{Information Assurance Workshop, 2005. IAW'05.
  Proceedings from the Sixth Annual IEEE SMC}.\hskip 1em plus 0.5em minus
  0.4em\relax IEEE, 2005, pp. 42--49.

\bibitem{rencher2003methods}
A.~C. Rencher, \emph{Methods of multivariate analysis}.\hskip 1em plus 0.5em
  minus 0.4em\relax John Wiley \& Sons, 2003, vol. 492.

\bibitem{Owezarski:OriginalAlgo}
\BIBentryALTinterwordspacing
J.~Mazel, P.~Casas, and P.~Owezarski, ``Sub-space clustering and evidence
  accumulation for unsupervised network anomaly detection,'' in
  \emph{Proceedings of the Third International Conference on Traffic Monitoring
  and Analysis}, ser. TMA'11.\hskip 1em plus 0.5em minus 0.4em\relax Berlin,
  Heidelberg: Springer-Verlag, 2011, pp. 15--28. [Online]. Available:
  \url{http://dl.acm.org/citation.cfm?id=1986282.1986285}
\BIBentrySTDinterwordspacing

\bibitem{Agrawal:SubspaceRule}
\BIBentryALTinterwordspacing
R.~Agrawal, J.~Gehrke, D.~Gunopulos, and P.~Raghavan, ``Automatic subspace
  clustering of high dimensional data for data mining applications,''
  \emph{SIGMOD Rec.}, vol.~27, no.~2, pp. 94--105, Jun. 1998. [Online].
  Available: \url{http://doi.acm.org/10.1145/276305.276314}
\BIBentrySTDinterwordspacing

\bibitem{gregio2007evaluation}
A.~Gr{\'e}gio, R.~Santos, and A.~Montes, ``Evaluation of data mining techniques
  for suspicious network activity classification using honeypots data,'' in
  \emph{Defense and Security Symposium}.\hskip 1em plus 0.5em minus 0.4em\relax
  International Society for Optics and Photonics, 2007, pp. 657\,006--657\,006.

\bibitem{Haykin1998Book}
S.~Haykin, \emph{Neural Networks: A Comprehensive Foundation}, 2nd~ed.\hskip
  1em plus 0.5em minus 0.4em\relax Upper Saddle River, NJ, USA: Prentice Hall
  PTR, 1998.

\bibitem{Quinlan1993Book}
J.~R. Quinlan, \emph{C4.5: Programs for Machine Learning}.\hskip 1em plus 0.5em
  minus 0.4em\relax San Francisco, CA, USA: Morgan Kaufmann Publishers Inc.,
  1993.

\bibitem{Kreibich2004Honeycomb}
\BIBentryALTinterwordspacing
C.~Kreibich and J.~Crowcroft, ``Honeycomb: Creating intrusion detection
  signatures using honeypots,'' \emph{SIGCOMM Comput. Commun. Rev.}, vol.~34,
  no.~1, pp. 51--56, Jan. 2004. [Online]. Available:
  \url{http://doi.acm.org/10.1145/972374.972384}
\BIBentrySTDinterwordspacing

\bibitem{ukkonen1995suffix}
E.~Ukkonen, ``On-line construction of suffix trees,'' \emph{Algorithmica},
  vol.~14, no.~3, pp. 249--260, 1995.

\bibitem{newsome2005polygraph}
J.~Newsome, B.~Karp, and D.~Song, ``Polygraph: Automatically generating
  signatures for polymorphic worms,'' in \emph{Security and Privacy, 2005 IEEE
  Symposium on}.\hskip 1em plus 0.5em minus 0.4em\relax IEEE, 2005, pp.
  226--241.

\bibitem{tang2005doubleHP}
Y.~Tang and S.~Chen, ``Defending against internet worms: A signature-based
  approach,'' in \emph{INFOCOM 2005. 24th Annual Joint Conference of the IEEE
  Computer and Communications Societies. Proceedings IEEE}, vol.~2.\hskip 1em
  plus 0.5em minus 0.4em\relax IEEE, 2005, pp. 1384--1394.

\bibitem{mohammed2010polymorphic}
M.~M. Mohammed, H.~A. Chan, N.~Ventura, M.~Hashim, I.~Amin, and E.~Bashier,
  ``Detection of zero-day polymorphic worms using principal component
  analysis,'' in \emph{Networking and Services (ICNS), 2010 Sixth International
  Conference on}.\hskip 1em plus 0.5em minus 0.4em\relax IEEE, 2010, pp.
  277--281.

\bibitem{dacier2004attack}
M.~Dacier, F.~Pouget, and H.~Debar, ``Attack processes found on the internet,''
  DTIC Document, Tech. Rep., 2004.

\bibitem{kohlrausch2009experiences}
J.~Kohlrausch, ``Experiences with the noah honeynet testbed to detect new
  internet worms,'' in \emph{IT Security Incident Management and IT Forensics,
  2009. IMF'09. Fifth International Conference on}.\hskip 1em plus 0.5em minus
  0.4em\relax IEEE, 2009, pp. 13--26.

\bibitem{alata2007lessons}
E.~Alata, V.~Nicomette, M.~Dacier, M.~Herrb \emph{et~al.}, ``Lessons learned
  from the deployment of a high-interaction honeypot,'' \emph{arXiv preprint
  arXiv:0704.0858}, 2007.

\bibitem{pouget2008understanding}
F.~Pouget, M.~Dacier, and V.-H. Pham, ``Understanding threats: a prerequisite
  to enhance survivability of computing systems,'' \emph{International Journal
  of Critical Infrastructures}, vol.~4, no. 1-2, pp. 153--171, 2008.

\bibitem{li2008survey}
P.~Li, M.~Salour, and X.~Su, ``A survey of internet worm detection and
  containment,'' \emph{Communications Surveys \& Tutorials, IEEE}, vol.~10,
  no.~1, pp. 20--35, 2008.

\bibitem{portokalidis2007sweetbait}
G.~Portokalidis and H.~Bos, ``Sweetbait: Zero-hour worm detection and
  containment using low-and high-interaction honeypots,'' \emph{Computer
  Networks}, vol.~51, no.~5, pp. 1256--1274, 2007.

\bibitem{HoneypotLegalIssues}
{Lance Spitzner}, ``The value of honeypots, part two: Honeypot solutions and
  legal issues,''
  \url{http://www.symantec.com/connect/articles/value-honeypots-part-two-honeypot-solutions-and-legal-issues},
  last accessed on January 2016.

\bibitem{TKGGermany}
{Bundesministerium der Justiz und für Verbraucherschutz},
  ``Telekommunikationsgesetz (tkg), 100, stoerungen von
  telekommunikationsanlagen und missbrauch von telekommunikationsdiensten,''
  \url{https://www.gesetze-im-internet.de/tkg_2004/__100.html}, last accessed
  July 2016.

\bibitem{BGHAZ}
{Bundesgerichtshof}, ``Urteil, az iii zr 391/13, 3. juli 2014,''
  \url{http://juris.bundesgerichtshof.de/cgi-bin/rechtsprechung/document.py?Gericht=bgh&Art=en&nr=68350&pos=0&anz=},
  last accessed July 2016.

\bibitem{Dantu2007FWC}
\BIBentryALTinterwordspacing
R.~Dantu, J.~W. Cangussu, and S.~Patwardhan, ``Fast worm containment using
  feedback control,'' \emph{IEEE Trans. Dependable Secur. Comput.}, vol.~4,
  no.~2, pp. 119--136, Apr. 2007. [Online]. Available:
  \url{http://dx.doi.org/10.1109/TDSC.2007.1002}
\BIBentrySTDinterwordspacing

\bibitem{HackerParagraph}
{German laws.}, ``Deutsch strafgesetzbuch - german criminal law code.''
  \url{https://dejure.org/gesetze/StGB/202c.html}, last accessed December 2015.

\end{thebibliography}

\end{document}